\documentclass[letterpaper,11pt]{article}
\usepackage{fullpage}
\usepackage{times}

\usepackage{doi}

\usepackage{amsmath}
\usepackage{mathtools}
\usepackage{bm}
\usepackage{cases}
\usepackage{booktabs}
\usepackage[inline]{enumitem}
\usepackage{float}
\usepackage{graphicx}
\usepackage{ifsym}
\usepackage{ifthen}
\usepackage[setrelation]{math}
\usepackage{makecell}
\usepackage{multirow}
\usepackage[bracketinterpret,seqinfers,footnotecalculus,abbrseqcontext]{logic}
\usepackage[pretest]{progreg}
\usepackage[bracketinterpret]{dL}
\usepackage{pgfplots}
\usepackage{prettyref}
\usepackage{rotating}
\usepackage{subcaption}
\usepackage{tabularx}
\usepackage{tikz}
\usepackage{wrapfig}
\usepackage{xspace}
\usepackage{fancyhdr}
\usepackage[square,numbers,sort&compress]{natbib}
\usepackage{wrapfig}

\usepackage{titling}
\usetikzlibrary{decorations.text}
\usetikzlibrary{shapes}
\usetikzlibrary{arrows}
\usetikzlibrary{calc}
\usetikzlibrary{fit}
\usetikzlibrary{patterns}

\usepackage{amsthm}

\theoremstyle{plain}

\newtheorem{remark}{Remark}

\usepackage{prettyref}
\newcommand{\rref}[2][]{\prettyref{#2}}
\newrefformat{sec}{Section\,\ref{#1}}
\newrefformat{appendix}{Appendix\,\ref{#1}}
\newrefformat{def}{Def.\,\ref{#1}}
\newrefformat{thm}{Theorem\,\ref{#1}}
\newrefformat{prop}{Proposition\,\ref{#1}}
\newrefformat{lem}{Lemma\,\ref{#1}}
\newrefformat{cor}{Corollary\,\ref{#1}}
\newrefformat{ex}{Example\,\ref{#1}}
\newrefformat{tab}{Table\,\ref{#1}}
\newrefformat{fig}{Fig.\,\ref{#1}}
\newrefformat{rem}{Remark\,\ref{#1}}
\newrefformat{model}{Model\,\ref{#1}}

\newcommand*{\old}{\text{old}}

\DeclareRobustCommand{\Abbreviation}[2]{#1.\,#2.\xspace}
\newcommand{\eg}{\Abbreviation{e}{g}}
\newcommand{\ie}{\Abbreviation{i}{e}}

\newcommand{\wrt}{w.r.t.\xspace}

\newcommand{\qdl}{Q\dL}
\newcommand{\safe}{\textit{safe}}

\newcommand{\dyn}{\textit{dyn}}

\newcommand{\ctrl}{\textit{ctrl}}
\newcommand{\ctrlo}{\textit{ctrl}_{\textit{o}}}
\newcommand{\ctrlr}{\textit{ctrl}_{\textit{r}}}
\newcommand{\dw}{\textit{dw}}

\newcommand{\curve}{\textit{curve}}

\newcommand{\Visible}{\textit{Visible}}
\newcommand{\inView}{\textit{isVisible}}
\newcommand{\drive}{\textit{drive}}

\newcommand{\pr}{\ensuremath p}
\newcommand{\prx}{\ensuremath \pr_x}
\newcommand{\pry}{\ensuremath \pr_y}
\newcommand{\po}{\ensuremath o}
\newcommand{\pox}{\ensuremath \po_x}
\newcommand{\poy}{\ensuremath \po_y}
\newcommand{\vr}{\ensuremath s}
\newcommand{\ar}{\ensuremath a}
\newcommand{\omegar}{\ensuremath \omega}
\newcommand{\dr}{d}
\newcommand{\drx}{\dr_x}
\newcommand{\dry}{\dr_y}
\newcommand{\pc}{\ensuremath c}
\newcommand{\pcx}{\ensuremath \pc_x}
\newcommand{\pcy}{\ensuremath \pc_y}
\newcommand{\pix}{\ensuremath x}
\newcommand{\pixr}{\ensuremath \pix_r}
\newcommand{\pixo}{\ensuremath \pix_o}
\newcommand{\rcurve}{\ensuremath r}
\newcommand{\vo}{\ensuremath v}
\newcommand{\vox}{\ensuremath \vo_x}
\newcommand{\voy}{\ensuremath \vo_y}
\newcommand{\vdiff}{\ensuremath \Delta_s}
\newcommand{\vmax}{\ensuremath \hat{\vr} + \vdiff}
\newcommand{\pg}{\ensuremath g}
\newcommand{\bo}{\ensuremath b_o}
\newcommand{\ao}{\ensuremath a_o}
\newcommand{\dobst}{\ensuremath d_o}
\newcommand{\dox}{\ensuremath d_{ox}}
\newcommand{\doy}{\ensuremath d_{oy}}

\definecolor{lsblue}{HTML}{16303A}
\definecolor{lslightblue}{HTML}{2E6579}
\definecolor{lsverylightblue}{HTML}{4699B9}
\definecolor{lsgreen}{HTML}{5ECEF9}
\definecolor{lslightgreen}{HTML}{54B9DF}

\definecolor{lsred}{HTML}{B94D5D}
\definecolor{lslightred}{HTML}{F16579}
\definecolor{lsdarkred}{HTML}{3A181D}

\floatstyle{ruled}
\newfloat{model}{tbp}{model}
\floatname{model}{Model}
\newtheorem{thm}{Theorem}

\graphicspath{{./}{fig/}}

\usepackage{ifpdf}
\ifpdf
\fi

\title{Formal Verification of Obstacle Avoidance and\\ Navigation of Ground Robots\thanks{Stefan Mitsch, Khalil Ghorbal, David Vogelbacher, Andr{\`e} Platzer, Formal verification of obstacle avoidance and navigation of ground robots, International Journal of Robotics Research (Vol. 36, Issue 12) pp. 1312-1340. Copyright © 2017 (The Authors). DOI: 10.1177/0278364917733549}}

\author{Stefan Mitsch\thanks{
 Computer Science Department, Carnegie Mellon University, Pittsburgh, USA
 {smitsch@cs.cmu.edu}
}, Khalil Ghorbal\thanks{
 INRIA, Rennes, France
 {kghorbal@cs.cmu.edu}
}, David Vogelbacher\thanks{
 Karlsruhe Institute of Technology, Karlsruhe, Germany
 {uaghg@student.kit.edu}
}, Andr\'e Platzer\thanks{
 Computer Science Department, Carnegie Mellon University, Pittsburgh, USA
 {aplatzer@cs.cmu.edu}
}}
\date{}

\begin{document}

\maketitle

\allowdisplaybreaks
\thispagestyle{empty}

\begin{abstract}
This article answers fundamental safety questions for ground robot navigation:
Under which circumstances does which control decision make a ground robot safely avoid obstacles?
Unsurprisingly, the answer depends on the exact formulation of the safety objective as well as the physical capabilities and limitations of the robot and the obstacles.
Because uncertainties about the exact future behavior of a robot's environment make this a challenging problem, we formally verify corresponding controllers and provide rigorous safety proofs justifying why they can never collide with the obstacle in the respective physical model.
To account for ground robots in which different physical phenomena are important, we analyze a series of increasingly strong properties of controllers for increasingly rich dynamics and identify the impact that the additional model parameters have on the required safety margins.

We analyze and formally verify:
\begin{enumerate*}[label=\emph{(\roman*)}]
\item \emph{static safety}, which ensures that no collisions can happen with stationary obstacles,
\item \emph{passive safety}, which ensures that no collisions can happen with stationary or moving obstacles while the robot moves, 
\item the stronger \emph{passive friendly safety} in which the robot further maintains sufficient maneuvering distance for obstacles to avoid collision as well, and
\item \emph{passive orientation safety}, which allows for imperfect sensor coverage of the robot, \ie, the robot is aware that not everything in its environment will be visible.
\end{enumerate*}
We formally prove that safety can be guaranteed despite sensor uncertainty and actuator perturbation.
We complement these provably correct safety properties with \emph{liveness} properties: we prove that provably safe motion is flexible enough to let the robot navigate waypoints and pass intersections.
In order to account for the mixed influence of discrete control decisions and the continuous physical motion of the ground robot, we develop corresponding \emph{hybrid system} models and use \emph{differential dynamic logic} theorem proving techniques to formally verify their correctness.
Since these models identify a broad range of conditions under which control decisions are provably safe, our results apply to any control algorithm for ground robots with the same dynamics.
As a demonstration, we, thus, also synthesize provably correct runtime monitor conditions that check the compliance of any control algorithm with the verified control decisions.
\\[\medskipamount]
\end{abstract}

\renewcommand{\thefootnote}{\alph{footnote}}

\section{Introduction}
\label{sec:intro}

Autonomous ground robots are increasingly promising as consumer products, ranging from today's autonomous household appliances \cite{DBLP:journals/arobots/FioriniP00} to the driverless cars of the future being tested on public roads\footnote{\url{http://www.nytimes.com/2010/10/10/science/10google.html?_r=0}}.
With the robots leaving the tight confounds of a lab or a locked-off industrial production site, robots face an increased need for ensuring safety, both for the sake of the consumer and the manufacturer.
At the same time, less tightly structured environments outside a limited-access factory increase the flexibility and uncertainty of what other agents may do.
This complicates the safety question, because it becomes even harder to achieve sufficiently exhaustive coverage of all possible behaviors.

Since the design of robot control algorithms is subject to many considerations and tradeoffs, the most useful safety results provide a broad characterization of the set of control decisions that are safe in each of the states of the system.
The control algorithms can then operate freely within the safe set of control decisions to optimize considerations such as reaching a goal or achieving secondary objectives without having to worry about safety.
The resulting characterization of safe control actions serves as a ``safety net'' underneath any control algorithm, which isolates the safety question and provides strong safety guarantees for any ground robot following the respective physical dynamics.

One of the most important and challenging safety considerations in mobile robotics is to ensure that the robot does not collide with any obstacles \cite{DBLP:journals/arobots/BouraineFS12,DBLP:journals/arobots/TaubigFHLMVW12,DBLP:journals/arobots/WuH12}.
Which control actions are safe under which circumstance crucially depends on the physical capabilities and limitations of the robot and moving obstacles in the environment.
It also crucially depends on the exact formulation of the safety criterion, of which there are many for mobile robots \cite{Macek2009}.
We capture the former in a physical model describing the differential equations of continuous motion of a ground robot as well as a description of what discrete control actions can be chosen.
This mix of discrete and continuous dynamics leads to a \emph{hybrid system}.
The safety criteria are formalized unambiguously in \emph{differential dynamic logic} \dL \cite{DBLP:journals/jar/Platzer08,DBLP:conf/lics/Platzer12a,DBLP:journals/jar/Platzer17}.

In order to justify the safety of the so-identified set of control decisions in the respective physical model, we formally verify the resulting controller and provide a rigorous proof in the \dL theorem prover \KeYmaeraX \cite{DBLP:conf/cade/FultonMQVP15}.
This proof provides undeniable mathematical evidence for the safety of the controllers, reducing safety of the robot to the question whether the appropriate physical model has been chosen for the robot and its environment.
Due to the uncertainties in the exact behavior of the robot and the agents in its environment, a range of phenomena are important in the models.

We consider a series of models with static obstacles at fixed positions, dynamic obstacles moving with bounded velocities, sensors with limited field of vision, sensor uncertainties, and actuator disturbances.
We identify the influence of each of those on the required design of safe controllers.
We also consider a series of safety criteria that account for the specific features of these models, since one of the subtle conceptual difficulties is what safety even means for an autonomous robot.
We would want it to be always collision-free, but that requires other vehicles to be reasonable, \eg, not actively try to run into our robot when it is just stopped in a corner.
One way of doing that is to assume stringent constraints on the behavior of obstacles~\cite{DBLP:conf/fm/LoosPN11,DBLP:journals/arobots/BouraineFS12}.

In this article, we refrain from doing so and allow obstacles with an arbitrary continuous motion respecting a known upper bound on their velocity.
Then our robot is safe, intuitively, if no collision can ever happen \emph{where the robot is to blame}.
For static obstacles, the situation is easy, because the robot is to blame for every collision that happens, so our safety property and its proof show that the robot will never collide with any static obstacle (\emph{static safety}).
For dynamic obstacles, safety is subtle, because other moving agents might actively try to ruin safety and cause collisions even if our robot did all it could to prevent them.
We analyze \emph{passive safety} \cite{Macek2009}, which requires that the robot does not actively collide, \ie, collisions only happen when a moving obstacle ran into the robot while the robot was stopped.
Our proofs guarantee passive safety with minimal assumptions about obstacles.
The trouble with passive safety is that it still allows the robot to stop in unsafe places, creating unavoidable collision situations in which an obstacle has no control choices left that would prevent a collision.
\emph{Passive friendly safety} \cite{Macek2009} addresses this challenge with more careful robot decisions that respect the dynamic limitations of moving obstacles (\eg, their braking capabilities).
A passive-friendly robot not only ensures that it is itself able to stop before a collision occurs, but it also maintains sufficient maneuvering room for obstacles to avoid a collision as well.
Finally, we introduce \emph{passive orientation safety}, which restricts the responsibility of the robot to avoid collisions to only parts of the robot's surroundings (\eg, the robot is responsible for collisions with obstacles to its front and sides, but obstacles are responsible when hitting the robot from behind).
We complement these safety notions with liveness proofs to show that our provably safe controllers are flexible enough to let the robot navigate waypoints and cross intersections.

All our models use symbolic bounds so our proofs hold \emph{for all} choices of the bounds.
As a result, we can account for uncertainty in several places (\eg, by instantiating upper bounds on acceleration or time with values including uncertainty).
We show how further uncertainty that cannot be attributed to such bounds (in particular location uncertainty, velocity uncertainty, and actuator uncertainty) can be modeled and verified \emph{explicitly}. 

The class of control algorithms we consider is inspired by the \emph{dynamic window} algorithm \cite{DBLP:journals/ram/FoxBT97}, but is equally significant for other control algorithms when combining our results of provable safety with verified runtime validation~\cite{DBLP:journals/fmsd/MitschP16}.
Unlike related work on obstacle avoidance (\eg, \cite{DBLP:journals/arobots/AlthoffKWB12,DBLP:journals/ijrr/PanZM12,DBLP:journals/arobots/TaubigFHLMVW12,DBLP:journals/arobots/SewardPA07,DBLP:journals/ijrr/BergAG11}), we use \emph{hybrid system} models and verification techniques that describe and verify the robot's discrete control choices \emph{along with its continuous, physical motion}.

In summary, our contributions are 
\begin{enumerate*}[label=\emph{(\roman*)}]
\item hybrid system models of navigation and obstacle avoidance control algorithms of robots, 
\item safety proofs that guarantee static safety, passive safety, passive friendly safety, and passive orientation safety in the presence of stationary and moving obstacles despite sensor uncertainty and actuator perturbation, and
\item liveness proofs that the safety measures are flexible enough to allow the robot to reach a goal position and pass intersections.
\end{enumerate*}
The models and proofs of this article are available\footnote{\url{http://web.keymaeraX.org/show/ijrr/robix.kyx}} in the theorem prover \KeYmaeraX~\cite{DBLP:conf/cade/FultonMQVP15} unless otherwise noted. 
They are also cross-verified with our previous prover \KeYmaera~\cite{DBLP:conf/cade/PlatzerQ08}.
This article extends our previous safety analyses \cite{DBLP:conf/rss/MitschGP13} with orientation safety for less conservative driving, as well as with liveness proofs to guarantee progress.
In order to take the vagaries of the physical environment into account, these guarantees are for hybrid system models that include discrete control decisions, reaction delays, differential equations for the robot's physical motion, bounded sensor uncertainty, and bounded actuator perturbation.

\section{Related Work}
\label{sec:relatedwork}

Isabelle has recently been used to formally verify that a C program implements the specification of the dynamic window algorithm \cite{DBLP:journals/arobots/TaubigFHLMVW12}.
We complement such effort by formally verifying the correctness of the dynamic window algorithm while considering continuous physical motion. 

\textsc{PassAvoid} \cite{DBLP:journals/arobots/BouraineFS12} is a navigation scheme designed to operate in unknown environments by stopping the robot before it collides with obstacles (passive safety).
The validation was however only based on simulations. 
In this work, we provide formal guarantees while proving the stronger passive friendly safety ensuring that the robot does not create unavoidable collision situations by stopping in unsafe places.

\citet{DBLP:journals/arobots/WuH12} assume unpredictable behavior for obstacles with known forward speed and maximum turn rate. 
The robot's motion is however explicitly excluded from their work which differs from the models we prove. 

We generalize the safety verification of straight line motions \cite{DBLP:conf/fm/LoosPN11,DBLP:conf/iccps/MitschLP12} and 
the two-dimensional planar motion with constant velocity \cite{DBLP:conf/hybrid/LoosRP13,DBLP:conf/fm/PlatzerC09} by allowing translational and rotational accelerations. 

\citet{DBLP:journals/ijrr/PanZM12} proposes a method to smooth the trajectories produced by sampling-based planners in a collision-free manner.
Our article proves that such trajectories are indeed safe when considering the control choices of a robot and its continuous dynamics.

LQG-MP \cite{DBLP:journals/ijrr/BergAG11} is a motion planning approach that takes into account the sensors, controllers, and motion dynamics of a robot while working with uncertain information about the environment.
The approach attempts to select the path that decreases the collision probability. 
\citet{DBLP:journals/arobots/AlthoffKWB12} use a probabilistic approach to rank trajectories according to their collision probability.
They propose a collision cost metric to refine the ranking based on the relative speeds and masses of the collision objects.
\citet{DBLP:journals/arobots/SewardPA07} try to avoid potentially hazardous situations by using Partially Observable Markov Decision Processes. 
Their focus, however, is on a user-definable trade-off between safety and progress towards a goal.
Safety is not guaranteed under all circumstances.
We rather focus on formally proving collision-free motions under reasonable assumptions of the environment. 

It is worth noting that formal methods were also used for other purposes in the hybrid systems context.  
For instance, in \cite{DBLP:journals/fmsd/PlakuKV09,DBLP:journals/sttt/PlakuKV13}, the authors combine model checking and motion planning to efficiently falsify a given 
property. 
Such lightweight techniques could be used to increase the trust in the model but are not designed to prove the property. 
LTLMoP \cite{DBLP:conf/rss/SaridXK12} enables the user to specify high-level behaviors (\eg, visit all rooms) when the environment is continuously updated.
The approach synthesizes plans, expressed in linear temporal logic, of a hybrid controller, whenever new map information is discovered while 
preserving the state and task completion history of the desired behavior. 
In a similar vein, the automated synthesis of controllers restricted to straight-line motion and satisfying a given property formalized in linear temporal logic has been recently 
explored in \cite{DBLP:journals/trob/Kress-GazitFP09}, and adapted to discrete-time dynamical systems in \cite{DBLP:conf/icra/WolffTM14}. 
\citet{DBLP:conf/amcc/KaramanF12} explore optimal trajectory synthesis from specifications in deterministic $\mu$-calculus.  

\section{Preliminaries: Differential Dynamic Logic}
\label{sec:dl}

A robot and the moving obstacles in its environment form a \emph{hybrid system}:
they make discrete control choices (\eg, compute the actuator set values for acceleration, braking, or steering), which in turn influence their actual physical behavior (\eg, slow down to a stop, move along a curve).
In test-driven approaches, simulators or field tests provide insight into the expected physical effects of the control code.
In formal verification, hybrid systems provide joint models for both discrete and continuous behavior, since verification of either component alone does not capture the full behavior of a robot and its environment.
In this section, we first give an overview of the relationship between testing, simulation, and formal verification, before we introduce the syntax and semantics of the specification language that we use for formal verification.

\subsection{Testing, Simulation, and Formal Verification}
Testing, simulation, and formal verification complement each other.
Testing helps to make a system robust under real-world conditions, whereas simulation lets us execute a large number of tests in an inexpensive manner (at the expense of a loss of realism). 
Both, however, show correctness for the finitely many tested scenarios only.
Testing and simulation discover the presence of bugs, but cannot show their absence. 
Formal verification, in contrast, provides precise and undeniable guarantees for \emph{all} possible executions of the modeled behavior. Formal verification either discovers bugs if present, or shows the absence of bugs in the model, but, just like simulation, cannot show whether or not the model is realistic.
In \rref{sec:monitoring}, we will see how we can use runtime monitoring to bridge both worlds.
Testing, simulation, and formal verification all base on similar ingredients, but apply different levels of rigor as follows.

\paragraph{Software.} Testing and simulation run a specific control algorithm with specific parameters (\eg, run a specific version of an obstacle avoidance algorithm with maximum velocity \(V=2\, \text{m/s}\)).
Formal verification can specify symbolic parameters and nondeterministic inputs and effects and, thereby, capture entire families of algorithms and many scenarios at once (\eg, verify all velocities \(0\leq v \leq V\) for any maximum velocity \(V \geq 0\) at once).

\paragraph{Hardware and physics.}
Testing runs a real robot in a real environment. 
Both simulation and formal verification, in contrast, work with \emph{models} of the hardware and physics to provide sensor values and compute how software decisions result in real-world effects.

\paragraph{Requirements.}
Testing and simulation can work with informal or semi-formal requirements (\eg, a robot should not collide with obstacles, which leaves open the question whether a slow bump is considered a collision or not).
Formal verification uses mathematically precise formal requirements expressed as a logical formula (without any ambiguity in their interpretation distinguishing precisely between correct behavior and faults).

\paragraph{Process.}
In testing and simulation, requirements are formulated as test conditions and expected test outcomes. 
A test procedure then runs the robot several times under the test conditions and one manually compares the actual output with the expected outcome (\eg, run the robot in different spaces, with different obstacles, various software parameters, and different sensor configurations to see whether or not any of the runs fail to avoid obstacles).
The test protocol serves as correctness evidence and needs to be repeated when anything changes.
In formal verification, the requirements are formulated as a logical formula. 
A theorem prover then creates a mathematical proof showing that \emph{all} possible executions---usually infinitely many---of the model are correct (safety proof), or showing that the model has a way to achieve a goal (liveness proof).
The mathematical proof is the correctness certificate.

\subsection{Differential Dynamic Logic}

This section briefly explains the language that we use for formal verification. 
It explains \emph{hybrid programs}, which is a program notation for describing hybrid systems, and \emph{differential dynamic logic} \dL~\cite{DBLP:journals/jar/Platzer08,Platzer10,DBLP:conf/lics/Platzer12a,DBLP:journals/jar/Platzer17}, which is the logic for specifying and verifying correctness properties of hybrid programs. 
Hybrid programs can specify how a robot and obstacles in the environment make decisions and move physically.
With differential dynamic logic we specify formally which behavior of a hybrid program is considered correct.
\dL allows us to make statements that we want to be true for all runs of a hybrid program (safety) or for at least one run (liveness).

One of the many challenges of developing robots is that we do not know the behavior of the environment exactly.
For example, a moving obstacle may or may not slow down when our robot approaches it.
In addition to programming constructs familiar from other languages (\eg, assignments and conditional statements), hybrid programs, therefore, provide nondeterministic operators that allow us to describe such unknown behavior of the environment concisely. 
These nondeterministic operators are also useful to describe parts of the behavior of our own robot (\eg, we may not be interested in the exact value delivered by a position sensor, but only that it is within some error range), which then corresponds to verifying an entire family of controllers at once.
Using nondeterminism to model our own robot has the benefit that later optimization (\eg, mount a better sensor or implement a faster algorithm) does not necessarily require re-verification since variations are already covered.

\rref{tab:hybridprograms} summarizes the syntax of hybrid programs together with their informal semantics.
Many of the operators will be familiar from regular expressions, but the discrete and continuous operators are crucial to describe robots.
A common and useful assumption when working with hybrid systems is that time only passes in differential equations, but discrete actions do not consume time (whenever they do consume time, it is easy to transform the model to reflect this just by adding explicit extra delays). 

We now briefly describe each operator with an example.
Assignment $\humod{x}{\theta}$ instantaneously assigns the value of the term $\theta$ to the variable $x$ (\eg, let the robot choose maximum braking).
Nondeterministic assignment $\humod{x}{*}$ assigns an arbitrary real value to $x$ (\eg, an obstacle may choose any acceleration, we do not know which value exactly).
Sequential composition $\alpha;\beta$ says that $\beta$ starts after $\alpha$ finishes (\eg, $\humod{\ar}{3};~\humod{\rcurve}{*}$ first let the robot choose acceleration to be $3$, then choose any steering angle). 
The nondeterministic choice $\pchoice{\alpha}{\beta}$ follows either $\alpha$ or $\beta$ (\eg, the obstacle may slow down or speed up).
The nondeterministic repetition operator $\prepeat{\alpha}$ repeats $\alpha$ zero or more times (\eg, the robot may encounter obstacles over and over again, or wants to switch between the options of a nondeterministic choice, but we do not know exactly how often).
The continuous evolution $\D{x}=\theta ~\&~ Q$ evolves $x$ along the differential equation \(\D{x}=\theta\) for any arbitrary amount of time within the evolution domain $Q$ (\eg, the velocity of the robot decreases along \m{\D{v}={-}b~\&~v\geq0} according to the applied brakes $-b$, but does not become negative since hitting the brakes won't make the robot drive backwards).
The test $\ptest{F}$ checks that the formula $F$ holds, and aborts the run if it does not (\eg, test whether the distance to an obstacle is large enough to continue driving).
Other nondeterministic choices may still be possible if one run fails, which explains why an execution of hybrid programs with backtracking is a good intuition.

A typical pattern with nondeterministic assignment and tests is to limit the assignment of arbitrary values to known bounds (\eg, limit an arbitrarily chosen acceleration to the physical limits of the robot, as in \m{\humod{a}{*};~ \ptest{(a\leq A)}}, which says $a$ is any value less or equal $A$).
Another useful pattern is a nondeterministic choice with complementary tests $\pchoice{(\ptest{P};\alpha)}{(\ptest{\neg P};\beta)}$, which models an if-then-else statement $\text{if }(P)~ \alpha \text{ else } \beta$.

\begin{table}[t]
\small\sf\centering
  \newcommand{\foform}{F\xspace}
  \caption{Hybrid program representations of hybrid systems.}
  \label{tab:hybridprograms}
  \begin{tabularx}{\columnwidth}{l@{\hspace{.5em}}X}
    \toprule
       \textbf{Statement}
    & \textbf{Effect}
    \\
    \midrule 
    $x:=\theta$ & assign current value of term $\theta$ to variable $x$ (discrete assignment)\\
    $x:=*$ & assign arbitrary real number to variable $x$\\
    $\alpha;~\beta$ & sequential composition, first run~$\alpha$, then $\beta$  \\
    $\alpha~\cup~\beta$ 
    	& nondeterministic choice, follow either~$\alpha$ or~$\beta$\\
    $\alpha^*$ & nondeterministic repetition repeats~$\alpha$ any \m{n\geq 0} number of times \\
    $\ptest{F}$ & check that a condition $F$ holds in the current state, and abort run if it does not\\
    \multirow{2}{*}{$\begin{aligned}&\bigl(\D{x_1}=\theta_1,\dots,\\&\phantom{\bigl(}\D{x_n}=\theta_n ~\&~ Q\bigr)\end{aligned}$}
     & evolve $x_i$ along differential equation system $\D{x_i} = \theta_i$ for any amount of time restricted to maximum evolution domain~$Q$\\
    \bottomrule
  \end{tabularx}
\end{table}

The \dL~formulas can be formed according to the following grammar
(where~\m{{\sim}} is any comparison operator in \m{\{<,\leq,=,\geq,>,\neq\}} and~$\theta_1,\theta_2$ are arithmetic expressions in~\m{+,-,\cdot,/} over the reals):
\begin{equation*}
\phi ::= \theta_1 \sim \theta_2 \mid \neg \phi \mid \phi \wedge \psi \mid \phi \vee \psi \mid \phi \rightarrow \psi \mid \lforall{x}{\phi} \mid \dbox{\alpha}{\phi} \mid \ddiamond{\alpha} \phi
\end{equation*}

Further operators, such as Euclidean norm $\norm{\theta}$ and infinity norm $\norm{\theta}{_\infty}$ of a vector $\theta$, are definable from these.
The formula \(\dbox{\alpha}{\phi}\) is true in a state if and only if \emph{all} runs of hybrid program $\alpha$ from that state lead to states in which formula $\phi$ is true.
The formula \(\ddiamond{\alpha}{\phi}\) is true in a state if and only if there is at least one run of hybrid program $\alpha$ to a state in which formula $\phi$ is true.

In particular, \dL formulas of the form $F\limply \dibox{\alpha}{G}$ mean that if $F$ is true in the initial state, then all executions of the hybrid program $\alpha$ only lead to states in which formula $G$ is true. 
Dually, formula $F \limply \ddiamond{\alpha}{G}$ expresses that if $F$ is true in the initial state then there is a state reachable by the hybrid program~$\alpha$ that satisfies formula~$G$.

\subsection{Proofs in Differential Dynamic Logic}

Differential dynamic logic comes with a verification technique to prove correctness properties \cite{DBLP:journals/jar/Platzer08,Platzer10,DBLP:conf/lics/Platzer12a,DBLP:journals/jar/Platzer17}.
The underlying principle behind a proof in \dL is to symbolically decompose a large hybrid program into smaller and smaller pieces until the remaining formulas no longer contain the actual programs, but only their logical effect. 
For example, the effect of a simple assignment \m{\humod{x}{1+1}} in a proof of formula \m{\dbox{\humod{x}{1+1}}{x=2}} results in the proof obligation $1+1=2$.
The effects of more complex programs may of course not be as obviously true.
Still, whether or not these remaining formulas in real arithmetic are valid is decidable by a procedure called \emph{quantifier elimination} \cite{DBLP:conf/automata/Collins75}.

Proofs in \dL consist of three main aspects:
\begin{enumerate*}[label=\emph{(\roman*)}]
\item find invariants for loops and differential equations,
\item symbolically execute programs to determine their effect, and finally
\item verify the resulting real arithmetic with external solvers for quantifier elimination.
\end{enumerate*}
High modeling fidelity becomes expensive in the arithmetic parts of the proof, since real arithmetic is decidable but of high complexity \cite{DBLP:journals/jsc/DavenportH88}.
As a result, proofs of high-fidelity models may require arithmetic simplifications (e.g., reduce the number of variables by abbreviating complicated terms, or by hiding irrelevant facts) before calling external solvers.

The reasoning steps in a \dL proof are justified by \dL axioms.
For example, the equivalence axiom \m{\pmb{\dbox{\pchoice{\alpha}{\beta}}{\phi}} \lbisubjunct \dbox{\alpha}{\phi} \land \dbox{\beta}{\phi}} allows us to prove safety about a program with a nondeterministic choice $\pchoice{\alpha}{\beta}$ by instead proving safety of the program $\alpha$ in $\dbox{\alpha}\phi$ and separately proving safety of the program $\beta$ in $\dbox{\beta}\phi$.
Reducing all occurrences of \m{\dbox{\pchoice{\alpha}{\beta}}{\phi}}
to corresponding conjunctions
\m{\dbox{\alpha}{\phi} \land \dbox{\beta}{\phi}}, which are handled separately, successively decomposes safety questions for a hybrid program of the form $\pchoice{\alpha}{\beta}$ into safety questions for simpler subsystems.

The theorem prover \KeYmaeraX \cite{DBLP:conf/cade/FultonMQVP15} implements a uniform substitution proof calculus for \dL \cite{DBLP:journals/jar/Platzer17} that checks all soundness-critical side conditions during a proof.
\KeYmaeraX also provides significant automation by bundling axioms into larger \emph{tactics} that perform multiple reasoning steps at once.
For example, when proving safety of a program with a loop $A \limply \dbox{\prepeat{\alpha}}{S}$, a tactic for loop induction tries to find a loop invariant $J$ to split the proof into three separate, smaller pieces: one branch to show that the invariant is true in the beginning ($A\limply J$), one branch to show that running the program $\alpha$ without loop once preserves the invariant ($J \limply \dbox{\alpha}J$), and another branch to show that the invariant is strong enough to guarantee safety ($J \limply S$).
If an invariant $J$ cannot be found automatically, users can still provide their own guess or knowledge about $J$ as input to the tactic.
Differential invariants provide a similar inductive reasoning principle for safety proofs about differential equations ($A \limply \dbox{\D{x}=\theta}{S}$) without requiring symbolic solutions, so they can be used to prove properties about non-linear differential equations, such as for robots.
Differential invariants can be synthesized for certain classes of differential equations \cite{DBLP:conf/vmcai/SogokonGJP16}.

The tactic language \cite{DBLP:conf/itp/FultonMBP17} of \KeYmaeraX can also be used by users for scripting proofs to provide human guidance when necessary.
We performed all proofs in this paper in the verification tool \KeYmaeraX \cite{DBLP:conf/cade/FultonMQVP15} and/or its predecessor \KeYmaera \cite{DBLP:conf/cade/PlatzerQ08}.
While all our proofs ship with \KeYmaera, we provide all but one proof also in its successor \KeYmaeraX, which provides rigorous verification from a small soundness-critical core, comes with high-assurance correctness guarantees from cross-verification results \cite{DBLP:conf/cpp/BohrerRVVP17} in the theorem provers Isabelle and Coq, and enables us to provide succinct tactics that produce the proofs and facilitate easier reuse of our verification results.
Along with the fact that \KeYmaeraX supports hybrid systems with nonlinear discrete jumps and nonlinear differential equations, these advantages make \KeYmaeraX more readily applicable to robotic verification than other hybrid system verification tools.
SpaceEx \cite{DBLP:conf/cav/FrehseGDCRLRGDM11}, for example, focuses on (piecewise) linear systems.
\KeYmaeraX implements automatic proof strategies that decompose hybrid systems symbolically.
This compositional verification principle helps scaling up verification, because \KeYmaeraX verifies a big system by verifying properties of subsystems.
Strong theoretical properties, including relative completeness, have been shown for \dL \cite{DBLP:journals/jar/Platzer08,DBLP:conf/lics/Platzer12b,DBLP:journals/jar/Platzer17}.

\section{Preliminaries: Obstacle Avoidance with the Dynamic Window Approach}
\label{sec:dynamicwindow}

The robotics community has come up with an impressive variety of robot designs, which differ not only in their tool equipment, but also (and more importantly for the discussion in this article) in their kinematic capabilities.
This article focuses on wheel-based ground vehicles.
In order to make our models applicable to a large variety of robots, we use only limited control options (\eg, do not move sideways to avoid collisions since Ackermann drive could not follow such evasion maneuvers).
We consider robots that drive forward (non-negative translational velocity) in sequences of arcs in two-dimensional space.
If the radius of such a circle is large, the robot drives (forward) on an approximately straight line.
Such trajectories can be realized by robots with single-wheel drive, differential drive (wheels may rotate in opposite directions), Ackermann drive (front wheels steer), synchro-drive (all wheels steer), or omni-directional drive (wheels rotate in any direction) \cite{Braeunl2006}.
In a nutshell, \emph{in order to stay on the safe side, our models conservatively underestimate the capabilities of our robot while conservatively overestimating the dynamic capabilities of obstacles}.

Many different navigation and obstacle avoidance algorithms have been proposed for such robots, \eg \emph{dynamic window} \cite{DBLP:journals/ram/FoxBT97}, \emph{potential fields} \cite{DBLP:conf/icra/Khatib85}, or \emph{velocity obstacles} \cite{DBLP:journals/ijrr/FioriniS98}.
For an introduction to various navigation approaches for mobile robots, see \cite{DBLP:journals/jirs/Bonin-FontOO08,Choset2005}.
The inspiration for the algorithm we consider in this article is the dynamic window algorithm \cite{DBLP:journals/ram/FoxBT97}, which is derived from the motion dynamics of the robot and thus discusses all aspects of a hybrid system (models of discrete and continuous dynamics).
But other control algorithms including path planners based on RRT \cite{DBLP:journals/ijrr/LaValleK01} or A$^*$ \cite{DBLP:journals/tssc/HartNR68} are compatible with our results when their control decisions are checked with a runtime verification approach \cite{DBLP:journals/fmsd/MitschP16} against the safety conditions we identify for the motion here.

The dynamic window algorithm is an obstacle avoidance approach for mobile robots equipped with synchro drive \cite{DBLP:journals/ram/FoxBT97} but can be used for other drives too \cite{DBLP:conf/icra/BrockK99}.
It uses circular trajectories that are uniquely determined by a translational velocity $v$ together with a rotational velocity $\omega$, see \rref{sec:dynmodel} below for further details.
The algorithm is organized into two steps: 
\begin{enumerate*}[label=\emph{(\roman*)}]
    \item The range of all possible pairs of translational and rotational velocities is reduced to admissible ones that result in safe trajectories (\ie, avoid collisions since those trajectories allow the robot to stop before it reaches the nearest obstacle) as follows~\cite[(14)]{DBLP:journals/ram/FoxBT97}:
$V_a{=}\left\{(v,\omega) \mid v \leq \sqrt{2 \text{dist}(v,\omega)\D{v}_b}
\land \omega \leq \sqrt{2 \text{dist}(v,\omega)\D{\omega}_b} \right\}$ 
This definition of admissible velocities, however, neglects the reaction time of the robot.
Our proofs reveal the additional safety margin that is entailed by the reaction time needed to revise decisions.
The admissible pairs are further restricted to those that can be realized by the robot within a short time frame $t$ (the dynamic window) from current velocities $v_a$ and $\omega_a$ to account for acceleration effects despite assuming velocity to be a piecewise constant function in time \cite[(15)]{DBLP:journals/ram/FoxBT97}:
$V_d{=}\{(v,\omega) \mid v \in \left[v_a-\D{v}t,v_a+\D{v}t\right] \\ 
    \land \omega \in \left[\omega_a-\D{\omega}t,\omega_a+\D{\omega}t\right]\}$.
Our models, instead, control acceleration and describe the effect on velocity in differential equations.
If the set of admissible and realizable velocities is empty, the algorithm stays on the previous safe curve (such curve exists unless the robot started in an unsafe state).
\item \label{item:DWA-progress} Progress towards the goal is optimized by maximizing a goal function among the set of all admissible controls.
\end{enumerate*}
For safety verification, we can omit step \ref{item:DWA-progress} and verify the stronger property that \emph{all} choices fed into the optimization are safe.
Even if none is identified, the previous safe curve can still be continued.

\section{Robot and Obstacle Motion Model}
\label{sec:dynmodel}

This section introduces the robot and obstacle motion models that we are using throughout the article.
\rref{tab:robotobstaclevars} summarizes the model variables and parameters of both the robot and the obstacle for easy reference.
In the following subsections, we illustrate their meaning in detail.

\begin{table}[H]
\small\sf\centering
\caption{Parameters and state variables of robot and obstacle}
\label{tab:robotobstaclevars}
\begin{tabularx}{\columnwidth}{l@{\hspace{.5em}}lX}
\toprule
& \textbf{2D} & \textbf{Description}\\
\midrule
$\pr$ & $(\prx,\pry)$ & Position of the robot\\
$\vr$ & & Translational speed\\
$\ar$ & & Translational acceleration, s.t. $-b \leq \ar \leq A$\\
$\omegar$ & & Rotational velocity, s.t. $\omegar \rcurve = \vr$\\
$\dr$ & $(\drx,\dry)$ & Orientation of the robot, s.t. $\norm{\dr}=1$\\
$\pc$ & $(\pcx,\pcy)$ & Curve center, s.t. $\dr=(\pr-\pc)^\bot$\\
$\rcurve$ & & Curve radius, s.t. $\rcurve=\norm{\pr-\pc}$\\
\midrule
$\po$ & $(\pox,\poy)$ & Position of the obstacle\\
$\vo$ & $(\vox,\voy)$ & Translational velocity, including orientation, s.t. $\norm{\vo} \leq V$\\
\midrule
$A$ & & Maximum acceleration $A \geq 0$\\
$b$ & & Minimum braking $b > 0$\\
$\varepsilon$ & & Maximum control loop reaction delay $\varepsilon > 0$\\
$V$ & & Maximum obstacle velocity $V \geq 0$\\
$\Omega$& & Maximum rotational velocity $\Omega \geq 0$\\
\bottomrule
\end{tabularx}
\end{table}

\subsection{Robot State and Motion}

\begin{figure}[htb]
\centering
\begin{tikzpicture}[line cap=round,line join=round,>=triangle 45,x=1.0cm,y=1.0cm]
\clip(0,2) rectangle (8,5);
\draw [shift={(2,3.44)},color=lsgreen,fill=lsgreen,fill opacity=0.1] (0,0) -- (-20:0.7) arc (-20:28:0.7) -- cycle;
\draw (2,3.44)-- (4.44,2.62);
\draw [shift={(2.5,3.44)},very thick] plot[domain=-0.39990429646811876:0.5453617994034332,variable=\t]({1.0*2.106181378704123*cos(\t r)+-0.0*2.106181378704123*sin(\t r)},{0.0*2.106181378704123*cos(\t r)+1.0*2.106181378704123*sin(\t r)});
\draw [->] (4.44,2.62) -- (4.820218962734227,3.519542424029757);
\draw [line width=0.4pt,dash pattern=on 5pt off 5pt] (2,3.44)-- (4.28,4.52);
\draw [->,arrows={-latex'},line width=0.4pt,dash pattern=on 2pt off 2pt] (4.44,2.62) -- (4.820218962734227,2.62);
\draw [->,arrows={-latex'},line width=0.4pt,dash pattern=on 2pt off 2pt] (4.44,2.62) -- (4.44,3.519542424029757);
\draw [line width=0.4pt,dash pattern=on 2pt off 2pt] (4.44,3.519542424029757)-- (4.820218962734227,3.519542424029757);
\draw [line width=0.4pt,dash pattern=on 2pt off 2pt] (4.820218962734227,3.519542424029757)-- (4.820218962734227,2.62);
\draw [fill=lsblue] (4.44,2.62) circle (1.5pt);
\draw[color=black,anchor=east] (2,3.44) node {\((\pcx,\pcy)=\pc\)};
\draw[color=black] (3.7,2.4) node {\((\prx,\pry)=\pr\)};
\draw [fill=lsblue] (2,3.44) circle (1.5pt);
\draw [fill=lsblue] (4.28,4.52) circle (1.5pt);
\draw[color=black,anchor=south west] (4,4.52) node {\(\tilde{\pr}\) after time \(\varepsilon\)};
\draw[color=black,anchor=east] (3.2,2.9) node {\(\rcurve = \norm{\pr-\pc}\)};
\draw[color=black,anchor=west] (4.5,4.3) node {trajectory (length \(\vr \varepsilon\))};
\draw[color=black] (5.8,3.7) node {\(\dr=(\drx,\dry)\)};
\draw[color=black] (2.5,3.44) node {\(\omegar \varepsilon\)};
\draw[color=black] (5.7,2.6) node {\(\drx = \cos \theta\)};
\draw[color=black] (3.8,3.7) node {\(\sin \theta = \dry\)};
\end{tikzpicture}
\caption{State illustration of a robot on a two-dimensional plane. The robot has position \(\pr=(\prx,\pry)\), orientation \(\dr=(\drx,\dry)\), and drives on circular arcs (thick arc) of radius \(\rcurve\) with translational velocity \(\vr\), rotational velocity \(\omegar\) and thus angle \(\omegar \varepsilon\) around curve center points \(\pc=(\pcx,\pcy)\).
In time \(\varepsilon\) the robot will reach a new position \(\tilde{\pr}\), which is \(\vr \varepsilon\) away from the initial position \(\pr\) when measured along the robot's trajectory arc.}
\label{fig:robotmodel}
\end{figure}
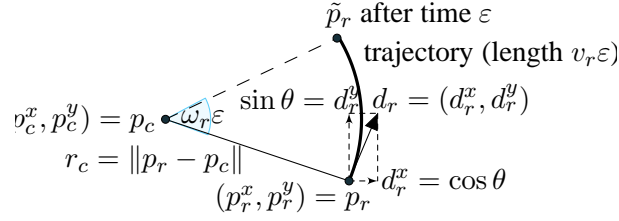

The dynamic window algorithm safely abstracts the robot's shape to a single point by increasing the (virtual) shapes of all obstacles correspondingly (cf. \cite{DBLP:journals/arobots/MinguezMS06} for an approach to attribute robot shape to obstacles).
We also use this abstraction to reduce the verification complexity.
\rref{fig:robotmodel} illustrates how we model the position \(\pr\), orientation \(\dr\), and trajectory of a robot.

The robot has state variables describing its current position $\pr=(\prx,\pry)$, translational velocity $\vr \geq 0$, translational acceleration $\ar$, orientation vector\footnote{As stated earlier, we study unidirectional motion: the robot moves along its direction, that is the vector $\dr$ gives the direction of the velocity vector.} $\dr=(\cos\theta,\sin\theta)$, and angular velocity\footnote{The derivative with respect to time is denoted by prime ($\D{}$).} $\D{\theta}=\omegar$.
The translational and rotational velocities are linked \wrt the rigid body planar motion by the formula $\rcurve \omegar = \vr$, where the curve radius $\rcurve=\norm{\pr-\pc}$ is the distance between the robot and the center of its current curve $\pc=(\pcx,\pcy)$.
The usual modeling approach with angle $\theta$ and trigonometric functions $\sin\theta$ and $\cos\theta$ to determine the position along a curve, however, results in undecidable arithmetic.
Instead, we encode sine and cosine functions in the dynamics using the extra variables $\drx = \cos\theta$ and $\dry=\sin\theta$ by \emph{differential axiomatization} \cite{DBLP:journals/logcom/Platzer10}.
The continuous dynamics for the dynamic window algorithm~\cite{DBLP:journals/ram/FoxBT97} can, thus, be described by the differential equation system of ideal-world dynamics of the planar rigid body motion: 
\[\D{\pr}=\vr \dr \syssep \D{\vr}=\ar \syssep \D{\dr}=\omegar \dr^\bot \syssep \D{(\rcurve\omegar)}=\ar\]
where 
\begin{itemize}[noitemsep]
\item $\D{\pr}=\vr \dr$ represents $\D{\pr}_x=\vr  \drx \syssep \D{\pr}_y=\vr \dry$ in vectorial notation, 
\item the condition $\D{\dr}=\omegar \dr^\bot$ is vector notation for the rotational dynamics $\D{\dr}_x=-\omegar \dry \syssep \D{\dr}_y=\omegar \drx$ where $^\bot$ is the orthogonal complement, and 
\item the condition $\D{(\rcurve \omegar)}=\ar$ encodes the rigid body planar motion $\rcurve \omegar = \vr$ that we consider.
\end{itemize}

The dynamic window algorithm assumes piecewise constant velocity $\vr$ between decisions despite accelerating, which is physically unrealistic. 
We, instead, control acceleration $\ar$ and do not perform instant changes of the velocity. 
Our model is closer to the actual dynamics of a robot.
The realizable velocities follow from the differential equation system according to the controlled acceleration $\ar$.

\begin{figure}[tb]
\centering
\begin{subfigure}[b]{.3\textwidth}
\includegraphics[width=\textwidth]{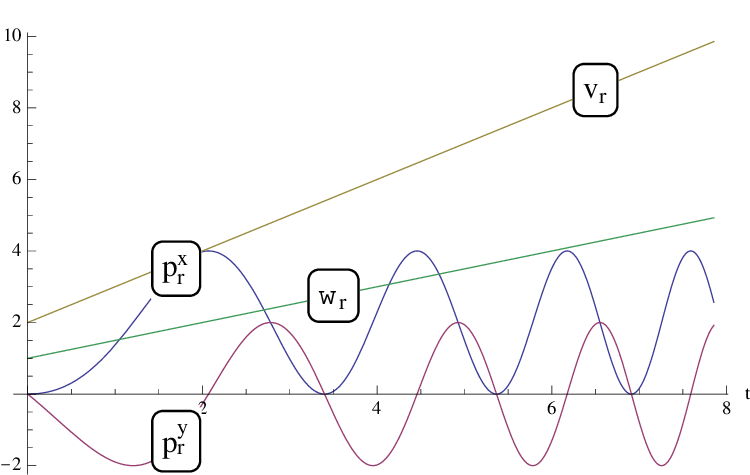}
\caption{Position ($p^x_r,p^y_r$), translational velocity $\vr$ and rotational velocity $\omegar$ for positive acceleration on a circle.}
\label{circlea_xyvw}
\end{subfigure}
\qquad
\begin{subfigure}[b]{.28\textwidth}
\includegraphics[width=\textwidth]{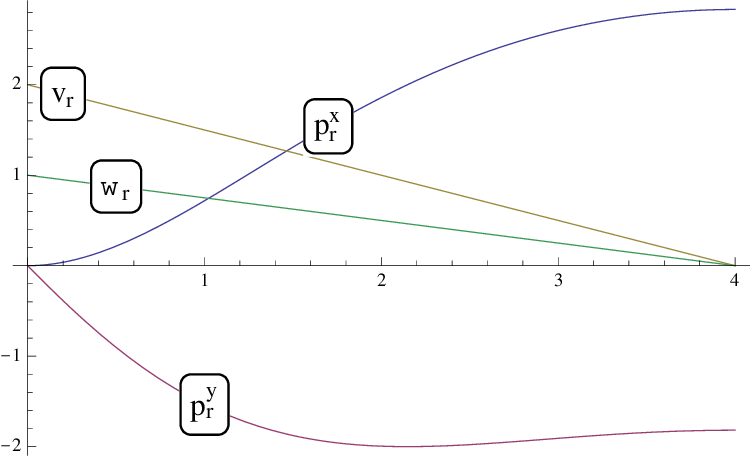}
\caption{Position ($p^x_r,p^y_r$), translational velocity $\vr$ and rotational velocity $\omegar$ for braking to a complete stop on a circle.}
\label{circle-b_xyvw}
\end{subfigure}
\qquad
\begin{subfigure}[b]{.28\textwidth}
\includegraphics[width=\textwidth]{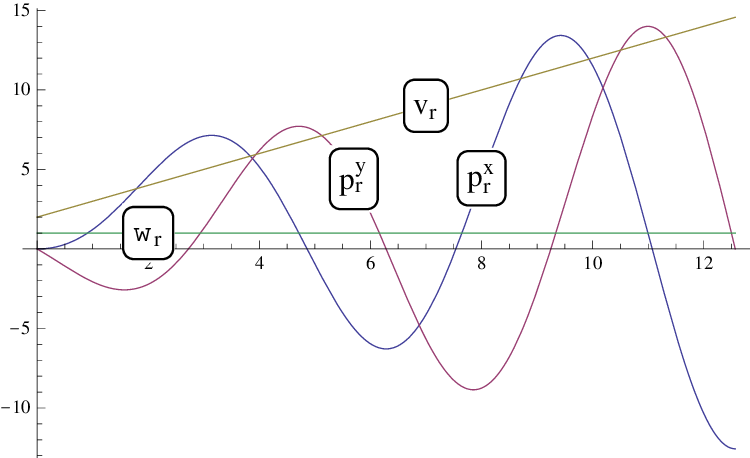}
\caption{Position ($p^x_r,p^y_r$), translational velocity $\vr$ and rotational velocity $\omegar$ for translational acceleration on a spiral.}
\label{spiral_xyvw}
\end{subfigure}
\\
\begin{subfigure}[b]{.28\textwidth}
\includegraphics[width=\textwidth]{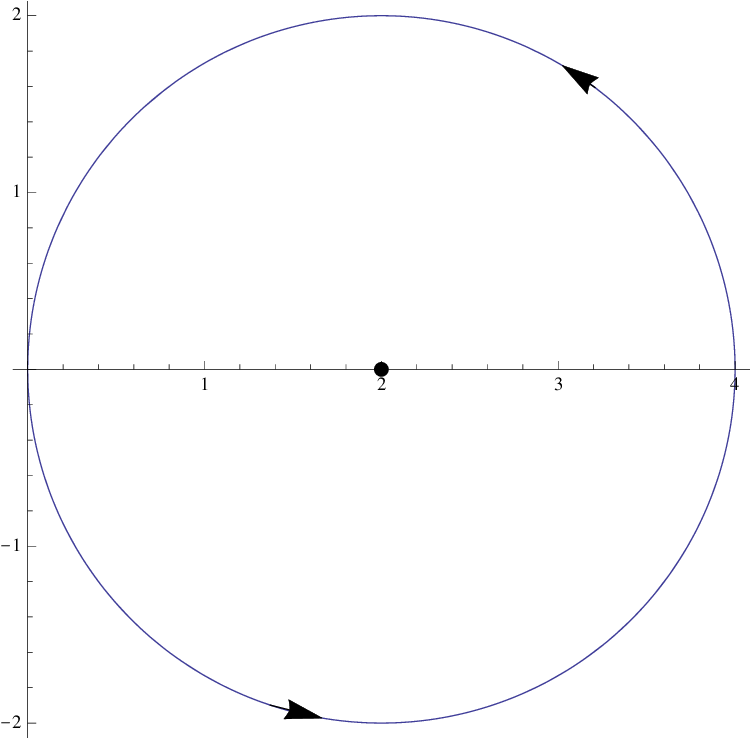}
\caption{$(\prx,\pry)$ motion plot for acceleration \protect\subref{circlea_xyvw}.}
\label{circlea_2d}
\end{subfigure}
\qquad
\begin{subfigure}[b]{.3\textwidth}
\includegraphics[width=\textwidth]{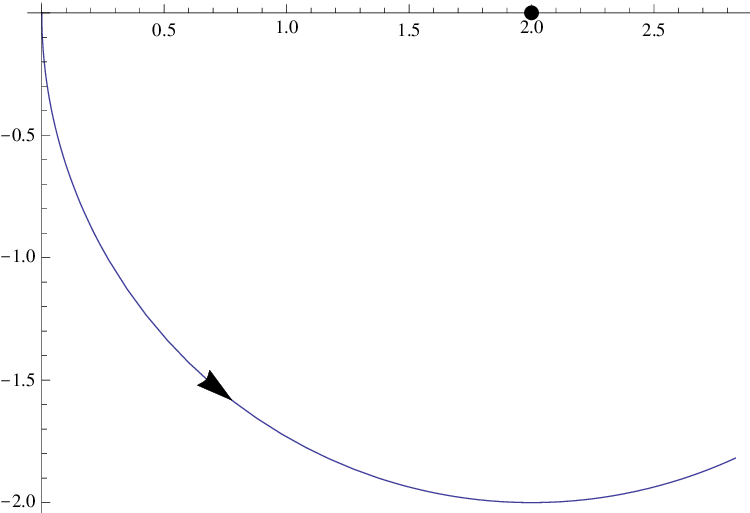}
\caption{$(\prx,\pry)$ motion plot for braking \protect\subref{circle-b_xyvw}.}
\label{circle-b_2d}
\end{subfigure}
\qquad
\begin{subfigure}[b]{.3\textwidth}
\includegraphics[width=\textwidth]{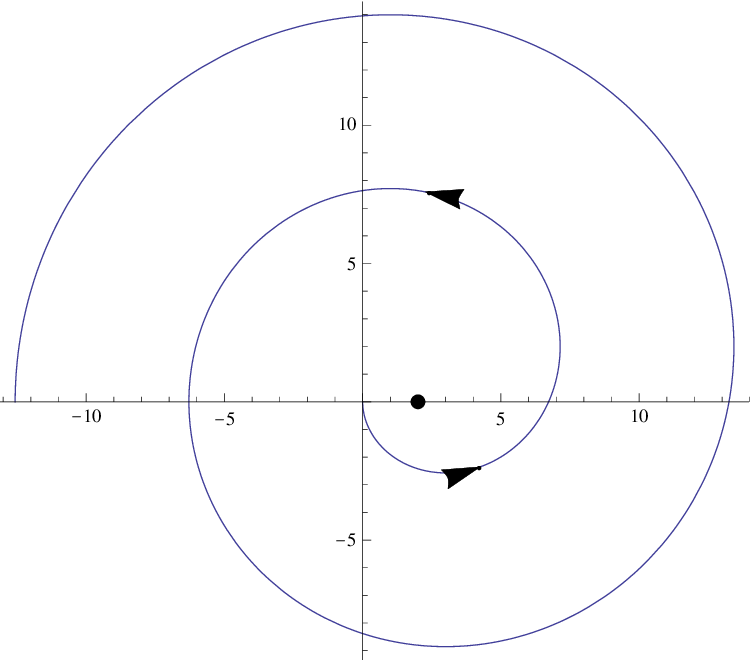}
\caption{$(\prx,\pry)$ motion plot for \protect\subref{spiral_xyvw}.}
\label{spiral_2d}
\end{subfigure}
\caption{Trajectories of the robot over time (top) or in planar space (bottom).}
\label{fig:motion}
\end{figure}

\rref{fig:motion}\subref{circlea_xyvw} depicts the position and velocity changes of a robot accelerating on a circle around a center point $\pc=(2,0)$. 
The robot starts at $\pr=(0,0)$ as initial position, with $\vr=2$ as initial translational velocity and $\omegar=1$ as initial rotational velocity;
\rref{fig:motion}\subref{circlea_2d} shows the resulting circular trajectory.
\rref{fig:motion}\subref{circle-b_xyvw} and \rref{fig:motion}\subref{circle-b_2d} show the resulting curve when braking (the robot brakes along the curve and comes to a complete stop before completing the circle).
If the rotational velocity is constant ($\D{\omegar}=0$), the robot drives an Archimedean spiral with the translational and rotational accelerations controlling the spiral's separation distance ($\ar / \omegar^2$). 
The corresponding trajectories are shown in Figures \ref{fig:motion}\subref{spiral_xyvw} and \ref{fig:motion}\subref{spiral_2d}.
Proofs for dynamics with spinning ($\rcurve = 0$, $\omegar\neq0$) and Archimedean spirals ($\D{\omegar}=0$, $\ar\neq0$) are available with \KeYmaera, but we do not discuss them here.

We assume bounds for the permissible acceleration $\ar$ in terms of a maximum acceleration $A \geq 0$ and braking power $b > 0$, as well as a bound $\Omega$ on the permissible rotational velocity $\omegar$. 
We use $\varepsilon$ to denote the upper bound for the control loop time interval (\eg, sensor and actuator delays, sampling rate, and computation time). 
That is, the robot might react quickly, but it can take no longer than time~$\varepsilon$ to react.
The robot would not be safe without such a time bound, because its control might then never run.
In our model, all these bounds will be used as symbolic parameters and not concrete numbers. 
Therefore, our results apply to \emph{all values} of these parameters and can be enlarged to include uncertainty.

\subsection{Obstacle State and Motion}

An obstacle has (vectorial) state variables describing its current position $\po=(\pox,\poy)$ and velocity $\vo=(\vox,\voy)$.
The obstacle model is deliberately liberal to account for many different obstacle behaviors. 
The only restriction about the dynamics is that the obstacle moves continuously with bounded velocity $\norm{\vo} \leq V$ while the physical system evolves for $\varepsilon$ time units. 
The original dynamic window algorithm considers the special case of $V=0$ (obstacles are stationary).
Depending on the relation of $V$ to $\varepsilon$, moving obstacles can make quite a difference, \eg, when other fast robots or the soccer ball meet slow communication-based virtual sensors as in RoboCup.\footnote{\url{http://www.robocup.org/}}

\section{Safety Verification of Ground Robot Motion}
\label{sec:models}

We want to prove motion safety of a robot whose controller tries to avoid obstacles.
Starting from a simplified robot controller, we develop increasingly more realistic models, and discuss different safety notions.
\emph{Static safety} describes a vehicle that never collides with stationary obstacles.
\emph{Passive safety}~\cite{Macek2009} considers a vehicle to be safe if no collisions happen while it moves (\ie, the vehicle does not itself collide with obstacles, so if a collision occurs at all then while the vehicle was stopped).
The intuition is that if collisions happen while our robot is stopped, then it must be the moving obstacle's fault.
Passive safety, however, puts some of the burden of avoiding collisions on other objects.
We, thus, also prove the stronger \emph{passive friendly safety} \cite{Macek2009}, which guarantees that our robot will come to a stop safely under all circumstances and  will leave sufficient maneuvering room for moving obstacles to avoid a collision.\footnote{The robot ensures that there is enough room for the obstacle to stop before a collision occurs. 
If the obstacle decides not to, then the obstacle is to blame and our robot is still considered safe.}
Finally, we prove \emph{passive orientation safety}, which accounts for limited sensor coverage of the robot and its orientation to reduce the responsibility of the robot in structured spaces, such as on roads with lanes.

\rref{tab:safetyoverview} gives an overview of the safety notions (both formally and informally) and the assumptions made about the robot and the obstacle in our models.
We consider all four models and safety properties to show the differences between the required assumptions and the safety guarantees that can be made.
The verification effort and complexity difference is quite instructive.
Static safety provides a strong guarantee with a simple safety proof, because only the robot moves.
Passive safety can be guaranteed by proving safety of all robot choices, whereas passive friendly safety requires additional liveness proofs for the obstacle. 
In the following sections, we discuss models and verification of the collision avoidance algorithm in detail.

\begin{table}[H]
\small\sf\centering
\caption{Overview of safety notions, responsibilities of the robot and its assumptions about obstacles}
\label{tab:safetyoverview}
\begin{tabularx}{\textwidth}{p{1.5cm}XX}
\toprule
\textbf{Safety} & \textbf{Responsibility of Robot} & \textbf{Assumptions about Obstacles}
\tabularnewline
\midrule
\multirow{3}{*}{\begin{minipage}{1.5cm}Static (\rref{model:dynamicwindowstatic})\end{minipage}} & Positive distance to all stationary obstacles &
Obstacles remain stationary and never move 
\tabularnewline
& $\norm{\pr - \po} > 0$ & $\vo = 0$ 
\tabularnewline
\cmidrule(ll){2-3}
& \multicolumn{2}{c}{Safety (\rref{thm:staticsafety}, feasible initial conditions $\phi_\text{ss}$):\hspace{1em} $\phi_\text{ss} \limply \dbox{\text{\rref{model:dynamicwindowstatic}}}\bigl(\norm{\pr - \po} > 0\bigr)$}
\tabularnewline
\midrule
\multirow{3}{*}{\begin{minipage}{1.5cm}Passive (\rref{model:dynamicwindowpassive})\end{minipage}}
& Positive distance to all obstacles while driving & Known maximum velocity \(V\) of obstacles
\tabularnewline
& $\vr \neq 0 \limply \norm{\pr - \po} > 0$ & $0 \leq \vo \leq V$ 
\tabularnewline
\cmidrule(ll){2-3}
& \multicolumn{2}{c}{Safety (\rref{thm:passivesafety}, feasible initial conditions $\phi_\text{ps}$):\hspace{1em} $\phi_\text{ps} \limply \dbox{\text{\rref{model:dynamicwindowpassive}}} \bigl(\vr \neq 0 \limply \norm{\pr - \po} > 0\bigr)$}
\tabularnewline
\midrule
\multirow{6}{*}{\begin{minipage}{1.5cm}Passive Friendly (\rref{model:dynamicwindowpassivefriendly2}+\ref{model:obstacle})\end{minipage}}
& Sufficient maneuvering space for obstacles & Known maximum velocity \(V\), minimum braking capability \(b_o\), and maximum reaction time \(\tau\) 
\tabularnewline
& $\vr \neq 0 \limply \norm{\pr - \po} > \frac{V^2}{2 b_o}+\tau V$ & $0 \leq \vo \leq V \land b_o > 0 \land \tau \geq 0$
\tabularnewline
\cmidrule(ll){2-3}
& \multicolumn{2}{c}{Safety (\rref{thm:passivefriendlysafety}, feasible initial conditions $\phi_\text{pfs}$):}\\ 
& \multicolumn{2}{l}{robot retains space
$\phi_\text{pfs} \limply  \dbox{\text{\rref{model:dynamicwindowpassivefriendly2}}} \big(\vr \neq 0 \limply \norm{\pr - \po} > \frac{V^2}{2 b_o}+\tau V\big)$}\\ & \multicolumn{2}{l}{obstacles can stop\hspace{1em}
$\phi_\text{pfs} \land \vr = 0 \land \norm{\pr - \po} > \frac{V^2}{2 b_o} + \tau V \limply \didia{\text{\rref{model:obstacle}}}\big(\norm{\pr - \po} > 0 \land \vo = 0\big)$}
\tabularnewline
\midrule
\multirow{6}{*}{\begin{minipage}{1.5cm}Passive Orientation (\rref{model:fieldOfView})\end{minipage}}
& Positive distance to all obstacles while driving, unless an invisible obstacle interfered with the robot while the robot cautiously stayed inside its observable region 
& Known maximum velocity \(V\) of obstacles
\tabularnewline
& $\begin{aligned}
\vr \neq 0 \limply \bigl(&\norm{\pr - \po} > 0\\ & \lor \left(\inView \leq 0 \land \abs{\beta} < \gamma\right) \bigr)
\end{aligned}$
& $0 \leq \vo \leq V$ 
\tabularnewline
\cmidrule(ll){2-3}
& \multicolumn{2}{c}{Safety (\rref{thm:passiveorientationsafety}):\hspace{1em} $\phi_\text{pos} \limply \dbox{\text{\rref{model:fieldOfView}}}\Big(\vr \neq 0 \limply \norm{\pr - \po} > 0 \lor \left(\inView \leq 0 \land \abs{\beta} < \gamma\right) \Big)$}
\tabularnewline
\bottomrule
\end{tabularx}
\end{table}

For the sake of clarity, we initially make the following simplifying assumptions to get an easier first model:
\begin{enumerate}[label=A{\arabic*}]
\item\label{assumption:maxab} in its decisions, the robot will use maximum braking or maximum acceleration, no intermediate controls, 
\item\label{assumption:forwardonly} the robot will not reverse its direction, but only drive smooth curves in forward direction, and
\item\label{assumption:nocirclecenter} the robot will not keep track of the center of the circle around which its current trajectory arc is taking it, but chooses steering through picking a curve radius.
\end{enumerate}
In \rref{sec:refinedsafety} we will see how to remove these simplifications again.

The subsections are structured as follows: we first discuss the rationale behind the model (see paragraphs \emph{Modeling}) and provide an intuition why the control choices in this model are safe (see paragraphs \emph{Identification of Safe Controls}).
Finally, we formally verify the correctness of the model, \ie, use the model in a correctness theorem and summarize the proof that the control choices indeed guarantee the model to satisfy the safety condition (see paragraphs \emph{Verification}).
Whether the model adequately represents reality is a complementary question that we discuss in \rref{sec:monitoring}.

\subsection{Static Safety with Maximum Acceleration}
\label{sec:models-staticmaxa}

In environments with only stationary obstacles, static safety ensures that the robot will never collide. 

\paragraph{Modeling}

The prerequisite for obtaining a formal safety result is to first formalize the system model in addition to its desired safety property.
We develop a model of the collision avoidance algorithm as a hybrid program, and express static safety as a safety property in \dL. 

As in the dynamic window algorithm, the collision avoidance controller uses the distance to the nearest obstacle for every possible curve to determine admissible velocities (\eg, compute distances in a loop and pick the obstacle with the smallest).
Instead of modeling the algorithm for searching the nearest obstacle and computing its closest perimeter point explicitly, our model exploits the power of nondeterminism to model this concisely.
It nondeterministically picks \emph{any} obstacle $\humod{\po}{(*,*)}$ and tests its safety.
Since the choice of the obstacle to consider was nondeterministic and the model is only safe if it is safe for \emph{all} possible ways of selecting \emph{any} obstacle nondeterministically, this includes safety for the closest perimeter point of the nearest obstacle (ties are included) and is thus safe for all possible obstacles.
Explicit representations of multiple obstacles will be considered in \rref{sec:models-nobstacles}.

In the case of non-point obstacles, $\po$ denotes the obstacle perimeter point that is closest to the robot (this fits naturally to obstacle point clouds delivered by radar and Lidar sensors, from which the closest point on the arc will be chosen).
In each controller run of the robot, the position $\po$ is updated nondeterministically (to consider any obstacle including the ones that now became closest).
In this process, the robot may or may not discover a new safe trajectory.
If it does, the robot can follow that new safe trajectory \wrt any nondeterministically chosen obstacle.
If not, the robot can still brake on the previous trajectory, which was shown to be safe in the previous control cycle for any obstacle, including the obstacle chosen in the current control cycle.

\rref{model:robotcontroller} summarizes the robot controller, which is parameterized with a $\drive$ action and a condition $\safe$ identifying when it is safe to take this $\drive$ action.
The formula $\safe$ is responsible for selecting control choices that keep the robot safe when executing control action $\drive$.

\begin{model}[htb]
\caption{Parametric robot controller model}
\label{model:robotcontroller}
\begin{align}
\ctrl_r&(\pmb{\drive}, \pmb{\safe}) \equiv \notag\\
\label{eq:robot:2-1} & \phantom{\cup}~~ (\humod{\ar}{-b})\\
\label{eq:robot:2-2} & \cup (\ptest{(\vr = 0)};~ \humod{\ar}{0};~\humod{\omegar}{0})\\
\label{eq:robot:2-3} & \cup (\pmb{\drive};~ \humod{\omegar}{*};~ \ptest{({-\Omega} \leq \omegar \leq \Omega)};~ \humod{\rcurve}{*};~\humod{\po}{(*,*)}; ~\ptest{(\curve \land \pmb{\safe})})\\
\label{eq:robot:3-1}\curve & \equiv \rcurve \neq 0 \land \rcurve \omegar = \vr
\end{align}
\end{model}

The robot is allowed to brake at all times since the assignment that assigns full braking to $\ar$ in \eqref{eq:robot:2-1} has no test. 
If the robot is stopped (\(\vr=0\)), it may choose to stay in its current spot without turning, cf. \eqref{eq:robot:2-2}.
Finally, if it is safe to accelerate, which is what formula parameter {\safe} determines, then the robot may choose a new safe curve in its dynamic window.
That is, it performs action {\drive} (e.g. maximum acceleration) and chooses any rotational velocity in the bounds, cf. \eqref{eq:robot:2-3} and computes the corresponding radius $r$ according to the condition \eqref{eq:robot:3-1}.
This corresponds to testing all possible rotational velocity values at the same time and choosing some that passes condition {\safe}.
An implementation in an imperative language would use loops to enumerate all possible values and all obstacles and test each pair $(\vr,\omegar)$ separately \wrt every obstacle, storing the admissible pairs in a data structure (as \eg, in \cite{DBLP:journals/arobots/TaubigFHLMVW12}).

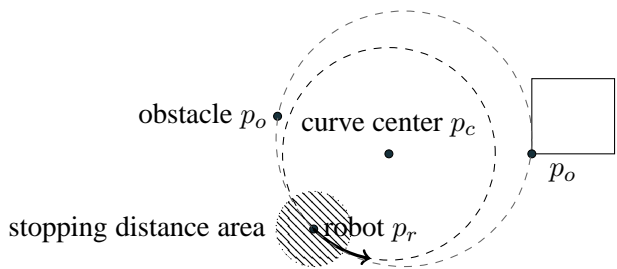
\begin{wrapfigure}{r}{.5\textwidth}
\centering
\begin{tikzpicture}
\draw[fill=white] (2.9,1) rectangle (4,2);
\draw[fill=lsblue] (2.9,1) circle (1.5pt);
\draw[color=black,anchor=north west] node at (3,1) {obstacle $\po$};
\draw[pattern=north west lines,style={dotted,thin}] (0,0) circle (0.5);
\draw[color=black,anchor=east,text width=2.2cm,align=right,xshift=-2em] node at (0,0) {stopping area\\ around robot $\pr$};
\draw[fill=lsblue] (0,0) circle (1.5pt);
\draw[->,color=black,dashed] (0,0) arc (225:320:1.2);
\draw[->,color=black,very thick] (0,0) arc (225:280:1.7);
\draw[->,color=black,dashed] (0,0) arc (225:320:2.3);
\draw[fill=lsblue] (1.2,1.2) circle (1.5pt);
\draw[color=black,anchor=south] node at (1,1.2) {curve center $\pc$};
\draw[color=black!60,dashed] (1.2,1.2) circle (1.7);
\draw[color=black] (0,0) -- node[anchor=south] {$\pr-\po$}  (2.9,1);
\draw[fill=white] (2.3,2.2) rectangle (2.8,2.7);
\end{tikzpicture}
\caption{Illustration of static safety: the robot must stop before reaching the closest obstacle on a curve (three of infinitely many curves illustrated).
Obstacles with shapes reduce to single points by considering the perimeter point that is closest to the robot.
}
\label{fig:staticsafetyillustration}
\end{wrapfigure}

The curve is determined by the robot following a circular trajectory of radius $\rcurve$ with angular velocity $\omegar$ starting in initial direction $\dr$, cf. \eqref{eq:robot:2-3}.
The trajectory starts at $\pr$ with translational velocity $\vr$ and rotational velocity $\omegar$, as defined by $\rcurve \omegar =\vr$ in \eqref{eq:robot:3-1}.
This condition ensures that we simultaneously pick an admissible angular velocity $\omegar$ according to \cite[(14)]{DBLP:journals/ram/FoxBT97} when choosing an admissible velocity $\vr$.
Together with the orientation $\dr$ of the robot, which is tangential to the curve, this also implicitly characterizes the rotation center $\pc$; see \rref{fig:staticsafetyillustration}.
We will explicitly represent the rotation center in \rref{appendix:models-distance} for more aggressive maneuvering. For starters, we only need to know how to steer by $\rcurve$.
For the sake of clarity we restrict the study to circular trajectories with non-zero radius ($\rcurve \neq 0$ so that the robot is not spinning on the spot). 
We do not include perfectly straight lines in our model, but instead mimic the control principle of a real robot that will control periodically to adjust for actuator perturbation and drift when trying to drive straight lines, so it suffices to approximate straight-line driving with large curve radii ($\rcurve$ approaches infinity).
The sign of the radius signifies if the robot follows the curve in clockwise ($\rcurve < 0$) or counter-clockwise direction ($\rcurve > 0$).
Since $\rcurve \neq 0$, the condition $\D{(\rcurve \omegar)}=\ar$ can be rewritten as differential equation $\D{\omegar}=\tfrac{\ar}{\rcurve}$. 
The distance to the nearest obstacle on that curve is measured by $\humod{\po}{(*,*)}$ in \eqref{eq:robot:2-3}.

\begin{model}[htb]
\caption{Dynamic window with static safety}
\label{model:dynamicwindowstatic}
\begin{align}
\label{eq:st:0}\textit{dw}_{\text{ss}}  & \equiv \prepeat{\bigl(\ctrl_r(\humod{\ar}{A}~,~ \safe_\text{ss});\dyn_\text{ss}\bigr)}\\
\label{eq:st:3-2}\safe_\text{ss} & \equiv 
\norm{\pr - \po}{_\infty} > \frac{\vr^2}{2 b} +\left(\frac{A}{b} + 1\right) \left(\frac{A}{2} \varepsilon^2 + \varepsilon \vr\right)\\
\label{eq:st:4-1}\dyn_\text{ss} & \equiv \humod{t}{0};~ \{ \D{t}=1 \syssep~ \D{{\pr}}=\vr \dr \syssep~ \D{\vr}=\ar \syssep \D{\dr}=\omegar \dr^\bot \syssep~ \D{\omegar}=\frac{\ar}{\rcurve} ~\&~ \vr \geq 0 \land t \leq \varepsilon \}
\end{align}
\end{model}

\rref{model:dynamicwindowstatic} represents the common controller-plant model: it repeatedly executes the robot control choices followed by dynamics, cf. \eqref{eq:st:0}.
Recall that the arbitrary number of repetitions is indicated by the $\prepeat{}$ at the end.
The continuous dynamics of the robot from \rref{sec:dynmodel} above is defined in \eqref{eq:st:4-1} of \rref{model:dynamicwindowstatic}.

\paragraph{Identification of Safe Controls} \label{par:staticsafety-control}
The most critical element of \rref{model:dynamicwindowstatic} is the choice of the formula \(\safe_\text{ss}\) in \eqref{eq:st:3-2} that we chose for parameter {\safe}.
This formula is responsible for selecting control choices that keep the robot safe.
While its ultimate justification will be the safety proof (\rref{thm:staticsafety}), this section explains intuitively why we chose the particular design in \eqref{eq:st:3-2}.
Generating such conditions is possible, see \cite{DBLP:journals/sttt/QueselMLAP16} for an approach how to phrase conjectures with unknown constraints in \dL and use theorem proving to discover constraints that make a formula provable.

A circular trajectory of radius $\rcurve$ ensures static safety if it allows the robot to stop before it collides with the nearest obstacle.
Consider the extreme case where the radius $\rcurve =\infty$ is infinitely large and the robot, thus, travels on a straight line.
In this case, the distance between the robot's current position $\pr$ and the nearest obstacle $\po$ must account for the following components: First, the robot  needs to be able to brake from its current velocity $\vr$ to a complete stop (equivalent to \cite[(14)]{DBLP:journals/ram/FoxBT97} characterizing admissible velocities), which takes time \(\tfrac{\vr}{b}\) and requires distance \(\tfrac{\vr^2}{2b}\): 

\begin{equation}\label{eq:st-stop}
\tfrac{\vr^2}{2 b} = \int_0^{\vr/b}(\vr-bt)dt \enspace .
\end{equation}

Second, it may take up to $\varepsilon$ time until the robot can take the next control decision. 
Thus, we must take into account the distance that the robot may travel \wrt the maximum acceleration $A$ and the distance needed for compensating its acceleration of $A$ during that reaction time with braking power $b$ (compensating for the speed increase \(A \varepsilon\) takes time \(\tfrac{A \varepsilon}{b}\)):

\begin{equation}\label{eq:st-compensate}
\left(\tfrac{A}{b}+1\right)\left(\tfrac{A}{2}\varepsilon^2+\varepsilon \vr\right)=\int_0^{\varepsilon}(\vr+At)dt + \int_0^{A\varepsilon/b}(\vr+A\varepsilon-bt)dt \enspace .
\end{equation}

The safety distance chosen for \(\safe_\text{ss}\) in \eqref{eq:st:3-2} of \rref{model:dynamicwindowstatic} is the sum of the distances \eqref{eq:st-stop} and \eqref{eq:st-compensate}.
The safety proof will have to show that this construction was indeed safe and that it is also safe for all other curved trajectories that the obstacle and robot could be taking in the model instead.

To simplify the proof's arithmetic, we measure the distance between the robot's position $\pr$ and the obstacle's position $\po$ in the infinity-norm $\norm{\pr-\po}{_\infty}$, \ie, either $\abs{\prx-\pox}$ or $\abs{\pry-\poy}$ must be safe. 
In the illustrations, this corresponds to replacing the circles representing reachable areas with outer squares.
This over-approximates the Euclidean norm distance $\norm{\pr-\po}{_2} = \sqrt{(\prx-\pox)^2+(\pry-\poy)^2}$ by a factor of at most $\sqrt{2}$.

\paragraph{Verification}
With the proof calculus of \dL~\cite{DBLP:journals/jar/Platzer08,DBLP:conf/lics/Platzer12a,Platzer10,DBLP:journals/jar/Platzer17},
we verify the safety of the control algorithm in \rref{model:dynamicwindowstatic}. 
The robot is safe, if it maintains positive distance $\norm{\pr-\po} > 0$ to (nondeterministic so any) obstacle $\po$ (see \rref{tab:safetyoverview}), \ie, it always satisfies:

\begin{equation}\label{eq:st-safe}
\psi_\text{ss} ~\equiv~ \norm{\pr-\po} > 0 \enspace .
\end{equation}

In order to guarantee $\psi_\text{ss}$ always holds, the robot must stay at a safe distance, which still allows the robot to brake to a complete stop before hitting any obstacle.
The following condition captures this requirement as an invariant $\varphi_\text{ss}$ that we prove to hold for all executions of the loop in \eqref{eq:st:0}:
\begin{equation}\label{eq:st-invariant}
\varphi_{\text{ss}} ~\equiv~ \norm{\pr-\po} > \frac{\vr^2}{2 b} \enspace .
\end{equation}
Formula~\eqref{eq:st-invariant} says that the robot and the obstacle are safely apart. 
In this case, the safe distance in the loop invariant coincides with \eqref{eq:st-stop}, which describes the stopping distance.

We prove that the property~\eqref{eq:st-safe} holds for all executions of \rref{model:dynamicwindowstatic} (so also all obstacles) under the assumption that we start in a state satisfying the symbolic parameter assumptions ($A\geq0$, $V\geq0$, $\Omega\geq0$, $b>0$, and $\varepsilon>0$) as well as the following initial conditions:
\begin{equation}
\label{eq:st-initial}
\phi_\text{ss} ~\equiv~ \vr=0 \land \norm{\pr-\po} > 0 \land \rcurve \neq 0 \land \norm{\dr} = 1 \enspace .
\end{equation}
The first two conditions of the conjunction formalize that the robot is stopped at a safe distance initially. 
The third conjunct states that the robot is not spinning initially. 
The last conjunct $\norm{\dr}=1$ says that the direction $\dr$ is a unit vector.
Any other formula $\phi_\text{ss}$ implying invariant $\varphi_\text{ss}$ is a safe starting condition as well (\eg, driving with sufficient space, so invariant $\varphi_\text{ss}$ itself).

\begin{thm}[Static safety]\label{thm:staticsafety}
Robots following \rref{model:dynamicwindowstatic} never collide with stationary obstacles as expressed by the provable \dL formula
\(\phi_\text{ss} \limply \dibox{\textit{dw}_\text{ss}}{\psi_\text{ss}} \enspace.\)
\end{thm}

\begin{proof}
We proved \rref{thm:staticsafety} for circular trajectories in \KeYmaeraX.
The proof uses the invariant $\varphi_\text{ss}$ \eqref{eq:st-invariant} for handling the loop.
It uses \emph{differential cuts} with \emph{differential invariants} \eqref{eq:ss-diffinvariants-time}--\eqref{eq:ss-diffinvariants-robposy}---an induction principle for differential equations \cite{DBLP:journals/lmcs/Platzer12}---to prove properties about \(\dyn\) without requiring symbolic solutions.

\begin{align}
t \geq 0 \label{eq:ss-diffinvariants-time}\\
\norm{\dr} = 1 \label{eq:ss-diffinvariants-orientation}\\
\vr = \old(\vr) + \ar t \label{eq:ss-diffinvariants-speed}\\
-t\left(\vr - \frac{\ar}{2}t\right) \leq \prx - \old(\prx) \leq t\left(\vr - \frac{\ar}{2}t\right) \label{eq:ss-diffinvariants-robposx}\\
-t\left(\vr - \frac{\ar}{2}t\right) \leq \pry - \old(\pry) \leq t\left(\vr - \frac{\ar}{2}t\right) \label{eq:ss-diffinvariants-robposy}
\end{align}

The differential invariants capture that time progresses \eqref{eq:ss-diffinvariants-time}, that the orientation stays a unit vector \eqref{eq:ss-diffinvariants-orientation}, that the new speed \(\vr\) is determined by the previous speed \(\old(\vr)\) and the acceleration \(\ar\) \eqref{eq:ss-diffinvariants-speed} for time $t$, and that the robot does not leave the bounding square of half side length \(t(\vr - \frac{\ar}{2}t)\) around its previous position \(\old(\pr)\) \eqref{eq:ss-diffinvariants-robposx}--\eqref{eq:ss-diffinvariants-robposy}. 
The function \(\old(\cdot)\) is shorthand notation for an auxiliary or ghost variable that is initialized to the value of \(\cdot\) before the ODE.
\end{proof}
\clearpage

\subsection{Passive Safety with Maximum Acceleration}
\label{sec:models-passivemaxa}

In the presence of moving obstacles, collision freedom gets significantly more involved, because, even if our robot is doing the best it can, other obstacles could still actively try to crash into it.

\begin{wrapfigure}{r}{.45\textwidth}
\centering
\begin{tikzpicture}
\coordinate (obstacle) at (-0.8,1.5);
\draw[fill=lsred] (obstacle) circle (1);
\draw[color=black,anchor=south east,text width=20ex] node at (-1.3,1.8) {obstacle reach area until robot stopped};
\draw[fill=lsblue] (obstacle) circle (1.5pt);
\draw[color=black,dashed,->] (obstacle) -- ($(obstacle)+(0.1,-1)$);
\draw[color=black,dashed,->] (obstacle) -- ($(obstacle)+(-0.5,-0.8)$);
\draw[color=black,dashed,->] (obstacle) -- ($(obstacle)+(-0.4,0.7)$);
\draw[color=black,dashed,->] (obstacle) -- ($(obstacle)+(0,0.4)$);
\draw[color=black,dashed,->] (obstacle) -- ($(obstacle)+(0.6,-0.7)$);
\draw[color=black,dashed,->] (obstacle) -- ($(obstacle)+(0.9,-0.3)$);
\draw[color=black,dashed,->] (obstacle) -- ($(obstacle)+(0.6,0.5)$);
\draw[color=black,anchor=east] node[fill=white] at ($(obstacle)+(-0.2,0)$) {obstacle $\po$};
\draw[pattern=north west lines,style={dotted,thin}] (0,0) circle (0.5);
\draw[color=black,anchor=east] node at (-0.5,0) {stopping area};
\draw[fill=lsblue] (0,0) circle (1.5pt);
\node[color=black,fill=white,anchor=west] at (0,0) {robot $\pr$};
\draw[color=black,dashed] (1,1) circle (1.414);
\draw[->,color=black,very thick] (0,0) arc (225:260:1.414);
\draw[fill=lsblue] (1,1) circle (1.5pt);
\draw[color=black,anchor=south,text width=1.5cm,align=center] node at (1,1.1) {curve center $\pc$};
\end{tikzpicture}
\caption{Illustration of passive safety: the area reachable by the robot until it can stop must not overlap with the area reachable by the obstacle during that time.}
\label{fig:passivesafetyillustration}
\end{wrapfigure}
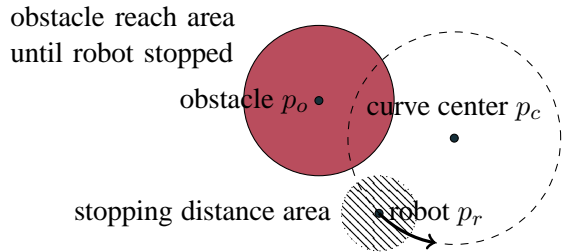

Passive safety, thus, considers the robot safe if no collisions can happen while it is driving. The robot, thus, needs to be able to come to a full stop before making contact with any obstacle, see \rref{fig:passivesafetyillustration}.
Intuitively, when every moving robot and obstacle follows passive safety then there will be no collisions.
Otherwise, if careless or malicious obstacles are moving in the environment, passive safety ensures that at least our own robot is stopped so that collision impact is kept small.
In this section, we will develop a robot controller that provably ensures passive safety.
We remove the restriction that obstacles cannot move, but the robot and the obstacle will decide on their next maneuver at the same time and they are still subject to the simplifying assumptions \ref{assumption:maxab}--\ref{assumption:nocirclecenter}.

\paragraph{Modeling}

\begin{model*}[tb]
\caption{Dynamic window with passive safety}
\label{model:dynamicwindowpassive}
\begin{align}
\label{eq:pf:0}\textit{dw}_{\text{ps}}  & \equiv \prepeat{(\ctrlo;\ctrlr(\humod{\ar}{A}~,~\safe_\text{ps}); \dyn_\text{ps})}\\
\label{eq:pf:1-1}\pmb{\ctrlo} & \pmb{\equiv \humod{\vo}{(*,*)};~ \ptest{\norm{\vo} \leq V}}\\
\label{eq:pf:3-2}\safe_\text{ps} & \equiv 
\norm{\pr - \po}{_\infty} > \frac{\vr^2}{2 b} + \pmb{V \frac{\vr}{b}} +\left(\frac{A}{b} + 1\right) \left(\frac{A}{2} \varepsilon^2 + \varepsilon (\vr\pmb{+V})\right)\\
\label{eq:pf:4-1}\dyn_\text{ps} & \equiv \humod{t}{0};~ \{ \D{t}=1 \syssep~ \pmb{\D{p}_o=\vo} \syssep~ \D{p}_r=\vr \dr \syssep~ \D{v}_r=\ar \syssep~ \D{d}_r=\omegar \dr^\bot \syssep~ \D{\omega}_r=\frac{\ar}{\rcurve} ~\&~ \vr \geq 0 \land t \leq \varepsilon \}
\end{align}
\end{model*}

We refine the collision avoidance controller and  model to include moving obstacles, and state its passive safety property in \dL. 
In the presence of moving obstacles all obstacles must be considered and tested for safety.
The main intuition here is that all obstacles will respect a maximum velocity $V$, so the robot is safe when it is safe for the worst-case behavior of the nearest obstacle.
Our model again exploits the power of nondeterminism to model this concisely by picking \emph{any} obstacle $\humod{\po}{(*,*)}$ and testing its safety.
In each controller run of the robot, the position $\po$ is updated nondeterministically (which includes the ones that are now closest because the robot and obstacles moved).
If the robot finds a new safe trajectory, then it will follow it (the velocity bound $V$ ensures that all obstacles will stay more distant than the worst-case of the nearest one chosen nondeterministically).
Otherwise, the robot will stop on the current trajectory, which was tested to be safe in the previous controller decision.

\rref{model:dynamicwindowpassive} follows a setup similar to \rref{model:dynamicwindowstatic}.
The continuous dynamics of the robot and the obstacle as presented in \rref{sec:dynmodel} above are defined in \eqref{eq:pf:4-1} of \rref{model:dynamicwindowpassive}.

The control of the robot is executed after the control of the obstacle, cf. \eqref{eq:pf:0}.
Both robot and obstacle only write to variables that are read in the dynamics, but not in the controller of the respective other agent.
Therefore, we could swap the controllers to $\ctrl_r;\textit{ctrl}_o$, or use a nondeterministic choice of $\pchoice{(\textit{ctrl}_o;\ctrl_r)}{(\ctrl_r;\textit{ctrl}_o)}$ to model independent parallel execution \cite{DBLP:conf/ifm/MullerMRSP16}.
Fixing one specific ordering $\textit{ctrl}_o;\ctrl_r$ reduces proof effort, because it avoids branching the proof into all the different possible execution orders (which in this case differ only in their intermediate computations but have the same effect on motion).

The obstacle may choose any velocity in any direction up to the maximum velocity $V$ assumed about obstacles ($\norm{\vo}\leq V$), cf. \eqref{eq:pf:1-1}.
This uses the modeling pattern from \rref{sec:dl}. 
We assign an arbitrary (two-dimensional) value to the obstacle's velocity ($\humod{\vo}{(*,*)}$), which is then restricted by the maximum velocity with a subsequent test ($\ptest{\norm{\vo}\leq V}$).
Overall, \eqref{eq:pf:1-1} allows obstacles to choose an arbitrary velocity in any direction, but at most of speed $V$.
Analyzing worst-case situations with a powerful obstacle that supports sudden direction and velocity changes is beneficial, since it keeps the model simple while it simultaneously allows \KeYmaeraX to look for unusual corner cases.

The robot follows the same control as in \rref{model:dynamicwindowstatic} but includes differential equations for the obstacle.
The main difference to \rref{model:dynamicwindowstatic} is the $\safe$ condition \eqref{eq:pf:3-2}, which now has to account for the fact that obstacles may move according to \eqref{eq:pf:4-1} while the robot tries to avoid collision.
The difference of \rref{model:dynamicwindowpassive} compared to \rref{model:dynamicwindowstatic} is highlighted in \textbf{boldface}.

\paragraph{Identification of Safe Controls} \label{par:passivesafety-control}
The most critical element is again the formula $\safe_\text{ps}$ that control choices need to satisfy in order to always keep the robot safe.
We extend the intuitive explanation from static safety to account for the additional obstacle terms in \eqref{eq:pf:3-2}, again considering the extreme case where the radius $\rcurve =\infty$ is infinitely large and the robot, thus, travels on a straight line.
The robot must account for the additional impact over the static safety margin \eqref{eq:st-compensate} from the motion of the obstacle.
During the stopping time ($\varepsilon + \tfrac{\vr+A\varepsilon}{b}$) entailed by \eqref{eq:st-stop} and \eqref{eq:st-compensate}, the obstacle might approach the robot, \eg, on a straight line with maximum velocity $V$ to the point of collision:

\begin{equation}\label{eq:pf-obsapproach}
V\left(\varepsilon + \frac{\vr+A\varepsilon}{b}\right) = V \left(\frac{\vr}{b} + \left(\frac{A}{b} + 1\right)\varepsilon\right) \enspace .
\end{equation}

The safety distance chosen for $\safe_\text{ps}$ in \eqref{eq:pf:3-2} of \rref{model:dynamicwindowpassive} is the sum of the distances \eqref{eq:st-stop}, \eqref{eq:st-compensate}, and \eqref{eq:pf-obsapproach}.
The safety proof will have to show that this construction was safe and that it is also safe for all other curved trajectories that the obstacle and robot could be taking instead.

\paragraph{Verification}
The robot in \rref{model:dynamicwindowpassive} is safe, if it maintains positive distance $\norm{\pr-\po} > 0$ to the obstacle while the robot is driving (see \rref{tab:safetyoverview}):

\begin{equation}\label{eq:ps-safe}
\psi_\text{ps} ~\equiv~ \vr \neq 0 \limply \left(\norm{\pr-\po} > 0 \right) \enspace .
\end{equation}

In order to guarantee $\psi_\text{ps}$, the robot must stay at a safe distance, which still allows the robot to brake to a complete stop before the approaching obstacle reaches the robot.
The following condition captures this requirement as an invariant $\varphi_\text{ps}$ that we prove to hold for all loop executions:
\begin{equation}\label{eq:ps-invariant}
\varphi_{\text{ps}} ~\equiv~ \vr \neq 0 \limply \left(\norm{\pr-\po} > \frac{\vr^2}{2 b} + V\frac{\vr}{b}\right) \enspace .
\end{equation}

Formula~\eqref{eq:ps-invariant} says that, while the robot is driving, the positions of the robot and the obstacle are safely apart. 
This accounts for the robot's braking distance $\tfrac{\vr^2}{2b}$ while the obstacle is allowed to approach the robot with its maximum velocity $V$ in time $\tfrac{\vr}{b}$.
We prove that formula~\eqref{eq:ps-safe} holds for all executions of \rref{model:dynamicwindowpassive} when started in a non-collision state as for static safety, \ie, $\phi_\text{ps} \equiv \phi_\text{ss}$ \eqref{eq:st-initial}. 

\begin{thm}[Passive safety]\label{thm:passivesafety}
Robots following \rref{model:dynamicwindowpassive} will never collide with static or moving obstacles while driving, as expressed by the provable \dL formula
\(\phi_\text{ps} \limply \dibox{\textit{dw}_\text{ps}}{\psi_\text{ps}} \enspace.\)
\end{thm}

\begin{proof}
The \KeYmaeraX proof uses invariant $\varphi_\text{ps}$ \eqref{eq:ps-invariant}.
It extends the differential invariants \eqref{eq:ss-diffinvariants-time}--\eqref{eq:ss-diffinvariants-robposy} for static safety with invariants \eqref{eq:ss-diffinvariants-obsposx} about obstacle motion.
\begin{equation}
-tV \leq \pox - \old(\pox) \leq tV \label{eq:ss-diffinvariants-obsposx}\qquad, \qquad -tV \leq \poy - \old(\poy) \leq tV
\end{equation}
Similar to the robot, the obstacle does not leave its bounding square of half side length \(tV\) around its previous position \(\old(\po)\).
\end{proof}

\subsection{Passive Friendly Safety of Obstacle Avoidance}
\label{sec:models-pfs}

In this section, we explore the stronger requirements of passive friendly safety, where the robot not only stops safely itself, but also allows for the obstacle to stop before a collision occurs.
Passive friendly safety requires the robot to take careful decisions that respect the dynamic capabilities of moving obstacles.
The intuition behind passive friendly safety is that our own robot should retain enough space for other obstacles to stop.
Unlike passive safety, passive friendly safety ensures that there will not be collisions, as long as every obstacle makes a corresponding effort to avoid collision when it sees the robot, even when some obstacles approach intersections carelessly and turn around corners without looking.
The definition of \citet{Macek2009} requires that the robot respects the worst-case braking time of the obstacle, which depends on its velocity and control capabilities.
In our model, the worst-case braking time is a consequence of the following assumptions.
We assume an upper bound $\tau$ on the obstacle's reaction time and a lower bound $b_o$ on its braking capabilities.
Then, $\tau V$ is the maximal distance that the obstacle can travel before beginning to react and $\tfrac{V^2}{2b_o}$ is the maximal distance for the obstacle to stop from the maximal velocity $V$ with an assumed minimum braking capability $b_o$.

\paragraph{Modeling}
\rref{model:dynamicwindowpassivefriendly2} uses the same basic obstacle avoidance algorithm as \rref{model:dynamicwindowpassive}.
The difference is reflected in what the robot considers to be a safe distance to an obstacle.
As shown in \eqref{eq:pfs:3-2} the safe distance not only accounts for the robot's own braking distance, but also for the braking distance $\tfrac{V^2}{2b_o}$ and reaction time $\tau$ of the obstacle. 
The verification of passive friendly safety is more complicated than passive safety as it accounts for the behavior of the obstacle discussed below.

\begin{model}[h]
\caption{Dynamic window with passive friendly safety}
\label{model:dynamicwindowpassivefriendly2}
\begin{align}
\label{eq:pfs:0}\text{\it dw}_{\text{pfs}}  & \equiv \prepeat{\bigl(\ctrl_o;\ctrl_r(\humod{\ar}{A}~,~ \safe_\text{pfs});\dyn_\text{ps}\bigr)}\\
\label{eq:pfs:3-2}\safe_\text{pfs} & \equiv 
\norm{\pr - \po}{_\infty} > \frac{\vr^2}{2 b} + V\frac{\vr}{b} + \pmb{\frac{V^2}{2 b_o} + \tau V} + \left(\frac{A}{b} + 1\right) \left(\frac{A}{2} \varepsilon^2 + \varepsilon (\vr+V)\right)
\end{align}
\end{model}

In \rref{model:dynamicwindowpassivefriendly2} the obstacle controller $ctrl_o$ is a coarse model given by equation~\eqref{eq:pf:1-1} from \rref{model:dynamicwindowpassive}, which only constrains its non-negative velocity to be less than or equal to $V$.
Such a liberal obstacle model is useful for analyzing the robot, since it requires the robot to be safe even in the presence of rather sudden obstacle behavior (\eg, be safe even if driving behind an obstacle that stops instantaneously or changes direction radically). 
However, now that obstacles must avoid collision once the robot is stopped, such instantaneous behavior becomes too powerful.
An obstacle that can stop or change direction instantaneously can trivially avoid collision, which would not tell us much about real vehicles that have to brake before coming to a stop.
Here, instead, we consider a more interesting refined obstacle behavior with braking modeled similar to the robot's braking behavior by the hybrid program \textit{obstacle} given in \rref{model:obstacle}.

\begin{model}[h]
\caption{Refined obstacle with acceleration control}
\label{model:obstacle}
\begin{align}
\label{eq:obstacle:0}\textit{obstacle}  & \equiv \prepeat{(\textit{ctrl}_{\tilde{o}};~ \textit{dyn}_{\tilde{o}})} \\
\label{eq:obstacle:1-1}\textit{ctrl}_{\tilde{o}} & \equiv \humod{\ao}{*};~ \ptest{\vo + \ao \tau \leq V}\\
\label{eq:obstacle:2}\textit{dyn}_{\tilde{o}} & \equiv \humod{t}{0};~ \{\D{t}=1 \syssep~ \D{\po}=\vo \dobst \syssep \D{\vo}=\ao ~\&~ t \leq \tau \land \vo \geq 0\}
\end{align}
\end{model}

The refined obstacle may choose any acceleration $a_o$, as long as it does not exceed the velocity bound $V$ \eqref{eq:obstacle:1-1}.
In order to ensure that the robot does not force the obstacle to avoid collision by steering (\eg, other cars at an intersection should not be forced to change lanes), we keep the obstacle's direction unit vector $d_o$ constant.
The dynamics of the obstacle are straight ideal-world translational motion in the two-dimensional plane with reaction time $\tau$, see \eqref{eq:obstacle:2}.

\paragraph{Verification}
We verify the safety of the robot's control choices as modeled in \rref{model:dynamicwindowpassivefriendly2}.
Unlike the passive safety case, the passive friendly safety property $\phi_{\text{pfs}}$ should guarantee that if the robot stops, moving obstacles (cf. \rref{model:obstacle}) still have enough time and space to avoid a collision. 
The conditions $\vo = \sqrt{\vox^2+\voy^2} \land \dox\vo=\vox \land \doy\vo=\voy$ link the combined velocity and direction vector $(\vox,\voy)$ of the abstract obstacle model from the robot safety argument to the velocity scalar $\vo$ and direction unit vector $(\dox,\doy)$ of the refined obstacle model in the liveness argument.
This requirement can be captured by the following \dL formula:
\begin{equation}\label{eq:Epfs}
\eta_{\text{pfs}} \equiv~
\bigl(\eta_\text{obs} \land 0 \leq \vo \land \vo = \sqrt{\vox^2+\voy^2} \land \vo\dox=\vox \land \vo\doy=\voy \bigr) \limply \ddiamond{\textit{obstacle}}\left(\norm{\pr-\po} > 0 \land \vo = 0\right)
\end{equation}
\noindent
where the property $\eta_{\text{obs}}$ accounts for the stopping distance of the obstacle:
\(
\eta_{\text{obs}} ~\equiv~ \norm{\pr-\po} > \frac{V^2}{2 b_o} + \tau V \enspace .
\)
Formula~\eqref{eq:Epfs} says that there exists an execution of the hybrid program \emph{obstacle}, (existence of a run is formalized by the diamond operator $\ddiamond{\textit{obstacle}}$ in \dL), that allows the obstacle to stop ($\vo = 0$) without having collided ($\norm{\pr-\po}>0$).
Passive friendly safety $\psi_\text{pfs}$ is now stated as 
\[\psi_{\text{pfs}} ~\equiv~ (\vr \neq 0 \limply \eta_{\text{obs}}) \land \eta_{\text{pfs}} \enspace .\]
We study passive friendly safety with respect to initial states satisfying the following property:
\begin{equation}
\label{eq:pfs-initial}
\phi_\text{pfs} ~\equiv~ \eta_{obs} \land \rcurve \neq 0 \land \norm{\dr} = 1 \enspace .
\end{equation}
Observe that, in addition to the condition $\eta_{\text{pfs}}$, the difference to passive safety is reflected in the special treatment of the case $\vr=0$. 
Even if the robot starts with speed $\vr=0$ (which is passively safe), $\eta_{\text{obs}}$ must be satisfied to prove passive friendly safety, since otherwise the obstacle may initially start out too close and thus unable to avoid collision.
Likewise, we are required to prove $\eta_{\text{pfs}}$ as part of $\psi_{\text{pfs}}$ to guarantee that obstacles can avoid collision after the robot came to a full stop.

\begin{thm}[Passive friendly safety]\label{thm:passivefriendlysafety}
Robots following \rref{model:dynamicwindowpassivefriendly2} will never collide while driving and will retain sufficient safety distance for others to avoid collision, as expressed by the provable \dL formula
\(\phi_\text{pfs} \limply \dibox{\textit{dw}_\text{pfs}} \psi_\text{pfs} \enspace.\)
\end{thm}

\begin{proof}
The proof in \KeYmaeraX splits into a safety argument for the robot and a liveness argument for the obstacle.
The loop and differential invariants in the robot safety proof are similar in spirit to passive safety, but account for the additional obstacle reaction time and stopping distance $\frac{V^2}{2 b_o}+\tau V$.
The obstacle liveness proof bases on \emph{loop convergence}, \ie, it uses conditions that describe how much progress the loop body of the hybrid program $\textit{obstacle}$ can make towards stopping.
Intuitively, the obstacle has made sufficient progress if either it is stopped already or can stop by braking $n$ times:

\[\vo - n \tau \bo \leq 0 \lor \vo =0 \enspace . \]

Additionally, the convergence conditions include the familiar bounds on the parameters ($\norm{\dobst}=1$, $\bo>0$, $\tau>0$, and $0\leq\vo\leq V$) and the remaining stopping distance $\norm{\pr-\po}>\frac{\vo^2}{2\bo}$.
\end{proof}

The symbolic bounds on velocity, acceleration, braking, and time in the above models represent uncertainty implicitly (\eg, the braking power $b$ can be instantiated with the minimum specification of the robot's brakes, or with the actual braking power achievable \wrt the current terrain). 
Whenever knowledge about the current state is available, the bounds can be instantiated more aggressively to allow efficient robot behavior. 
For example, in a rare worst case we may face a particularly fast obstacle, but right now there are only slow-moving obstacles around.
Or the worst case reaction time $\varepsilon$ may be slow when difficult obstacle shapes are computed, but is presently quick as circular obstacles suffice to find a path.
Theorems~\ref{thm:staticsafety}--\ref{thm:passivefriendlysafety} are verified for all those values.
\rref{sec:models-nobstacles} illustrates how to explicitly model different kinds of obstacles simultaneously in a single model.
Other aspects of uncertainty need explicit changes in the models and proofs, as discussed in subsequent sections.

\subsection{Passive Orientation Safety}
\label{sec:models-orientation}

So far, we did not consider orientation as part of the safety specification. 
The notion of passive safety requires the robot to stop to avoid imminent collision, which can be inefficient or even impossible when sensor coverage is not exhaustive. 
For example, if an obstacle is close behind the robot (see \rref{fig:fieldOfViewAOne}), the robot would have to stop to obey passive safety. 
This may be the right behavior in an unstructured environment like walking pedestrians but is not helpful when driving on the lanes of a road.
With a more liberal safety notion, the robot could choose a new curve that leads away from the obstacle. 

\begin{wrapfigure}[17]{r}{.5\textwidth}
\centering
\begin{tikzpicture}
\coordinate (obstacle) at (-2,0);
\draw[fill=lsred] (obstacle) circle (1);
\draw[color=black,text width=2.5cm,anchor=north,xshift=-0.5cm] node at ($(obstacle)+(0,-1)$) {obstacle area};
\draw[fill=lsblue] (obstacle) circle (1.5pt);
\draw[color=black,dashed,->] (obstacle) -- ($(obstacle)+(0.1,-1)$);
\draw[color=black,dashed,->] (obstacle) -- ($(obstacle)+(-0.5,-0.8)$);
\draw[color=black,dashed,->] (obstacle) -- ($(obstacle)+(0,0.4)$);
\draw[color=black,dashed,->] (obstacle) -- ($(obstacle)+(0.6,-0.7)$);
\draw[color=black,dashed,->] (obstacle) -- ($(obstacle)+(0.9,-0.3)$);
\draw[color=black,dashed,->] (obstacle) -- ($(obstacle)+(0.6,0.5)$);
\draw[color=black,anchor=east] node[fill=white] at ($(obstacle)+(-0.2,0)$) {obstacle $\po$};
\draw[pattern=north west lines,style={dotted,thin}] (0,0) circle (1.2);
\draw[color=black,text width=4cm,anchor=north,xshift=1cm,align=left] node at (0,-1.2) {area reachable\\ by robot};
\draw[fill=lsblue] (0,0) circle (1.5pt);
\node[color=black,fill=white,anchor=north west] at (0.1,0) {robot $\pr$};
\draw[color=black,dashed] (-1,1) circle (1.414);
\draw[->,color=black,very thick] (0,0) arc (315:350:1.414);
\draw[fill=lsblue] (-1,1) circle (1.5pt);
\draw[color=black,anchor=south] node at (-1,1.1) {curve center $\pc$};
\end{tikzpicture}
\caption{When ignoring orientation, passive safety requires the robot to stop when the robot's reachable area and the trajectory overlap with the obstacle area, even when moving away would increase the safety distance.}
\label{fig:fieldOfViewAOne}
\end{wrapfigure}

We introduce \emph{passive orientation safety} that only requires the robot to remain safe with respect to the obstacles in its \emph{orientation of responsibility}. 
Overall system safety depends on the sensor coverage of the robot and the obstacles.
For example, if two robots drive side-by-side with only very narrow sensor coverage to the front, they might collide when their paths cross.
Even with limited sensor coverage, if both robots can observe some separation markers in space (\eg, lane markers) that keeps their paths separated, then passive orientation safety ensures that there will not be collisions.
Likewise, passive orientation safety ensures that there will be no collisions when every robot and obstacle covers \(180^\circ\) in its orientation of responsibility, \ie, everyone is responsible for obstacles ahead, but not for those behind.

This notion of safety is suitable for structured spaces where obstacles can easily determine the trajectory and observable region of the robot (\eg, lanes on streets). 
The robot is responsible for collisions inside its observable area (``field of vision'', see \rref{fig:visibleregion}) and has to ensure that it can stop if needed before leaving the observable region, because it could otherwise cause collisions when moving into the blind spot just outside its observable area. 

\begin{figure}[htb]
\centering
\begin{tikzpicture}
\draw[fill=lsgreen] (-1,-0.414) circle (0.4);
\draw[color=black,anchor=north,text width=1.5cm,align=center] node at (-2,-0.5) {invisible obstacle behind};
\draw[fill=lsblue] (-1,-0.414) circle (1.5pt);
\node[color=black,anchor=south,text width=3cm,align=center] at (0.158,2.3) {invisible obstacle ahead};
\draw[fill=lsgreen] (0.158,1.811) circle (0.4);
\draw[fill=lsblue] (0.158,1.811) circle (1.5pt);
\draw[color=black,fill=lslightgreen,fill opacity=0.5] (0,0) -- (2,1) arc (25:70:2.23) -- cycle;
\draw[fill=lsred] (1.5,1.5) circle (0.4);
\draw[color=black,text width=1.5cm,anchor=west,align=center] node at (1.9,1.5) {visible obstacle ahead};
\draw[fill=lsblue] (1.5,1.5) circle (1.5pt);
\draw[pattern=north west lines,style={dotted,thin}] (0,0) circle (1.2);
\draw[color=black,anchor=north,xshift=1cm,align=center] node at (0,-1.2) {area reachable by robot};
\draw[fill=lsblue] (0,0) circle (1.5pt);
\node[color=black,fill=white,anchor=north west] at (0.1,0) {robot $\pr$};
\draw[color=black,dashed] (-1,1) circle (1.414);
\draw[->,color=black,very thick] (0,0) arc (315:350:1.414);
\end{tikzpicture}
\caption{Passive orientation safety:
The area observable by the robot (circular sector centered at robot): the distance to all visible obstacles must be safe.
The robot must also ensure that it can stop inside its current observable area, since an obstacle might sit just outside the observable area.
Obstacles outside the visible area are responsible for avoiding collision with the robot until they become visible, \ie, obstacles are assumed to not blindside the robot.
}
\label{fig:visibleregion}
\end{figure}

\begin{model*}
\begin{align}
\label{eq:dw6:model} \textit{dw}_\text{pos} & \equiv \prepeat{\bigl(\ctrl_o;\ctrl_r(\humod{\ar}{A};\pmb{\humod{\beta}{0};\humod{\Visible}{*}}~,~ \safe_\text{pos} \land \pmb{\textit{cda}});\dyn_\text{pos}\bigr)} \\
\label{eq:dw6:asa}\safe_\text{pos} & \equiv 
\pmb{\Visible > 0 \limply} \norm{\pr - \po}{_\infty} > \frac{\vr^2}{2 b} + V \frac{\vr}{b} +\left(\frac{A}{b} + 1\right) \left(\frac{A}{2} \varepsilon^2 + \varepsilon (\vr+V)\right)\\
\label{eq:dw6:infov} \pmb{\textit{cda}} & \pmb{\equiv \gamma \abs{\rcurve} >\frac{\vr^2}{2b} + \left(\frac{A}{b} + 1\right)\left(\frac{A}{2} \varepsilon^2 + \varepsilon \vr\right)} \\
\label{eq:dw6:dyn1}  \dyn_\text{pos} & \equiv \humod{t}{0};~
 \{\D{p}_r = \vr \dr,~ \D{d}_r = \omegar \dr^\bot,~ \D{v}_r = \ar,~ \pmb{\D{\beta} = \omegar},~ \D{\omega}_r = \frac{\ar}{r},~\D{p}_o = \vo,~ \D{t} = 1 ~\&~ \vr \geq 0 \land t \leq \varepsilon\}
\end{align}
\caption{Passive orientation safety}
\label{model:fieldOfView}
\end{model*}

The robot does not make guarantees for obstacles that it cannot see. 
If an obstacle starts outside the observable region and subsequently hits the robot, then it is considered the fault of the obstacle.
If the robot guarantees passive orientation safety and every obstacle outside the observable region guarantees that it will not interfere with the robot, a collision between the robot and an obstacle never happens while the robot is moving. 
In fact, collisions can be avoided when obstacles do not cross the trajectory of the robot. 
Any obstacles inside the observable region can drive with passive safety restrictions (\ie, guarantee not to exceed a maximum velocity) because the robot will brake or choose a new curve to avoid collisions. 
Obstacles that start outside the observable region can rely on the robot to only enter places it can see (\ie the robot will be able to stop before it drives to places that it did not see when evaluating the safety of a curve).

\paragraph{Modeling}
To express that an obstacle was invisible to the robot when it chose a new curve, in \rref{model:fieldOfView} we introduce a variable $\Visible$ with \(\Visible>0\) indicating that an obstacle was visible to the robot when it chose a new curve. 
The observable region is aligned with the orientation of the robot and extends symmetrically to the left and right of the orientation vector $\dr$ by a constant design parameter $\gamma$ that captures the angular width of the field of vision.
The robot can see everything within angle $\frac{\gamma}{2}$ to its left or right. 
With these, passive orientation safety can be expressed as:  
\begin{equation*}
\psi_{\text{pos}} \equiv \vr \neq 0 \limply \norm{\pr - \po} > 0
\lor (\Visible \leq 0 \land \abs{\beta} < \gamma)
\end{equation*}

\begin{figure}[H]
\centering
\begin{tikzpicture}
\draw[color=black,fill=lslightgreen!50] (0,0) -- (1.67,2.5) arc (56.31:123.69:3) -- cycle;
\node[color=black,anchor=south] at (0,3) {observable area};
\draw[fill=lsblue] (-1.23,1.85) circle (1.5pt);
\node[color=black,anchor=south west] at (-1.23,1.85) {curve exit};
\draw[color=black] (-2,0) -- (-1.23,1.85) -- (0,0) -- cycle;
\draw[color=black,fill=lsblue,fill opacity=1] (0,0) -- (-0.28,0.41) arc (123.69:180:0.5) -- cycle;
\node[color=black,anchor=east] at (-0.4,0.2) {$\kappa$};
\draw[color=black,fill=lsblue,fill opacity=1] (-1.23,1.85) -- (-1.42,1.39) arc (247.38:303.69:0.5) -- cycle;
\node[color=black,anchor=north] at (-1.1,1.39) {$\lambda$};
\draw[color=black,fill=lsgreen,fill opacity=1] (-2,0) -- (-1.3,0) arc (0:67.38:0.7) -- cycle;
\node[color=black,anchor=west] at (-1.5,0.5) {$\beta$};
\draw[fill=lsblue] (0,0) circle (1.5pt);
\node[color=black,fill=white,anchor=north west] at (0,0) {robot $\pr$};
\draw[color=black,fill=lsgreen] (0,0) -- (0.39,0.58) arc (56.31:123.69:0.7) -- cycle;
\node[color=black,anchor=south] at (0.3,0.6) {$\frac{\gamma}{2}$};
\draw[color=black,dashed] (0,0) arc (0:190:2);
\draw[->,color=black] (0,0) -- (0,1);
\node[color=black,anchor=south] at (0,1) {$\dr$};
\draw[->,color=black,very thick] (0,0) arc (0:30:2);
\draw[fill=lsblue] (-2,0) circle (1.5pt);
\draw[color=black,anchor=north] node at (-2,0) {curve center $\pc$};
\end{tikzpicture}
\caption{Determining the point where the curve exits the observable region of angular width \(\gamma\) by keeping track of the angular progress \(\beta\) along the curve:
$\kappa = 90^\circ-\frac{\gamma}{2}$ because $\gamma$ extends equally to both sides of the orientation $\dr$, which is perpendicular to the line from the robot to $\pc$ (because $\dr$ is tangential to the curve). 
$\lambda = \kappa$ because the triangle is isosceles. 
Thus, $\beta = 180^\circ - \kappa - \lambda  = \gamma$ at exactly the moment when the robot would leave the observable region. \label{fig:thales}}
\end{figure}
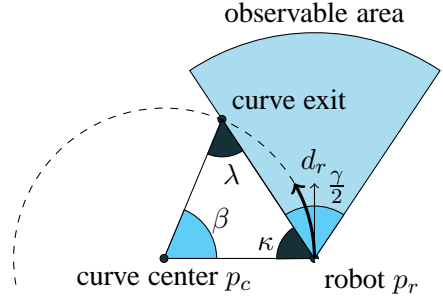

This means that, when the robot is driving (\(\vr\neq0\)), every obstacle is either sufficiently far away or it came from outside the observable region (so \(\Visible \leq 0\)) while the robot stayed inside \(\abs{\beta} < \gamma\). 
For determining whether or not the robot stayed inside the observable region, we compare the robot's angular progress \(\beta\) along the curve with the angular width \(\gamma\) of the observable region, see \rref{fig:thales} for details. 
The angular progress $\beta$ is reset to zero when the robot chooses a new curve in \eqref{eq:dw6:model} and evolves according to $\beta'=\omegar$ when the robot moves \eqref{eq:dw6:dyn1}. 
Thus, $\beta$ always holds the value of the angle on the current curve between the current position of the robot and its position when it chose the curve. 
Passive safety is a special case of passive orientation safety for $\gamma = \infty$.
The model does not take advantage of the fact that $360^\circ$ already subsumes unrestricted visibility.
Passive orientation safety restricts admissible curves to those where the robot can stop before $\abs{\beta}>\gamma$.

The new robot controller now only takes obstacles in its observable region into account (modeled by variable $\Visible$ to distinguish between obstacles that the sensors can see and those that are invisible) when computing the safety of a new curve in $\safe_\text{pos}$ \eqref{eq:dw6:asa}. 
In an implementation of the model, $\Visible$ is naturally represented since sensors only deliver distances to visible obstacles anyway.
It chooses curves such that it can stop before leaving the observable region, \ie, it ensures a clear distance ahead (\(\textit{cda}\)): such a curve is characterized by the braking distance of the robot being less than $\gamma \abs{\rcurve}$, which is the length of the arc between the starting position when choosing the curve and the position where the robot would leave the observable region, cf. \rref{fig:thales}. 
In the robot's drive action \eqref{eq:dw6:model} for selecting a new curve, the angular progress $\beta$ along the curve is reset and the status of the obstacle (\ie whether or not it is visible) is stored in variable $\Visible$ so that the visibility state is available when checking the safety property.

\paragraph{Verification}

Passive orientation safety  (\rref{thm:passiveorientationsafety}) is proved in \KeYmaeraX.

\begin{thm}[Passive orientation safety]\label{thm:passiveorientationsafety}
Robots following \rref{model:fieldOfView} will never collide with the obstacles in sight while driving, and will never drive into unobservable areas, as expressed by the provable \dL formula
\(
\phi_\text{pos} \limply \dibox{\textit{dw}_\text{pos}}\psi_\text{pos} \enspace .
\)
\end{thm}

\begin{proof}
The proof in \KeYmaeraX extends the loop invariant conditions for passive safety so that the robot not only maintains the familiar stopping distance $\frac{\vr^2}{2 b}$ to all obstacles, but also to the border of the visible region in case the nearest obstacle is invisible:
\begin{align*}
\vr>0 &\limply \phantom{\lor} {\norm{\pr-\po}}_\infty > \frac{\vr^2}{2 b} \\
& \phantom{\limply} \lor \Visible\leq 0 \land \abs{\rcurve\gamma} - \abs{\rcurve \beta} > \frac{\vr^2}{2 b} \enspace .
\end{align*}
Here, we characterize the angular progress $\beta$ with the differential invariant $\beta = \old{(\beta)} + \frac{1}{\rcurve}\left(\old{(\vr)}t + \frac{\ar}{2}t^2\right)$, in addition to the differential invariants for passive safety used in the proof of \rref{thm:passivesafety}.
\end{proof}

\section{Refined Models for Safety Verification}
\label{sec:refinedsafety}

The models used for safety verification so far made simplifying assumptions to focus on the basics of different safety notions.
In this section, we discuss how to create more realistic models with different accelerations, measurement uncertainty, actuator disturbance, asynchronous control of obstacle and robot, and explicit representation of arbitrary many obstacles.
We introduce the model extensions for passive safety (\rref{model:dynamicwindowpassive}) as an example.
The extensions apply to static safety and passive friendly safety in a similar fashion by adapting $\safe_\text{ss}$ and $\safe_\text{pfs}$; passive orientation safety needs to account for the changes both in the translational safety margin $\safe_\text{pos}$ and the angular progress $\textit{cda}$.

\subsection{Passive Safety with Actual Acceleration}
\label{sec:models-psar}

\rref{model:dynamicwindowpassive} uses the robot's maximum acceleration $A$ in its safety requirement \eqref{eq:pf:3-2} when it determines whether or not a new curve will be safe. 
This condition is conservative, since the robot of \rref{model:dynamicwindowpassive} can only decide between maximum acceleration (\(\humod{\ar}{A}\)) or maximum braking (\(\humod{\ar}{-b}\) from \rref{model:robotcontroller}).
If \eqref{eq:pf:3-2} does not hold (which is independent from the chosen curve, \ie the radius $\rcurve$), then \rref{model:dynamicwindowpassive} forces a driving robot to brake with maximum deceleration ${-b}$, even if it might be sufficiently safe to coast or slightly brake or just not accelerate in full. 
As a result, \rref{model:dynamicwindowpassive} is passively safe but lacks efficiency in that it may take the robot longer to reach a goal because it can only decide between extreme choices. 
Besides efficiency concerns, extreme choices are undesirable for comfort reasons (\eg, decelerating a car with full braking power should be reserved for emergency cases). 

\begin{figure}[htb]
\centering
\begin{tikzpicture}
\coordinate (obstacle) at (-0.8,1.5);
\draw[fill=lsred] (obstacle) circle (1);
\draw[color=black,anchor=south east] node at (-1.8,1.8) {obstacle area};
\draw[fill=lsblue] (obstacle) circle (1.5pt);
\draw[color=black,dashed,->] (obstacle) -- ($(obstacle)+(0.1,-1)$);
\draw[color=black,dashed,->] (obstacle) -- ($(obstacle)+(-0.5,-0.8)$);
\draw[color=black,dashed,->] (obstacle) -- ($(obstacle)+(-0.4,0.7)$);
\draw[color=black,dashed,->] (obstacle) -- ($(obstacle)+(0,0.4)$);
\draw[color=black,dashed,->] (obstacle) -- ($(obstacle)+(0.6,-0.7)$);
\draw[color=black,dashed,->] (obstacle) -- ($(obstacle)+(0.9,-0.3)$);
\draw[color=black,dashed,->] (obstacle) -- ($(obstacle)+(0.6,0.5)$);
\draw[color=black,anchor=east] node[fill=white] at ($(obstacle)+(-0.2,0)$) {obstacle $\po$};
\draw[pattern=north west lines,style={solid,thin}] (0,0) circle (1);
\draw[color=black,anchor=east] node at (-1.2,0) {unsafe accelerations $\leq A$};
\draw[fill=lsblue] (-0.85,0) circle (1.5pt);
\draw[color=black,solid] (-1.2,0)--(-0.85,0);
\draw[fill=lsgreen,style={solid,thin}] (0,0) circle (0.7);
\draw[fill=white,style={dotted,thin}] (0,0) circle (0.4);
\node[color=black,fill=white,anchor=east] at (-1,-0.7) {maximum braking $-b$};
\draw[fill=lsblue] (-0.35,-0.2) circle (1.5pt);
\draw[color=black,solid] (-1,-0.7)--(-0.35,-0.2);
\node[color=black,fill=white,anchor=north] at (1,-1.1) {safe accelerations};
\draw[fill=lsblue] (0.1,-0.55) circle (1.5pt);
\draw[color=black,solid] (0.1,-0.55)--(1,-1.1);
\draw[fill=lsblue] (0,0) circle (1.5pt);
\node[color=black,fill=white,anchor=west] at (0,0) {robot $\pr$};
\draw[color=black,dashed] (1,1) circle (1.414);
\draw[->,color=black,very thick] (0,0) arc (225:260:1.414);
\draw[fill=lsblue] (1,1) circle (1.5pt);
\node[color=black,anchor=south,text width=2cm,align=center] at (1,1.1) {curve\\ center $\pc$};
\end{tikzpicture}
\caption{Passive safety with actual acceleration: the actual acceleration choice $-b \leq \ar \leq A$ must not take the robot into the area reachable by the obstacle.
Dotted circle around robot position \(\pr\): earliest possible stop with maximum braking \(-b\); solid blue area between dotted circle and dashed area: safe \(\ar\); dashed area: reachable with unsafe accelerations.}
\label{fig:passivesafetyactualaccillustration}
\end{figure}
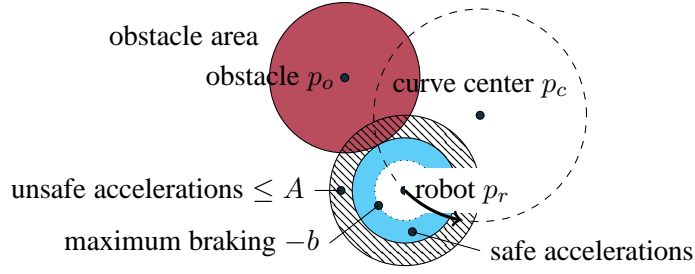

\rref{fig:passivesafetyactualaccillustration} illustrates how safety constraint \eqref{eq:pf:3-2} represents the maximally conservative choice: 
it forces the robot to brake (the outermost circle around the robot \(\pr\) intersects with the obstacle), even though many points reachable with $-b \leq \ar < A$ would have been perfectly safe (solid blue area does not intersect with the obstacle). 

\paragraph{Modeling}

\rref{model:dynamicwindowpassiveactuala} refines  \rref{model:dynamicwindowpassive} to work with the actual acceleration, \ie, in the acceleration choice \eqref{eq:pfar:0} the robot picks any arbitrary acceleration \(\ar\) within the physical limits \(-b \leq \ar \leq A\) instead of just maximum acceleration.

\begin{model}[h]
\caption{Passive safety with actual acceleration}
\label{model:dynamicwindowpassiveactuala}
\begin{align}
\label{eq:pfar:0}
\textit{dw}_\text{psa} & \equiv 
\bigl(\textit{ctrl}_o;\ctrl_r(\pmb{\humod{\ar}{*};\pmb{\ptest{-b \leq \ar \leq A}}~,}~ \safe_\text{psa});\dyn_\text{ps}\bigr)^* \\
\label{eq:pfar:3-2}\safe_\text{psa} & \equiv \norm{\pr-\po}{_\infty} > 
\pmb{\begin{cases}
\textit{dist}_\geq & \text{if } \vr + \ar \varepsilon \geq 0\\
\textit{dist}_< & \text{otherwise}
\end{cases}}
\end{align}
\end{model}

This change requires us to adapt the control condition \eqref{eq:pfar:3-2} that keeps the robot safe.
We first give the intuition behind condition \eqref{eq:pfar:3-2}, then justify its correctness with a safety proof.

\paragraph{Identification of Safe Constraints}
Following~\cite{DBLP:conf/itsc/LoosWSP13} we relax constraint \eqref{eq:pf:3-2} so that the robot can choose any acceleration $-b \leq \ar \leq A$ and checks this actual acceleration $\ar$ for safety.
That way, it only has to fall back to the emergency braking branch $\humod{\ar}{-b}$ if there is no other safe choice available.
We distinguish two cases:
\begin{itemize}[noitemsep]
\item $\vr + \ar\varepsilon \geq 0$: the acceleration choice $-b \leq \ar \leq A$ always keeps a nonnegative velocity during the full cycle duration $\varepsilon$.
\item $\vr + \ar\varepsilon < 0$: the acceleration choice $-b \leq \ar < 0$ cannot be followed for the full duration $\varepsilon$ without stopping the evolution to prevent a negative velocity.
\end{itemize}
In the first case, we continue to use formula \eqref{eq:pf:3-2} with actual $\ar$ substituted for $A$ to compute the safety distance:
\begin{equation}\label{eq:pfar:safe1}
\begin{aligned}
\textit{dist}_{\geq} & = \frac{\vr^2}{2 b} + V \frac{\vr}{b} + \left(\frac{\ar}{b} + 1\right) \left(\frac{\ar}{2} \varepsilon^2 + \varepsilon (\vr+V)\right)
\end{aligned}
\end{equation}

In the second case, distance \eqref{eq:pfar:safe1} is unsafe, because the terminal velocity when following \(\ar\) for \(\varepsilon\) time is negative (unlike in case 1).
Thus, the robot may have collided at a time before $\varepsilon$, while the term in \eqref{eq:pfar:safe1} only indicates that it will no longer be in a collision state at time~$\varepsilon$ after having moved backwards.
Consider the time $t_b$ when the robot's velocity becomes zero ($\vr+\ar t_b = 0$) so that its motion stops (braking does not make the robot move backwards but merely stop). 
Hence, $t_b = -\frac{\vr}{\ar}$ since case 1 covers \m{\ar=0}. 
Within duration $t_b$ the robot will drive a total distance of $dist_r=-\frac{\vr^2}{2 \ar}=\int_0^{t_b} \vr+\ar t\, dt$. 
The obstacle may drive up to $dist_o=V t_b$ until the robot is stopped.
Thus, we compute the distance using \eqref{eq:pfar:safe2} to account for the worst case that both robot and obstacle drive directly towards each other (note that $-b \leq \ar < 0$). 

\begin{equation}\label{eq:pfar:safe2}
\textit{dist}_{<} ~=~ -\frac{\vr^2}{2 \ar} - V \frac{\vr}{\ar}
\end{equation}

\paragraph{Verification}
\label{par:passivesafetyactual-proof}

We verify the safety of the actual acceleration control algorithm as modeled in \rref{model:dynamicwindowpassiveactuala} in \KeYmaeraX.

\begin{thm}[Passive safety with actual acceleration]
\label{thm:passivesafetyactuala}
Robots following \rref{model:dynamicwindowpassiveactuala} to base their safety margins on the current acceleration choice instead of worst-case acceleration will never collide while driving,
as expressed by the provable \dL formula \(\phi_\text{ps} \limply \dibox{\textit{dw}_\text{psa}}{\psi_\text{ps}} \enspace .\)
\end{thm}

Even though the safety constraint $\safe_\text{psa}$ now considers the actual acceleration instead of the maximum possible acceleration when estimating the required safety margin, it can still be conservative when the robot makes sharp turns.
During sharp turns, the straight-line distance from the origin is shorter than the distance along the circle, which can be exploited when computing the safety margin.
This extension is in \rref{appendix:models-distance}.

\subsection{Passive Safety for Sharp Turns}
\label{appendix:models-distance}

Models \ref{model:dynamicwindowpassive} and \ref{model:dynamicwindowpassiveactuala} used a safety distance in supremum norm ${\norm{\cdot}}_\infty$ for the safety constraints, which conservatively overapproximates the actual trajectory of the robot by a box around the robot.
For example, recall the safety distance \eqref{eq:pf:3-2} of \rref{model:dynamicwindowpassive} 
\begin{equation}
\norm{\pr - \po}{_\infty} > \frac{\vr^2}{2 b} + V \frac{\vr}{b} + \left(\frac{A}{b} + 1\right) \left(\frac{A}{2} \varepsilon^2 + \varepsilon (\vr+V)\right)
\tag{\ref{eq:pf:3-2}*}
\end{equation}
which needs to be large enough in either one axis, irrespective of the actual trajectory that the robot will be taking.
This constraint is safe but inefficient when the robot chooses a trajectory that will keep it close to its current position (\eg, when driving along a small circle, meaning it makes a sharp turn).
For example, a robot with constant velocity $\vr=4$ and reaction time $\varepsilon = 1$ will traverse a small circle with radius $\rcurve=\frac{1}{\pi}$ and corresponding circumference $2 \pi \rcurve = 2$ twice within time $\varepsilon$. 
Safety constraint \eqref{eq:pf:3-2} required the total distance of $4$ as a safety distance between the robot and the obstacle, because it overapproximated its actual trajectory by a box. 
However, the robot never moves away more than $\frac{2}{\pi}$ from its original position then because it moves on a circle (cf.\ \rref{fig:oldNotNewYes}). 
With full 360$^\circ$ sensor coverage the robot can exploit that the closest obstacle does not cross its trajectory, which makes this extension suitable for passive safety and passive friendly safety, but not for passive orientation safety.

\begin{figure}[H]
\centering
\begin{subfigure}{.48\columnwidth}
\begin{tikzpicture}
\coordinate (obstacle) at (-2,0);
\draw[fill=lsred] (obstacle) circle (1);
\draw[color=black,text width=2.5cm,anchor=north,xshift=-0.5cm] node at (-2,-1) {obstacle area: $V\left(\varepsilon + \frac{\vr + A\varepsilon}{b}\right)$};
\draw[fill=lsblue] (obstacle) circle (1.5pt);
\draw[color=black,dashed,->] (obstacle) -- ($(obstacle)+(0.1,-1)$);
\draw[color=black,dashed,->] (obstacle) -- ($(obstacle)+(-0.5,-0.8)$);
\draw[color=black,dashed,->] (obstacle) -- ($(obstacle)+(-0.4,0.7)$);
\draw[color=black,dashed,->] (obstacle) -- ($(obstacle)+(0,0.4)$);
\draw[color=black,dashed,->] (obstacle) -- ($(obstacle)+(0.6,-0.7)$);
\draw[color=black,dashed,->] (obstacle) -- ($(obstacle)+(0.9,-0.3)$);
\draw[color=black,dashed,->] (obstacle) -- ($(obstacle)+(0.6,0.5)$);
\draw[color=black,anchor=east] node[fill=white] at ($(obstacle)+(-0.2,0)$) {obstacle $\po$};
\draw[pattern=north west lines,style={dotted,thin}] (0,0) circle (1.2);
\draw[color=black,text width=4cm,anchor=north,xshift=1cm] node at (0,-1.2) {area reachable by robot: $\frac{\vr^2}{2b} + \left(\frac{A}{b}+1\right)\left(\frac{A}{2}\varepsilon + \varepsilon \vr \right)$};
\draw[fill=lsblue] (0,0) circle (1.5pt);
\node[color=black,fill=white,anchor=west] at (0.1,0) {robot $\pr$};
\draw[color=black,dashed] (1,1) circle (1.414);
\draw[->,color=black,very thick] (0,0) arc (225:260:1.414);
\draw[fill=lsblue] (1,1) circle (1.5pt);
\draw[color=black,anchor=south] node at (1,1.1) {curve center $\pc$};
\draw[very thick,<->] (-2,0) -- (-0.34,0.56);
\draw[dashed] (-0.34,0.56) -- (1,1);
\node[draw,color=black,fill=white,anchor=east,rectangle callout,text width=3.5cm,inner sep=2pt,callout absolute pointer={(-1.5,0.2)}] at (-1,1.7) {distance to trajectory: $\bigl\lvert \abs{\rcurve} - \norm{\po-\pc} \bigr\rvert$}; 
\end{tikzpicture}
\caption{Safe since the obstacle area does not overlap the dashed trajectory.}
\label{fig:oldNotNewYes}
\end{subfigure}
\begin{subfigure}{.48\columnwidth}
\centering
\begin{tikzpicture}
\coordinate (obstacle) at (-0.8,1.5);
\draw[fill=lsred] (obstacle) circle (1);
\draw[color=black,anchor=south east] node at (-1.8,1.8) {obstacle area};
\draw[fill=lsblue] (obstacle) circle (1.5pt);
\draw[color=black,dashed,->] (obstacle) -- ($(obstacle)+(0.1,-1)$);
\draw[color=black,dashed,->] (obstacle) -- ($(obstacle)+(-0.5,-0.8)$);
\draw[color=black,dashed,->] (obstacle) -- ($(obstacle)+(-0.4,0.7)$);
\draw[color=black,dashed,->] (obstacle) -- ($(obstacle)+(0,0.4)$);
\draw[color=black,dashed,->] (obstacle) -- ($(obstacle)+(0.6,-0.7)$);
\draw[color=black,dashed,->] (obstacle) -- ($(obstacle)+(0.9,-0.3)$);
\draw[color=black,dashed,->] (obstacle) -- ($(obstacle)+(0.6,0.5)$);
\draw[color=black,anchor=east] node[fill=white] at ($(obstacle)+(-0.2,0)$) {obstacle $\po$};
\draw[pattern=north west lines,style={dotted,thin}] (0,0) circle (0.3);
\draw[color=black,anchor=east] node at (-0.3,0) {area reachable by robot};
\draw[fill=lsblue] (0,0) circle (1.5pt);
\node[color=black,fill=white,anchor=west,xshift=1em] at (0,0) {robot $\pr$};
\draw[color=black,dashed] (1,1) circle (1.414);
\draw[->,color=black,very thick] (0,0) arc (225:260:1.414);
\draw[fill=lsblue] (1,1) circle (1.5pt);
\draw[color=black,anchor=south,text width=1.5cm,align=center] node at (1,1.1) {curve center $\pc$};
\end{tikzpicture}
\caption{Safe since obstacle area and dotted robot area do not overlap.}
\label{fig:newBad}
\end{subfigure}
\caption{Two different reasons for safe robot trajectories}
\label{fig:safetrajectories}
\end{figure}

\paragraph{Modeling}

\begin{model*}[htb]
\begin{align}
\label{eq:dw7:model} \textit{dw}_\text{psdm} & \equiv \prepeat{(\textit{ctrl}_o; \ctrl_r; \textit{dyn})}\\
\textit{ctrl}_o & \equiv \text{see \rref{model:dynamicwindowpassive}}\\
\label{eq:dw7:ctrlr} \ctrl_r &\equiv \phantom{\cup} (\humod{\ar}{-b}) \\
\label{eq:dw7:c2}&\phantom{\equiv}  \cup \bigl(\ptest{\vr=0};~\humod{\ar}{0};~\humod{w_r}{0};~ \left(\pchoice{\humod{\dr}{-\dr}}{\humod{\dr}{\dr}}\right);~\humod{\rcurve}{*};~\humod{\pc}{(*,*)};~\ptest{\textit{curve}} \bigr) \\
&\phantom{\equiv}  \cup (\humod{\ar}{*}; ~\ptest{-b\leq \ar\leq A};~ \humod{\omegar}{*};~ \ptest{-\Omega \leq \omegar \leq \Omega};\\ &\phantom{\equiv \cup (} \humod{\rcurve}{*};~\humod{\pc}{(*,*)};~ \humod{\po}{(*,*)};~\ptest{\textit{curve} \land \textit{safe}} ) \\
\label{eq:dw7:curve} \curve &\equiv \rcurve \neq 0 \land \abs{\rcurve} = \norm{\pr - \pc} \land \dr = \frac{(\pr-\pc)^\bot}{\rcurve} \land \rcurve \omegar = \vr\\
\begin{split}
\label{eq:dw7:safe} \textit{safe} & \equiv \phantom{\lor}
\left(\norm{\pr-\po}{_\infty} > 
\begin{cases}
\textit{dist}_\geq & \text{if } \vr + \ar \varepsilon \geq 0\\
\textit{dist}_< & \text{otherwise}
\end{cases}\right)
\\
& \phantom{\equiv} \lor 
 \left(\abs{ \abs{\rcurve} - \norm{\po - \pc}} >
\begin{cases}
V  \left( \varepsilon+ \frac{\vr+\ar \varepsilon}{b} \right) & \text{if } \vr + \ar \varepsilon \geq 0\\
-V \frac{\vr}{\ar} & \text{otherwise}
\end{cases}\right)
\end{split}\\
 \dyn_\text{psdm} &\equiv \text{ see \rref{model:dynamicwindowpassive}}
\end{align}
\caption{Passive safety when considering the trajectory of the robot in distance measurement, extends \rref{model:dynamicwindowpassiveactuala}}
\label{model:newSafe}
\end{model*}

We change the robot controller to improve its efficiency.
One choice would be to explicitly express circular motion in terms of sine and cosine and then compute all possible positions of the robot explicitly. 
However, besides being vastly inefficient in a real controller, this introduces transcendental functions and would leave decidable real arithmetic. 
Hence, we will use the distance of the obstacle to the trajectory itself in the control conditions.
Such a distance computation requires that we adapt the constraint $\curve$ to express the curve center explicitly in \eqref{eq:dw7:curve}. 
So far, the curve was uniquely determined by the radius $\rcurve$ and the orientation $\dr$ of the robot. 
Now that we need the curve center explicitly for distance calculation to the obstacle, the controller chooses the curve center $\pc$ such that:
\begin{itemize}
\item $(\pr-\pc)$ is perpendicular to the robot orientation $\dr$, \ie, $\dr$ is tangential to the curve, and 
\item $(\pr-\pc)$ is located correctly to the left or right of the robot, so that it fits to the clockwise or counter-clockwise motion indicated by the sign of $\rcurve$.
\end{itemize}
Thus, the condition $\curve$ \eqref{eq:dw7:curve} in \rref{model:newSafe} now checks if the chosen curve and the direction of the robot are consistent, \ie, $\abs{\rcurve}=\norm{\pr-\pc}$ and $\dr = \frac{(\pr-\pc)^\bot}{\rcurve}$. 
Additionally, we augment the robot with a capability to turn on the spot when stopped ($\vr=0$). 
For this, \eqref{eq:dw7:c2} is extended with a choice of either turning around ($\humod{\dr}{-\dr}$) or remaining oriented as is ($\humod{\dr}{\dr}$) when stopped, and the corresponding choice of a curve center $\pc$ such that the curve variables remain consistent according to the subsequent test $\ptest{\curve}$.

\paragraph{Identification of Safe Controls}
With the changes in distance measurement introduced above, we relax the control conditions that keep the robot safe.
The distance of the obstacle to the trajectory can be described in two steps:
\begin{enumerate}
\item Calculate the distance of the obstacle to the circle: $\bigl\lvert  \abs{\rcurve} - \norm{\po - \pc} \bigr\rvert$, which is the absolute value of the radius minus the distance between the obstacle and the circle center.
\item Calculate the maximum distance that the obstacle can drive until the robot comes to a stop. This distance is equal to the distances calculated in the previous models, \ie in the case $\vr + \ar \varepsilon \leq 0$ it is $-V \frac{\vr}{\ar}$ and in the case $\vr + \ar \varepsilon \geq 0$ it is $V \left( \varepsilon+ \frac{\vr+\ar \varepsilon}{b} \right)$.
\end{enumerate}

If the distance between the obstacle and the circle describing the robot's trajectory is greater than the sum of those distances, then the robot can stop before hitting the obstacle.
Then choosing the new curve is safe, which leads us to choose the following safety condition:

\begin{equation}\label{eq:ps-trajectorysafe}
\bigl\lvert \abs{\rcurve} - \norm{\po - \pc} \bigr\rvert > 
\begin{cases}
V  \left( \varepsilon+ \frac{\vr+\ar \varepsilon}{b} \right) & \text{if } \vr + \ar \varepsilon \geq 0\\
-V \frac{\vr}{\ar} & \text{otherwise}
\end{cases}
\end{equation}

We use condition \eqref{eq:ps-trajectorysafe}, which now uses the Euclidean norm \(\norm{\cdot}\), for choosing a new curve in \rref{model:newSafe}. 
With this new constraint, the robot is allowed to choose the curve in \rref{fig:oldNotNewYes}. 
However, constraint \eqref{eq:ps-trajectorysafe}  has drawbacks when the trajectory of the robot is slow along a large circle and the obstacle is close to the circle, as illustrated in \rref{fig:newBad}. 
In this case the robot is only allowed to choose very small accelerations because the obstacle is very close to the circle. 
Formula \eqref{eq:dw7:safe} in \rref{model:newSafe} follows the more liberal of the two constraints---\ie, \eqref{eq:pfar:3-2} $\lor$ \eqref{eq:ps-trajectorysafe}---to provide the best of both worlds.

\paragraph{Verification}

We verify the safety of the robot's control choices in \KeYmaeraX.

\begin{thm}[Passive safety for sharp turns]
\label{thm:passivesafety-trajectorydistance}
Robots using trajectory distance measurement according to \rref{model:newSafe} in addition to direct distance measurement guarantee passive safety, as expressed by the provable \dL formula
\(\phi_\text{ps} \limply \dibox{\textit{dw}_\text{psdm}}{\psi_\text{ps}} \enspace.\)
\end{thm}

\begin{proof}
The most important condition in the loop invariant of the proof guarantees that the robot either maintains the familiar safe stopping distance ${\norm{\pr-\po}}_\infty>\frac{\vr^2}{2 b} $, or that the obstacle cannot reach the robot's curve until the robot is stopped:
\[\vr>0 \limply {\norm{\pr-\po}}_\infty>\frac{\vr^2}{2 b} \lor \bigl\lvert \abs{\rcurve} - \norm{\po - \pc} \bigr\rvert > V\frac{\vr}{b} \enspace .\]
\end{proof}

\subsection{Passive Safety Despite Uncertainty}

Robots have to deal with uncertainty in almost every aspect of their interaction with the environment, ranging from sensor inputs (\eg, inaccurate localization, distance measurement) to actuator effects (\eg, uncertain wheel slip depending on the terrain).
In this section, we show how the three most important classes of uncertainty can be handled explicitly in the models.
First, we allow localization uncertainty, so that the robot knows its position only approximately, which has a considerable impact on uncertainty over time.
We then consider imperfect actuator commands, which means that the effective physical braking and acceleration will differ from the controller's desired output.
Finally, we allow velocity uncertainty, so the robot knows its velocity only approximately, which also has an impact over time.
We use nondeterministic models of uncertainty as intervals around the real position, acceleration, and velocity, without any probabilistic assumptions about their distribution.\footnote{Other error models are supported, as long as they are clipped to guaranteed intervals, because in the safety proof we have to analyze all measured values, regardless of their probability. 
For an advanced analysis technique considering probabilities, see stochastic \dL~\cite{DBLP:conf/cade/Platzer11}.}
Such intervals are instantiated, \eg, according to sensor or actuator specification (\eg, GPS error), or \wrt experimental measurements.\footnote{Instantiation with probabilistic bounds means that the symbolically guaranteed safety is traded for a probability of safety.}

\subsubsection{Location Uncertainty}

\rref{model:positionuncertainty} introduces location uncertainty. 
It adds a location measurement $\hat{\pr}$ before the control decisions are made such that the controller only bases its decisions on the most recent location measurement $\hat{\pr}$, which can deviate from the true location $\pr$.
This location measurement may deviate from the real position $\pr$ by no more than the symbolic parameter $\Delta_p \geq 0$, cf. \eqref{eq:ups:1}.
The measured location $\hat{\pr}$ is used in all control decisions of the robot (\eg, in \eqref{eq:ups:3-2} to compute whether or not it is safe to change the curve).
The robot's physical motion still follows the real position $\pr$ even if the controller does not know it.

\begin{model}[h]
\caption{Passive safety despite location uncertainty, extends \rref{model:dynamicwindowpassive}}
\label{model:positionuncertainty}
\begin{align}
\label{eq:ups:0} \textit{dw}_\text{pslu}  & \equiv \bigl(\pmb{\textit{locate}}; \ctrl_o; \ctrl_r(\humod{\ar}{A}~,~ \safe_\text{pslu}); \dyn_\text{ps}\bigr)^*\\
\label{eq:ups:1}\pmb{\textit{locate}} & \pmb{\equiv \humod{\hat{\pr}}{(*,*)};~ \ptest{\norm{\hat{\pr}-\pr}\leq \Delta_p}}\\
\label{eq:ups:3-2}\safe_\text{pslu} & \equiv 
\norm{\pmb{\hat{\pr}} - \po}{_\infty} > \frac{\vr^2}{2 b} + V\frac{\vr}{b} + \pmb{\Delta_p} + \left(\frac{A}{b} + 1\right) \left(\frac{A}{2} \varepsilon^2 + \varepsilon (\vr+V)\right)
\end{align}
\end{model}

\begin{thm}[Passive safety despite location uncertainty]
\label{thm:passivesafety-locationuncertainty}
Robots computing their safety margins from location measurements with maximum uncertainty $\Delta_p$ by \rref{model:positionuncertainty} will never collide while driving, as expressed by the provable \dL formula
\(\phi_\text{ps} \land \Delta_p \geq 0 \limply \dibox{\textit{dw}_\text{pslu}}{\psi_\text{ps}} \enspace.\)
\end{thm}

Uncertainty about the obstacle's position is already included in the nondeterministic behavior of previous models by increasing the shapes according to uncertainty.

\subsubsection{Actuator Perturbation}

\rref{model:motionuncertainty} introduces actuator perturbation between control and dynamics, cf. \eqref{eq:uas:0}.
Actuator perturbation affects the acceleration by a damping factor $\delta_a$, known to be at most a maximum damping $\Delta_a$, \ie, $\delta_a \in \left[\Delta_a,1\right]$, cf. \eqref{eq:uas:1}.
Note that the damping factor $\delta_a$ can change arbitrarily often, but is assumed to be constant during the continuous evolution that takes a maximum of $\varepsilon$ time units.
The perturbation may cause the robot to now have full acceleration ($\delta_a=1$) but later fully reduced braking ($\delta_a=\Delta_a$). 
This combination results in the largest possible stopping distance (for a certain speed $\vr$). 
For instance, the robot accelerates on perfect terrain, but is unlucky enough to be on slippery terrain again when it needs to brake.
The robot considers this worst-case scenario during control in its safety constraint \eqref{eq:uas:2}.

\begin{model}[h]
\caption{Passive safety despite actuator perturbation, extends \rref{model:dynamicwindowpassive}}
\label{model:motionuncertainty}
\begin{align}
\label{eq:uas:0} \text{\it dw}_\text{psap}  & \equiv \bigl(\ctrl_o; \ctrl_r(\humod{\ar}{A}~,~ \safe_\text{psap}); \pmb{\textit{act}}; \dyn_\text{ps}\bigr)^*\\
\label{eq:uas:1}\pmb{\textit{act}} & \pmb{\equiv \humod{\delta_a}{*};~\ptest{(0 < \Delta_a \leq \delta_a \leq 1)};~ \humod{\tilde{\ar}}{\delta_a \ar}}\\
\label{eq:uas:2}\safe_\text{psap} & \equiv 
\norm{\pr - \po}{_\infty} > \frac{\vr^2}{2 b \pmb{\Delta_a}} + V\frac{\vr}{b \pmb{\Delta_a}} + \left(\frac{A}{b \pmb{\Delta_a}} + 1\right)\left(\frac{A}{2} \varepsilon^2 + \varepsilon (\vr+V)\right)\\
\notag \dyn_\text{ps} & \text{ of \rref{model:dynamicwindowpassive} with } \ar \text{ replaced by disturbed } \pmb{\tilde{\ar}}
\end{align}
\end{model}

\begin{thm}[Passive safety despite actuator perturbation]
\label{thm:passivesafety-actuatorperturbation}
Robots with inaccurate actuation being subject to maximum disturbance $\Delta_a$ according to \rref{model:motionuncertainty} will never collide while driving,
as expressed by the provable \dL formula
\[\phi_\text{ps} \land \Delta_a > 0 \limply \dibox{\textit{dw}_\text{psap}}{\psi_\text{ps}}\enspace.\]
\end{thm}

\subsubsection{Velocity Uncertainty}

\begin{model}[htb]
\begin{align}
\label{eq:velUncert:model} \dw_\text{psvu} &\equiv \bigl(\pmb{\textit{sense}};\ctrl_o; \ctrl_r(\humod{\ar}{A}~,~ \safe_\text{psvu}); \dyn_\text{ps}\bigr)^*\\
\label{eq:ctlr:velUncert} \pmb{\textit{sense}} & \pmb{\equiv  \humod{\hat{\vr}}{*};~ \ptest{(\hat{\vr} \geq 0 \land \vr-\vdiff \leq \hat{\vr} \leq \vr+\vdiff)}}\\
\safe_\text{psvu} & \equiv \norm{\pr-\po}{_\infty} > 
\frac{(\pmb{\vmax})^2}{2b} + V\frac{\pmb{\vmax}}{b} + \left( \frac{A}{b}+1 \right) \left( \frac{A}{2} \varepsilon^2  + \varepsilon (\pmb{\vmax} + V)\right)\notag
\end{align}
\caption{Passive safety despite velocity uncertainty, extends \rref{model:dynamicwindowpassive}}
\label{model:velocityuncertainty}
\end{model}

\rref{model:velocityuncertainty} introduces velocity uncertainty. 
To account for the uncertainty, at the beginning of its control phase the robot reads off a (possibly inexact) measurement $\hat{\vr}$ of its speed $\vr$. 
It knows that the measured speed $\hat{\vr}$ deviates by at most a measurement error $\vdiff$ from the actual speed $\vr$, see \eqref{eq:ctlr:velUncert}. 
Also, the robot knows that its speed is non-negative. 
Thus, we can assume that $\hat{\vr}$ is always equal to or greater than zero by transforming negative measurements. 
In order to stay safe, the controller has to make sure that the robot stays safe even if its true speed is maximally larger than the measurement, \ie $\vr=\hat{\vr}+ \vdiff$. 
The idea is now that the controller makes all control choices with respect to the maximal speed $\vmax$ instead of the actual speed $\vr$. 
The continuous evolution, in contrast, still uses the actual speed $\vr$, because the robot's physics will not be confused by a sensor measurement error. 

Since we used the maximal possible speed when considering the safety of new curves in the controller we can prove that the robot will still be safe. 
A modeling subtlety arises when using $\hat{\vr}$ instead of $\vr$ in the second branch \eqref{eq:ctlr:velUncert} of $\ctrlr$: 
Because of the velocity uncertainty we no longer know if $\vr$ is zero (\ie the robot is stopped). 
However, the branch for stopped situations models discrete physics rather than a conscious robot decision (even if a real robot controller chooses to hit the brakes, as soon as the robot is stopped physics turns this decision into $\ar=0$), so we still use the test $\ptest{(\vr=0)}$ instead of \(\ptest(\hat{\vr}=0)\). 

\begin{thm}[Passive safety despite velocity uncertainty]
\label{thm:passivesafety-velocityuncertainty}
Robots computing their safety margins from velocity measurements with maximum uncertainty $\Delta_v$ according to \rref{model:velocityuncertainty} will never collide while driving, as expressed by the provable \dL formula \(\phi_\text{ps} \land \Delta_v \geq 0 \limply \dibox{\textit{dw}_\text{psvu}}{\psi_\text{ps}} \enspace .\)
\end{thm}

\subsection{Asynchronous Control of Obstacle and Robot}
\label{sec:models-nonsync}

In the models so far, the controllers of the robot and the obstacle were executed synchronously, \ie, the robot and the obstacle made their control decisions at the same time. 
While the obstacle could always choose its previous control choices again if it does not want to act, the previous models only allowed the obstacle to decide when the robot made a control decision, too.\footnote{Note that \dL follows the common assumption that discrete actions do not take time; time only passes in ODEs. 
So all discrete actions happen at the same real point in time, even though they are ordered sequentially.} 
This does not reflect reality perfectly, since we want liberal obstacle models without assumptions about when an obstacle makes a control decision. 
So, we ensure that the robot remains safe regardless of how often and at which times the obstacles change their speed and orientation.

\begin{model}[h]
\caption{Asynchronous obstacle and robot control, extends \rref{model:dynamicwindowpassive}}
\label{model:mchange1}
\begin{align}
\label{eq:dw2:ctrlplant2}  \textit{dw}_\text{psns} &\equiv  \prepeat{\bigl(\ctrlr(\humod{\ar}{A}~,~\safe_\text{ps});~\humod{t}{0};~\pmb{\prepeat{(\ctrlo;~\dyn_\text{ps})}}\bigr)}
\end{align}
\end{model}

In \rref{model:mchange1} we now model the control of the obstacle \(\ctrl_o\) in an inner loop around the continuous evolution $\dyn$ in  \eqref{eq:dw2:ctrlplant2} so that the obstacle control can interrupt continuous evolution at any time to make a decision, and then continue the dynamics immediately without giving the robot's controller a chance to run. 
This means that the obstacle can make as many control decisions as it wants without the robot being able to react every time. 
The controller \(\ctrl_r\) of the robot is still guaranteed to be invoked after at most time $\varepsilon$ has passed, as modeled with the evolution domain constraint $t \leq \varepsilon$ in \(\dyn_\text{ps}\). 

\begin{thm}[Passive safety for asynchronous controllers]
\label{thm:passivesafety-nonsync}
Robots following \rref{model:mchange1} will never collide while driving, even if obstacles change their direction arbitrary often and fast, as expressed by the provable \dL formula \(\phi_\text{ps} \limply \dibox{\textit{dw}_\text{psns}}{\psi_\text{ps}} \enspace .\)
\end{thm}

\begin{proof}
The \KeYmaeraX proof of \rref{thm:passivesafety-nonsync} uses $\phi_\text{ps}$ as an invariant for the outer loop, whereas the invariant for the inner loop additionally preserves the differential invariants used for handling the dynamics \(\dyn_\text{ps}\).
\end{proof}

\subsection{Arbitrary Number of Obstacles}
\label{sec:models-nobstacles}

The safety proofs so far modeled obstacles with a sensor system that nondeterministically delivers the position of any obstacle, including the nearest obstacle, to the control algorithm.
In this section, we also explicitly analyze how that sensor system lets the robot avoid collision with each one of many obstacles.
In order to prevent duplicating variables for each of the objects, which is undesirable even for a very small, known number of objects, we need a way of referring to countably many objects concisely.

\paragraph{Quantified Differential Dynamic Logic}

With \emph{quantified differential dynamic logic} \qdl \cite{DBLP:conf/csl/Platzer10,DBLP:journals/lmcs/Platzer12b}, we can explicitly refer to each obstacle individually by using quantification over objects of a sort (here all objects of the sort $O$ of obstacles).
\qdl is an extension of \dL suited for verifying distributed hybrid systems by quantifying over sorts.
\QdL extends hybrid programs to \emph{quantified hybrid programs}, which can describe the dynamics of distributed hybrid systems with any number of agents. 
Instead of using a single state variable $\pox :\reals$ to describe the $x$ coordinate of one obstacle, we can use a function term $\pox : O \rightarrow \reals$ in \qdl to denote that obstacle $i$ has $x$-coordinate $\pox(i)$, for each obstacle $i$ of obstacle sort $O$.
Likewise, instead of a single two-dimensional state variable $\po : \reals^2$ to describe the planar position of one obstacle, we can use a function term $\po : O \rightarrow \reals^2$ in \qdl to denote that obstacle $i$ is at position $\po(i)$, for each obstacle $i$.
We use a \emph{non-rigid} function symbol $\po$, which means that the value of all $\po(i)$ may change over time (\eg, the position $\po(\textit{car})$ of an obstacle named $\textit{car}$).
Other function symbols are \emph{rigid} if they do not change their values over time (\eg, the maximum velocity $V(i)$ of obstacle $i$ never changes).
Pure differential dynamic logic \dL uses the sort $\reals$. 
\qdl formulas can use quantifiers to make statements about all obstacles of sort $O$ with $\forall i \in O$ and $\exists i\in O$, similar to the quantifiers for the special sort $\mathbb{R}$ that \dL already provides.

\qdl allows us to explicitly track properties of all obstacles simultaneously.
Of course, it is not just the position data that is important for obstacles, but also that the model allows all moving obstacles to change their positions according to their respective differential equations.
Quantified hybrid programs allow the evolution of properties expressed as non-rigid functions for all objects of the same sort simultaneously (so all obstacles move simultaneously).

\rref{tab:qdl} lists the additional statements that quantified hybrid programs add to those of hybrid programs \cite{DBLP:conf/csl/Platzer10,DBLP:journals/lmcs/Platzer12b}.

\begin{table}[h]
  \newcommand{\foform}{F\xspace}
\small\sf\centering
  \caption{Statements of quantified hybrid programs}
  \label{tab:qdl}
  \begin{tabularx}{\columnwidth}{lX}
    \toprule
       \textbf{Statement}
    & \textbf{Effect}
    \\
    \midrule 
    $\forall i{\in}C\, \humod{x(i)}{\theta}$ & Assigns the current value of term $\theta$ to $x(i)$ simultaneously for all objects of sort $C$.\\
    \multirow{2}{*}{$\begin{aligned}\forall i{\in}C\, \bigl(\D{x(i)}=\theta(i)\\ ~\&~ Q\bigr)\end{aligned}$} & Evolves all $x(i)$ for any $i$ along differential equations $\D{x(i)} = \theta(i)$ restricted to evolution domain~$Q$\\
    \bottomrule
  \end{tabularx}
\end{table}

\begin{model*}[htb]
\begin{align}
\label{eq:dw5:model} \textit{dw}_\text{nobs} & \equiv \prepeat{\bigl(\ctrlo;~\ctrlr(\humod{\ar}{*};\ptest{(-b \leq \ar \leq A)}~,~\safe_\text{nobs});~\dyn_\text{nobs}\bigr)}\\
\label{eq:dw5:ctrlo} \ctrlo &\equiv \prepeat{\bigl(\pmb{\humod{i}{*}};~ \humod{\pmb{\vo(i)}}{(*,*)};~ \pmb{\ptest{\norm{\vo(i)} \leq V(i)}} \bigr)} \\
\safe_\text{nobs} &\equiv \pmb{\forall i{\in}O}~ \norm{\pr-\pmb{\po(i)}}{_\infty} >
\begin{cases}
\pmb{-\frac{\vr^2}{\ar} - \pmb{V(i)} \frac{\vr}{\ar}} & \pmb{\text{ if } \vr + \ar \varepsilon < 0}\\
\frac{\vr^2}{2b} +\pmb{V(i)}\frac{\vr}{b} + \left( \frac{\ar}{b}+1 \right) \left( \frac{\ar}{2} \varepsilon^2  + \varepsilon (\vr + \pmb{V(i)})\right) & \text{ otherwise}
\end{cases}\\
\label{eq:dw5:dyn1} \dyn_\text{nobs} &\equiv \pmb{\forall i{\in}O}~ (\D{t} = 1,~ \D{\pr} = \vr \dr,~ \D{\dr} = -\omegar \dr^\bot,~ \D{\vr} = \ar,~ \D{\omegar} = \frac{\ar}{r},~ \pmb{\D{\po}(i) = \vo(i)} ~\&~ \vr \geq 0 \land t \leq \varepsilon)
\end{align}
\caption{Explicit representation of countably many obstacles, extends
\rref{model:dynamicwindowpassiveactuala}}
\label{model:multipleObstacles}
\end{model*}

We can use \QdL to look up characteristics of specific obstacles, such as their maximum velocity, which allows an implementation to react to different kinds of obstacles differently if appropriate sensors are available.

\paragraph{Modeling}
In \rref{model:multipleObstacles} we move from hybrid programs to quantified hybrid programs for distributed hybrid systems \cite{DBLP:conf/csl/Platzer10,DBLP:journals/lmcs/Platzer12b}, \ie, systems that combine distributed systems aspects (lots of obstacles) with hybrid systems aspects (discrete control decisions and continuous motion).
We introduce a sort $O$ representing obstacles so that arbitrarily many obstacles can be represented in the model simultaneously. 
Each obstacle $i$ of the sort $O$ has a maximum velocity $V(i)$, a current position $\po(i)$, and a current vectorial velocity $\vo(i)$. 
We use non-rigid function symbols $\po:O \rightarrow \mathbb{R}^2$, $\vo:O\rightarrow \mathbb{R}^2$, and $V:O\rightarrow \mathbb{R}$.
Both $\po(i)$ and $\vo(i)$ are two-dimensional vectors.

This new modeling paradigm also allows for another improvement in the model. 
So far, an arbitrary obstacle was chosen by picking any position nondeterministically in $\ctrlr$. 
Such a nondeterministic assignment includes the closest one. 
A controller implementation needs to compute which obstacle is actually the closest one (or consider them all one at a time). 
Instead of assigning the closest obstacle nondeterministically in the model, \QdL can consider all obstacles by quantifying over all obstacles of sort $O$.

In the obstacle controller $\ctrlo$ \eqref{eq:dw5:ctrlo} we use a loop to allow multiple obstacles to make a control decision. 
Each run of that loop selects one obstacle instance $i$ arbitrarily and updates its velocity vector (but no longer its position, since obstacles are now tracked individually). 
The loop can be repeated arbitrarily often, so any arbitrary finite number of obstacles can make control choices in \eqref{eq:dw5:ctrlo}.
In the continuous evolution, we quantify over \emph{all obstacles} $i$ of sort $O$ in order to express that all obstacles change their state simultaneously according to their respective  differential equations \eqref{eq:dw5:dyn1}.

Initial condition, safety condition, and loop invariants are as before \eqref{eq:ps-safe}--\eqref{eq:ps-invariant} except that they are now phrased for all obstacles $i\in O$.
Initially, our robot is assumed to be stopped and we do not need to assume anything about the obstacles initially because passive safety does not consider collisions when stopped:
\begin{align}
\label{eq:nobs-initial}
\phi_\text{nobs} & \equiv \vr=0 \land \rcurve \neq 0 \land \norm{\dr}=1\\
\intertext{
The safety condition is passive safety for all obstacles:
}
\label{eq:nobs-safety}
\psi_\text{nobs} & \equiv \vr \neq 0 \limply \forall i{\in}O\, \norm{\pr-\po(i)}{_\infty} > 0
\end{align}

\paragraph{Verification}

We use \QdL to prove passive safety in the presence of arbitrarily many obstacles.
Note that the controller condition $\safe_\text{nobs}$ for multiple obstacles needs to distinguish obstacles that will stop during the next control cycle from those that will not.

\begin{thm}[Passive safety for arbitrarily many obstacles]
\label{thm:passivesafety-qdl}
Robots tracking any number of obstacles of their respective maximum velocities by \rref{model:multipleObstacles} will never collide with any obstacle while driving, as expressed by the provable \QdL formula \(\phi_\text{nobs} \limply \dibox{\textit{dw}_\text{nobs}}\psi_\text{nobs} \enspace .\)
\end{thm}

\begin{proof}
Since \QdL is not yet implemented in \KeYmaeraX, we proved \rref{thm:passivesafety-qdl} with its predecessor \KeYmaera. 
The proof uses \eqref{eq:ps-invariant} with explicit \(\forall i \in O\) as loop invariant:
\begin{equation*}
\varphi_\text{nobs} \equiv \vr \neq 0 \limply \forall i{\in}O\, \norm{\pr-\po(i)}{_\infty} > \frac{\vr^2}{2 b} + V(i) \frac{\vr}{b}
\end{equation*}
The proof uses quantified differential invariants to prove the properties of the quantified differential equations \cite{DBLP:conf/hybrid/Platzer11}.
\end{proof}

\section{Liveness Verification of Ground Robot Navigation}
\label{sec:livenessmodels}

Safety properties formalize that a precisely-defined bad behavior (such as collisions) will never happen.
Liveness properties formalize that certain good things (such as reaching a goal) will ultimately happen.
It is easy to design a trivial controller that is only safe (just never moves) or only live (full speed toward the goal ignoring all obstacles).
The trick is to design robot controllers that meet both goals.
The safe controllers identified in the previous sections guarantee safety (no collisions) and still allow motion.
This combination of guaranteed safety under all circumstances (by a proof) and validated liveness under usual circumstances (validated only by some tests) is often sufficient for practical purposes.
Yet, without a liveness proof, there is no guarantee that the robot controller will reach its respective goal except in the circumstances that have been tested before.
In this section, we verify liveness properties, since the precision gained by formalizing the desired liveness properties as well as the circumstances under which they can be guaranteed are insightful.

Formalizing liveness properties is even more difficult and the resulting questions in practice much harder than safety (even if liveness can be easier in theory \cite{DBLP:journals/tocl/Platzer15}).
Both safety and liveness properties only hold when they are true in the myriad of situations with different environmental behavior that they conjecture.
They are diametrically opposed, because liveness requires motion but safety considerations inhibit motion.
For the safe robot models that we consider here, liveness is, thus, quite a challenge, because there are many ways that environmental conditions or obstacle behavior would force the robot to stop or turn around for safety reasons, preventing it from reaching its goal.
For example, an unrestricted obstacle could move around to block the robot's path and then, as the robot re-plans to find another trajectory, dash to block the new path too.
To guarantee liveness, one has to characterize \emph{all necessary conditions} that allow the robot to reach its goal, which are often prohibitively many.
Full adversarial behavior can be handled but is challenging \cite{DBLP:journals/tocl/Platzer15}.

For a liveness proof, we deem three conditions important:
\begin{description}[noitemsep]
\item[\textbf{Adversarial behavior.}] Carefully defines acceptable adversarial behavior that the robot can handle.
For example, sporadically crossing a robot's path might be acceptable in the operating conditions, but permanently trapping the robot in a corner might not.
\item[\textbf{Conflicting goals.}] Identifies conflicting goals for different agents. 
For example, if the goal of one robot is to indefinitely occupy a certain space and that of another is to reach this very space it is impossible for both to satisfy their respective requirements.
\item[\textbf{Progress.}] Characterizes progress formally. 
For example, in the presence of obstacles, a robot sometimes needs to move away from the goal in order to ultimately get to the goal.
But how far is a robot allowed to deviate on the detour?
\end{description}

Liveness properties that are actually true need to define some reasonable restrictions on the behavior of other agents in the environment. 
For example, a movable obstacle may block the robot's path for some limited amount of time, but not indefinitely. 
And when the obstacle moves on, it may not turn around immediately again.
Liveness conditions might define a compromise between reaching the goal and having at least invested \emph{reasonable effort} of trying to get to the goal, if unacceptable adversarial behavior occurred or goals conflicted, or progress is physically impossible.

In this section, we start with a stationary environment, so that we first can concentrate on finding a notion for progress for the robot itself.
Next, we let obstacles cross the robot's path and define what degree of adversarial behavior is acceptable for guaranteeing liveness.

\subsection{Reach a Waypoint on a Straight Lane}

As a first liveness property, we consider a stationary environment without obstacles, which prevents adversarial behavior as well as conflicting goals, so that we can concentrate on the conditions to describe how the robot makes progress without the environment interfering. 
We focus on low-level motion planning where the robot has to make decisions about acceleration and braking in order to drive to a waypoint on a straight line.
We want our robot to \emph{provably} reach the waypoint, so that a high-level planning algorithm knows that the robot will reliably execute its plan by stitching together the complete path from straight-line segments between waypoints.
To model the behavior at the final waypoint when the robot stops (because it reached its goal) and at intermediate waypoints in a uniform way, we consider a simplified version where the robot has to stop at each waypoint, before it turns toward the next waypoint.
That way, we can split a path into straight-line segments that make it easier to define progress, because they are describable with solvable differential equations when abstracted into one-dimensional space.

\paragraph{Modeling}

We say that the robot reached the waypoint when it stops inside a region of size $2\Delta_g$ around the waypoint.
That is:
\begin{enumerate*}[label=\emph{(\roman*)}]
\item at least one execution enters the goal region, and
\item all executions stop before exiting the goal region $\pg + \Delta_g$.
\end{enumerate*}
The liveness property $\psi_\text{wp}$ \eqref{eq:waypoint:liveness} characterizes these conditions formally.
\begin{equation}\label{eq:waypoint:liveness}
\psi_\text{wp} \equiv \didia{\textit{dw}_\text{wp}}(\pg - \Delta_g < \pr) \land \dibox{\textit{dw}_\text{wp}}(\pr < \pg + \Delta_g)
\end{equation}

\begin{model*}
\begin{align}
\textit{dw}_\text{wp} & \equiv \prepeat{(\textit{\ctrl};\textit{\dyn})}\\
\label{eq:wpmodel:brake} \textit{\ctrl} & \equiv \phantom{\cup} (\humod{\ar}{-b})\\
\label{eq:wpmodel:stopped}&\phantom{\equiv} \cup (\ptest{\vr = 0};~ \humod{\ar}{0})\\
\label{eq:wpmodel:acc} &\phantom{\equiv} \cup 	(\ptest{\pr + \frac{\vr^2}{2 b} + \left(\frac{A}{b} + 1\right)\left(\frac{A}{2}\varepsilon^2 + \varepsilon \vr\right) < \pg + \Delta_g \land \vr+A\varepsilon \leq V_g};~ \humod{\ar}{A})\\	
\label{eq:wpmodel:approach} & \phantom{\equiv} \cup (\ptest{\pr \leq \pg - \Delta_g \land \vr \leq V_g};~ \humod{\ar}{*};~ \ptest{-b \leq \ar \leq \frac{V_g - \vr}{\varepsilon} \leq A})\\
\label{eq:wpmodel:dyn} \textit{\dyn} & \equiv \humod{t}{0};~ \{ \D{\pr}=\vr,~ \D{\vr}=\ar,~ \D{t}=1~ \&~ t \leq \varepsilon \land \vr \geq 0 \}
\end{align}
\caption{Robot follows a straight line to reach a waypoint}
\label{model:waypoint}
\end{model*}

\begin{remark}
The liveness property $\psi_\text{wp}$ \eqref{eq:waypoint:liveness} is formulated as a conjunction of two formulas: at least one run enters the goal region $\didia{\textit{dw}_\text{wp}}\,\pg - \Delta_g < \pr$, while none exit the goal region on the other end $\dibox{\textit{dw}_\text{wp}}\,\pr < \pg + \Delta_g$.
In particular, there is a run that will stop inside the goal region, which, explicitly, corresponds to extending formula \eqref{eq:waypoint:liveness} to the following liveness property:
\begin{equation}\label{eq:waypoint:liveness2}
\begin{aligned}
\didia{\textit{dw}_\text{wp}}\big(\pg - \Delta_g < \pr \land 0 \leq \vr \leq V_g \land \didia{\textit{dw}_\text{wp}}\,\vr=0\big)\\ 
\land \dibox{\textit{dw}_\text{wp}}(\pr < \pg + \Delta_g)
\end{aligned}
\end{equation}

Formula \eqref{eq:waypoint:liveness2} means that there is an execution of model $\textit{dw}_\text{wp}$ where the robot enters the goal region without exceeding the maximum approach velocity $V_g$, and from where the model has an execution that will stop the robot $\didia{\textit{dw}_\text{wp}}\,\vr=0$.
The proof for formula \eqref{eq:waypoint:liveness2} uses the formula
$\vr = 0 \lor (\vr > 0 \land \vr - n \varepsilon b \leq 0)$ to characterize progress (\ie, braking for duration $n \varepsilon$ will stop the robot).
\end{remark}

\rref{model:waypoint} describes the behavior of the robot for approaching a goal region. 
In addition to the three familiar options from previous models of braking unconditionally \eqref{eq:wpmodel:brake}, staying stopped \eqref{eq:wpmodel:stopped}, or accelerating when safe \eqref{eq:wpmodel:acc}, the model now contains a fourth control option \eqref{eq:wpmodel:approach} to slowly approach the goal region, because nondeterministically big acceleration choices might overshoot the goal.

The liveness proof has to show that the robot will get to the goal under \emph{all} circumstances except those explicitly characterized as being assumed not to happen, \eg, unreasonably small goal regions, high robot velocity, or hardware faults, such as engine or brake failure.
Similar to safety proofs, these assumptions are often linked.
For example, what makes a goal region unreasonably small depends on the robot's braking and acceleration capabilities.
The robot cannot stop at the goal if accelerating just once from its initial position will already make it impossible for the robot to brake before shooting past the goal region.
In this case, both options of the robot will violate our liveness condition: it can either stay stopped and not reach the goal, or it can start driving and miss the goal.

Therefore, we introduce a maximum velocity \(V_g\) that the robot has to obey when it is close to the goal.
That velocity must be small enough so that the robot can stop inside the goal region and is used as follows.
While obeying the approach velocity $V_g$ outside the goal region \eqref{eq:wpmodel:approach}, the robot can choose any acceleration that will not let it exceed the maximum approach velocity.
The dynamics of the robot in this model follows a straight line, assuming it is already oriented directly towards the goal \eqref{eq:wpmodel:dyn}.

\paragraph{Identification of Live Controls}

Now that we know what the goal of the robot is, we provide the intuition behind the conditions that make achieving the goal possible. 
The robot is only allowed to adapt its velocity with controls other than full braking when those controls will not overshoot the goal region, see \(\pg+\Delta_g\) in \eqref{eq:wpmodel:acc} and \(\pg-\Delta_g\) in \eqref{eq:wpmodel:approach}.
Condition \(-b \leq \ar \leq \frac{V_g-\vr}{\varepsilon} \leq A\) in \eqref{eq:wpmodel:approach} ensures that the robot will only pick acceleration values that will never exceed the approach velocity \(V_g\) in the next \(\varepsilon\) time units, \ie, until it can revise its decision.
Once inside the goal region, the only remaining choice is to brake, which makes the robot stop reliably in the waypoint region.

The robot is stopped initially ($\vr=0$) outside the goal region ($\pr < \pg - \Delta_g$), its brakes $b>0$ and engine $A>0$ are working,\footnote{For safety, $A \geq 0$ was sufficient, but in order to reach a goal the robot must be able to accelerate to non-zero velocities.} and it has some known reaction time $\varepsilon>0$:
\begin{equation}\label{eq:waypoint:initial}
\phi_\text{wp} \equiv \vr = 0 \land \pr < \pg - \Delta_g \land b > 0 \land A > 0 \land \varepsilon > 0 \land 0 < V_g \land V_g \varepsilon + \frac{V_g^2}{2 b} < 2 \Delta_g
\end{equation}

Most importantly, the approach velocity $V_g$ and the size of the goal region $2\Delta_g$ must be compatible.
That way, we know that the robot has a chance to approach the goal with a velocity that fits to the size of the goal region.

\paragraph{Verification}

Similar to safety verification, for liveness verification we combine the initial condition $\phi_\text{wp}$ \eqref{eq:waypoint:initial}, the model $\textit{dw}_\text{wp}$ (\rref{model:waypoint}), and the liveness property $\psi_\text{wp}$ \eqref{eq:waypoint:liveness} in \rref{thm:waypoint}.

\begin{thm}[Reach waypoint]\label{thm:waypoint}
Robots following \rref{model:waypoint} can reach the goal area $\pg - \Delta_g < \pr$ and will never overshoot $\pr < \pg + \Delta_g$,
as expressed by the provable \dL formula $\phi_\text{wp} \limply \psi_\text{wp}$, \ie, with $\psi_\text{wp}$ from \eqref{eq:waypoint:liveness} expanded:
\begin{align*}
\phi_\text{wp} \limply \bigl(&\didia{\textit{dw}_\text{wp}}(\pg - \Delta_g < \pr)\\
&\land \dibox{\textit{dw}_\text{wp}}(\pr < \pg + \Delta_g)\bigr) \enspace .
\end{align*}
\end{thm}

\begin{proof}
We proved \rref{thm:waypoint} using \KeYmaeraX.
Instead of an invariant characterizing what does not change, we now need a variant characterizing what it means to make progress towards reaching the goal region \cite{DBLP:journals/jar/Platzer08,DBLP:conf/lics/Platzer12a}.
If the progress measure indicates the goal would be reachable with $n$ iterations of the main loop of \rref{model:waypoint}, then we have to show that by executing the loop once we can get to a state where the progress measure indicates the goal would be reachable in the remaining $n-1$ loop iterations.

Informally, the robot reaches the goal if it has a positive speed $\vr > 0$ and can enter the goal region by just driving for time $n \varepsilon$ with that speed, as summarized by the loop variant 
\(\varphi_\text{wp} \equiv 0 < \vr \leq V_g \land \pg - \Delta_g < \pr + n \varepsilon \vr\).
\end{proof}

After having proved how the robot can always reach its goal when it is on its own, we next analyze liveness in the presence of other moving agents.

\subsection{Cross an Intersection}

\begin{figure}[htb]
\centering
\begin{tikzpicture}
\draw[color=black] (-3,0)--(1,0);
\draw[fill=lsblue] (-2,0) circle (1.5pt);
\draw[->,color=black,very thick] (-2,0)--(-1,0);
\draw[color=black,anchor=south] node at (-2,0) {robot $\pr$};
\draw[color=lsred,dashed] (-0.5,1)--(0.5,-1);
\draw[fill=lsblue] (-0.4,0.8) circle (1.5pt);
\draw[->,color=black,very thick] (-0.4,0.8)--(-0.1,0.2);
\draw[color=black,anchor=west] node at (-0.4,0.8) {$\po$ obstacle};
\draw[fill=lsblue] (0,0) circle (1.5pt);
\draw[color=black,anchor=north west] node at (0.1,0) {$\pix=(\pixr,\pixo)$ intersection};
\end{tikzpicture}
\caption{Illustration of the paths of a robot (black solid line) and an obstacle (red dashed line) crossing an intersection at point $\pix$.}
\label{fig:cx-illustration}
\end{figure}
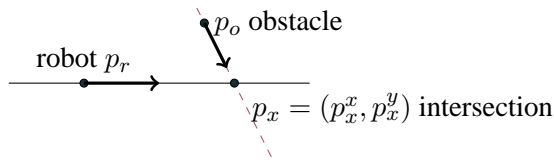

In this section, we prove liveness for scenarios in which the robot has to pass an intersection, while a moving obstacle may cross the robot's path, so that the robot may need to stop for safety reasons to let the obstacle pass.
We want to prove that it is always possible for the robot to successfully pass the intersection.
The model captures the general case of a point-intersection with two entering roads and two exits at the opposing side, so that it subsumes any scenario where a robot and an obstacle drive straight to cross an intersection, as illustrated in \rref{fig:cx-illustration}.

\paragraph{Modeling}

Since there is a moving obstacle, the robot needs to follow a collision avoidance protocol in order to safely cross the intersection.
We choose passive safety for simplicity.
Collision avoidance alone, however, will not reliably let the robot make progress.
Thus, we will model a robot that favors making progress towards the other side of the intersection, and only falls back to collision avoidance when the obstacle is too close to pass safely.

Intersections enable the obstacle to trivially prevent the robot from ever passing the intersection.
All that the obstacle needs to do is just block the entire intersection forever by stopping there (\eg, somebody built a wall so that the intersection disappeared).
Clearly, no one could demand the robot passes the intersection in such impossible cases.
We prove that the robot can pass the intersection when obstacles \emph{behave reasonably}, for a precisely defined characterization of what is reasonable for an obstacle to do.
We, therefore, include a restriction on how long the obstacle may reside at the intersection.
We choose a strictly positive minimum velocity $V_\textit{min}$ to prevent the obstacle from stopping.
Other fairness conditions (\eg, an upper bound on how long the intersection can be blocked, enforced with a traffic light) are representable in hybrid programs as well.

\begin{model}
\begin{align}
\textit{dw}_\text{cx} &\equiv \prepeat{(\ctrl_o;~\ctrl_r;~\dyn)}\\
\ctrl_o &\equiv \humod{a_o}{*};~\ptest{(-b \leq a_o \leq A)}\\
\ctrl_r &\equiv 
\begin{cases}
\humod{\ar}{*};~ \ptest{(-b \leq \ar \leq A)} & \text{if } \textit{AfterX}\\
\humod{\ar}{*};~ \ptest{(0 \leq \ar \leq A)} & \text{if } \textit{PassFaster}\\
\humod{\ar}{0} & \text{if } \textit{PassCoast}\\
\phantom{\cup~} (\humod{\ar}{-b}) \\ 
\cup~(\ptest{\vr = 0};~ \humod{\ar}{0}) & \text{otherwise \rref{model:dynamicwindowpassive}}\\ 
\cup~(\ptest{\textit{safe}};~ \text{\ldots}) \\
\end{cases}\\
\dyn &\equiv \humod{t}{0};~ \{\D{p}_r=\vr \syssep~ \D{v}_r=\ar \syssep \D{p}_o=\vo \syssep~ \D{v}_o=a_o,~ \D{t}=1 ~\&~ t \leq \varepsilon \land \vr \geq 0 \land \vo \geq V_\textit{min}\}
\end{align}
\caption{Robot safely crosses an intersection}
\label{model:crossintersection}
\end{model}

\paragraph{Identification of Live Controls}

For ensuring progress, the model uses three conditions ($\textit{AfterX}$, $\textit{PassFaster}$, and $\textit{PassCoast}$) that tell the robot admissible conditions for choosing its acceleration, depending on its own position and the obstacle position in relation to the intersection. 
The robot can choose any acceleration after it passed the intersection ($\pr>\pixr$) or after the obstacle passed ($\po>\pixo$):
\[\textit{AfterX} \equiv \pr > \pixr \lor \po > \pixo \enspace .\]
The robot is allowed to increase its speed if it manages to pass safely in front of the obstacle (even if the obstacle speeds up during the entire process), or if speeding up would still let the robot pass safely behind the obstacle (even if the obstacle drives with only minimum speed $V_\textit{min}$):
\begin{align*}
\textit{PassFaster} &\equiv \vr > 0 \land \left(\textit{PassFront} \lor \textit{PassBehind}\right)\\
\textit{PassFront} &\equiv \po + \vo\frac{\pixr - \pr}{\vr} + \frac{A}{2} \left(\frac{\pixr - \pr}{\vr}\right)^2 < \pixo \\
\textit{PassBehind} &\equiv \pixo < \po + V_\textit{min} \frac{\pixr - \pr}{\vr + A\varepsilon}
\end{align*}
The robot is allowed to just maintain its speed if it either passes safely in front or behind the obstacle with that speed:
\[\textit{PassCoast} \equiv \vr > 0 \land \pixo < \po + V_\textit{min} \frac{\pixr - \pr}{\vr} \enspace .
\]

In all other cases, the robot has to follow the collision avoidance protocol from \rref{model:dynamicwindowpassive} to choose its speed, modified accordingly for the one-dimensional variant here.

\paragraph{Verification}

As a liveness condition, we prove that the robot will make it past the intersection without colliding with the obstacle.

\begin{thm}[Pass Intersection]\label{thm:cx-liveness}
Robots following \rref{model:crossintersection} can pass an intersection while avoiding collisions with obstacles at the intersection, as expressed in the provable \dL formula
\begin{align*}
\phi_\text{cx} \limply \quad & \dibox{\textit{dw}_\text{cx}}\,\left(\pr=\pixr \limply \po \neq \pixo\right)\\ \land & \didia{\textit{dw}_\text{cx}}\,\left(\pr > \pixr\right) \enspace .
\end{align*}
\end{thm}

\begin{proof}
We proved \rref{thm:cx-liveness} in \KeYmaeraX.
In the loop invariant of the safety proof we combine the familiar stopping distance $\pr + \frac{\vr^2}{2b}< \pixr$ with the conditions $\textit{AfterX}$, $\textit{PassCoast}$, and $\textit{PassFront}$ that allow driving in its loop invariant:
\begin{equation*}
0 \leq \vr \land V_\text{min} \leq \vo \land \bigl(\pr+ \frac{\vr^2}{2b} < \pixr \lor \textit{AfterX} \lor \textit{PassCoast} \lor \textit{PassFront} \bigr) \enspace .
\end{equation*}

The main insight in the liveness proof is that achieving the goal can be split into two phases: first, the robot waits for the obstacle to pass; afterwards, the robot accelerates to pass the intersection.
We split the loop into these two phases with $\didia{\prepeat{\textit{dw}_\text{cx}}}(\pr>\pixr) \lbisubjunct \didia{\prepeat{\textit{dw}_\text{cx}}}\didia{\prepeat{\textit{dw}_\text{cx}}}(\pr>\pixr)$ so that we can analyze each of the resulting two loops with its own separate loop variant.
In the first loop, we know that the obstacle drives with at least speed $\vo \geq V_\text{min}$, so with $n$ steps it will pass the intersection, which is characterized in the loop variant $\po + n\varepsilon V_\text{min}>\pixo$.
This loop variant implies $\po > \pixo$ when $n \leq 0$.
Once the obstacle is past the intersection, in the second loop the robot controller can safely favor its $\textit{AfterX}$ control.
Since the robot might be stopped, we unroll the loop once to $\didia{\textit{dw}_\text{cx}}\didia{\prepeat{\textit{dw}_\text{cx}}}(\pr>\pixr)$ in order to ensure that the robot accelerates with $A$ to a positive speed.
The loop variant then exploits that the robot's speed is $\vr \geq A\varepsilon$ after accelerating once for time $\varepsilon$, so it will pass the intersection $\pixr$ with $n$ steps of duration $\varepsilon$ as follows: $\pr + n\varepsilon(A\varepsilon) > \pixr$.
\end{proof}

The liveness proofs show that the robot \emph{can} achieve a useful goal if it makes the right choices.
When the robot controller is modeled such that it \emph{always} makes the right choices, we prove that the controller will always safely make it to the goal within a specified time budget.
We discuss robot controllers that provably meet deadlines in \rref{appendix:livenessdeadlines}.

\subsection{Liveness with Deadlines}
\label{appendix:livenessdeadlines}

The liveness proofs in the article showed that the robot can achieve a useful goal if it makes the right choices.
The proofs neither guarantee that the robot will always make the right decisions, nor specify how long it will take until the goal will be achieved.
In this section, we prove that it always achieves its goals within a given reasonable amount of time.
Previously we showed that the robot \emph{can} do the right thing to ultimately get to the goal, while here we prove that it \emph{always} makes the right decisions that will take it to the waypoint or let it cross an intersection \emph{within a bounded amount of time}.
It is no longer enough to show existence of an execution that makes the robot achieve its goals.
Now we need to show that all possible executions do so in the given time.
This needs more deterministic controllers that only brake when necessary. 

We are going to illustrate two alternatives for modeling arrival deadlines: in \rref{sec:livenessdeadlines-waypoint} we use a countdown $T$ that is initialized to the deadline and expires when $T\leq0$, whereas in \rref{sec:livenessdeadlines-intersection} we use $T$ as a clock that is initialized to a starting value $T\leq0$ and counts up to a deadline $D>0$, so that two deadlines (crossing zero and exceeding $D$) can be represented with a single clock variable.

\subsubsection{Reaching a Waypoint}
\label{sec:livenessdeadlines-waypoint}

We start by defining a correctness condition for reaching a waypoint.
\begin{equation}\label{eq:waypointdl:correctness}
\psi_\text{wpdl} ~\mequiv~ \pr < \pg + \Delta_g \land (T \leq 0 \limply \vr = 0 \land \pg - \Delta_g < \pr)
\end{equation}

Formula \eqref{eq:waypointdl:correctness} expresses that the robot will never be past the goal region ($\pr < \pg + \Delta_g$), and after the deadline ($T \leq 0$, \ie after countdown $T$ expired) it will be stopped inside the goal region ($\vr = 0 \land \pg - \Delta_g < \pr$).

\paragraph{Modeling}

\begin{model*}
\begin{align}
\label{eq:wpdlmodel:model} \textit{dw}_\text{wpdl} & \equiv \prepeat{(\ctrl;\dyn)}\\
\label{eq:wpdlmodel:ctrl} \ctrl & \equiv
\begin{cases}
\pchoice{(\humod{\ar}{-b})}{(\ptest{\vr=0};\humod{\ar}{0})} & \pmb{\text{ if } \pg - \Delta_g < \pr} \\
\humod{\ar}{A} & \pmb{\text{ if } \pr + \frac{\vr^2 - V_g^2}{2 b} + \left(\frac{A}{b} + 1\right) \left(\frac{A}{2}\varepsilon^2 + \varepsilon \vr \right) \leq \pg - \Delta_g}\\
\humod{\ar}{*};~ \ptest{-b \leq \ar \leq \frac{V_g - \vr}{\varepsilon} \leq A} & \text{ otherwise}
\end{cases}\\
\label{eq:wpdlmodel:dyn} \textit{\dyn} & \equiv \humod{t}{0};~ \D{\pr}=\vr,~ \D{\vr}=\ar,~ \D{t}=1,~ \pmb{\D{T}=-1} ~\&~ t \leq \varepsilon \land \vr \geq 0
\end{align}
\caption{Robot reaches a waypoint before a deadline}
\label{model:waypointdeadline}
\end{model*}

\rref{model:waypointdeadline} is the familiar loop of control followed by dynamics \eqref{eq:wpdlmodel:model}.
Unlike in previous models, braking and staying put is no longer allowed unconditionally for the sake of reaching the waypoint reliably in time \eqref{eq:wpdlmodel:ctrl}.
The robot accelerates maximally whenever possible without rushing past the waypoint region, cf. \eqref{eq:wpdlmodel:ctrl}.
In all other cases, the robot chooses acceleration to control towards the approach velocity $V_g$ \eqref{eq:wpdlmodel:ctrl}.
The dynamics remain unchanged, except for the additional countdown $\D{T}=-1$ of the deadline in \eqref{eq:wpdlmodel:dyn}.

\paragraph{Identification of Live Controls}

In order to prove this model live, we need to set achievable deadlines.
The deadline has to be large enough 
\begin{enumerate*}[label=\emph{(\roman*)}]
\item for the robot to accelerate to velocity $V_g$, 
\item drive to the waypoint with that velocity, and 
\item once it is there, have sufficient time to stop.
\end{enumerate*}
It also needs a slack time $\varepsilon$, so that the robot has time to react to the deadline.
Finally, the conditions $\phi_\text{wp}$ from \eqref{eq:waypoint:initial}, which enable the robot to reach a waypoint at all, have to hold as well.
Formula \eqref{eq:waypointdl:achievabledeadline} summarizes these deadline conditions.
\begin{equation}\label{eq:waypointdl:achievabledeadline}
\phi_\text{wpdl} \equiv \phi_\text{wp} \land T > \underbrace{\frac{V_g - \vr}{A}}_\textit{(i)} + \underbrace{\frac{\pg - \Delta - \pr}{V_g}}_\textit{(ii)} + \underbrace{\frac{V_g}{b}}_\textit{(iii)} + \varepsilon
\end{equation}

\paragraph{Verification}

A proof of the robot always making the right choices is a combination of a safety and a liveness proof: we have to prove that \emph{all} choices of the robot reach the goal before the deadline expires (safety proof), and that \emph{there exists at least one} way of the robot reaching the goal before the deadline expires (liveness proof).
Both $\dbox{\cdot}$ and $\didia{\cdot}$ are needed to express that the robot always makes the right choices to get to the waypoint, since $\dbox{\cdot}$ alone does not guarantee existence of such a choice.

\begin{thm}[Reach waypoint with deadline]\label{thm:waypointdl}
Robots following \rref{model:waypointdeadline} will always reach the waypoint before the deadline expires, as expressed by the provable \dL formula 
\[\phi_\text{wpdl} \limply \bigl( \dibox{\textit{dw}_\text{wpdl}}\psi_\text{wpdl} \land \didia{\textit{dw}_\text{wpdl}}\psi_\text{wpdl} \bigr) \enspace . \]
\end{thm}

\begin{proof}
We proved \rref{thm:waypointdl} with \KeYmaeraX, using automated tactics to handle the solvable differential equation system.
The proof uses the following conditions as loop invariants:
\begin{align*}
\pr & + \frac{\vr^2}{2b} < \pg+\Delta_g \land 0 \leq \vr \leq V_g \land\\
& \begin{cases}
\vr=0 \lor T \geq \frac{\vr}{b} & \text{ if } \pg - \Delta_g < \pr\\
T > \frac{\pg-\Delta_g-\pr}{A\varepsilon} + \frac{V_g}{b} + \varepsilon & \text{ if } \pr \leq \pg - \Delta_g \land \vr \geq A\varepsilon \\
T > \varepsilon - \frac{\vr}{A} + \frac{\pg-\Delta_g-\pr}{A\varepsilon} + \frac{V_g}{b} + \varepsilon & \text{ if } \pr \leq \pg - \Delta_g \land \vr \leq A\varepsilon
\end{cases}
\end{align*}
The robot maintains sufficient margin to avoid overshooting the goal area and it respects the approach velocity $V_g$.
Reaching the goal is then split into increasingly critical cases: if the robot already is at the goal ($\pg-\Delta_g < \pr$) it is either stopped already or will manage to stop before the deadline expires.
If the robot is not yet at the goal, but at least already traveling with some non-zero speed $\vr \geq A\varepsilon$, then it still has sufficient time to drive to the goal with the current speed and stop.
Finally, if the robot is not yet traveling fast enough, it still has sufficient time to speed up.
\end{proof}

\subsubsection{Crossing an Intersection}
\label{sec:livenessdeadlines-intersection}

Crossing an intersection before a deadline is more complicated than reaching a waypoint, because the robot may need to wait for the intersection to clear so that the robot can cross it safely in the first place.

\paragraph{Modeling}

\rref{model:crossintersectiondeadline} remains almost identical to \rref{model:crossintersection}, except for the robot controller, which has an additional control branch: when the obstacle has already passed the intersection, we want the robot to pass as fast as it can by accelerating fully with maximum acceleration $A$ (no dawdling).

\begin{model}[h]
\begin{align}
\textit{dw}_\text{cxd} &\equiv \prepeat{(\ctrl_o;~\ctrl_r;~\dyn)}\\
\ctrl_o & \equiv \ctrl_o \text{ of \rref{model:crossintersection}}\\
\ctrl_r &\equiv 
\begin{cases}
\humod{\ar}{A} & \text{if } \po > \pixo \\
\ctrl_r \text{ of \rref{model:crossintersection}} & \text{otherwise}
\end{cases}\\
\dyn &\equiv \dyn \text{ of \rref{model:crossintersection}}
\end{align}
\caption{Crossing an intersection before a deadline}
\label{model:crossintersectiondeadline}
\end{model}

\paragraph{Identification of Live Controls}

Given the robot behavior of \rref{model:crossintersectiondeadline} above, we need to set a deadline that the robot can actually achieve, considering when and how much progress the robot can make while driving (recall that it should still not collide with the obstacle).
The deadline has to account for both the robot and the obstacle position relative to the intersection, as well as for how much the robot can accelerate.
We start with the easiest case for finding a deadline $D$: when the obstacle already passed the intersection, the robot simply has to accelerate with maximum acceleration until it itself passes the intersection. 
The obstacles are assumed to never turn back, so accelerating fully is also a safe choice.
The robot might be stopped. 
So, assuming we start a deadline timer $T$ at time $0$, the robot will drive a distance of $\frac{A}{2}D^2$ until the deadline $D$ expires (\ie, until $T=D$).
However, since we use a sampling interval of $\varepsilon$ in the robot controller, the robot may not notice that the obstacle already passed the intersection for up to time $\varepsilon$, which means it will only accelerate for time $D-\varepsilon$.
Formula~\eqref{eq:cxd-deadline1} summarizes this case.
\begin{equation}\label{eq:cxd-deadline1}
\eta_\text{cxd}^D ~\mequiv~ D \geq \varepsilon \land \pixr - \prx < \frac{A}{2}\left(D-\varepsilon\right)^2
\end{equation}

If unlucky, the robot determines that it cannot pass safely in front of the obstacle and will have to wait until the obstacle passed the intersection.
Hence, within the deadline we have to account for the additional time that the obstacle may need at most to pass the intersection.
We could increase $D$ with the appropriate additional time and still start the timer at $T=0$, if we were to rephrase the implicit definition of the deadline $\pixr - \pr < \frac{A}{2}(D-\varepsilon)^2$ in \eqref{eq:cxd-deadline1} to its explicit form.
In \eqref{eq:cxd-deadline2}, instead, we start the deadline timer with time  \footnote{Recall $\po \leq \pixo$ holds when the obstacle did not yet pass the intersection.} $T\leq0$, such that it becomes $T=0$ when the obstacle is located at the intersection.
\begin{equation}\label{eq:cxd-deadline2}
\eta_\text{cxd}^T ~\equiv~ T = \min \left(0, \frac{\po - \pixo}{V_\textit{min}}\right)
\end{equation}

\paragraph{Verification}

\rref{thm:cxd-liveness} uses the deadline conditions \eqref{eq:cxd-deadline1} and \eqref{eq:cxd-deadline2} in a liveness property for \rref{model:crossintersectiondeadline}.

\begin{thm}[Cross intersection before deadline\label{thm:cxd-liveness}]
Model $\textit{dw}_\text{cxd}$ has a run where the robot can drive past the intersection ($\pr>\pixr$).
For appropriate deadline choices, all runs of model $\textit{dw}_\text{cxd}$, such that when the deadline timer is expired ($T \geq D$) the robot is past the intersection ($\pr > \pixr$).
All runs prevent collision, i.e., robot and obstacle never occupy the intersection at the same time ($\pr=\pixr \limply \po \neq \pixo$).
\begin{equation*}
\phi_\text{cxd} \land \eta_\text{cxd}^D \land \eta_\text{cxd}^T \limply
 \didia{\textit{dw}_\text{cxd}}(\pr>\pixr) \land \dibox{\textit{dw}_\text{cxd}}\bigl(\left(T \geq D \limply \pr > \pixr\right) \land \left(\pr=\pixr \limply \po \neq \pixo\right)\bigr)
\end{equation*}
\end{thm}

\begin{proof}
We proved \rref{thm:cxd-liveness} with \KeYmaeraX.
Collision avoidance $\dibox{\textit{dw}_\text{cxd}}(\pr=\pixr \limply \po \neq \pixo)$ and liveness $\didia{\textit{dw}_\text{cxd}}\pr>\pixr$ follow the approach in \rref{thm:cx-liveness}.
The loop invariant used for proving that the robot always meets the deadline ensures that there is sufficient time remaining until the deadline expires.
Similar to the liveness proof in \rref{thm:cx-liveness}, the deadline is split into two phases, because the robot may not be able to pass safely in front of the obstacle, so it may need to let the obstacle pass first. 
Recall that $T \leq 0$ when the obstacle is not yet past the intersection, so we characterize the worst-case remaining time until the obstacle passed with minimum speed $V_\text{min}$ by $T \leq \frac{\po-\pixo}{V_\text{min}}$.
In case the obstacle is not yet past the intersection, the robot must be positioned such that it can pass in $D-\varepsilon$ time, so $T \leq 0 \land \pr+\frac{A}{2}(D-\varepsilon)^2>\pixr$.
Finally, once the obstacle passed, the robot has $D-T$ time left to pass itself, which is summarized in $T>0 \land \pr+\vr\max(0,D-T) + \frac{A}{2}\max(0,D-T)^2>\pixr$.
\end{proof}

\section{Interpretation of Verification Results}
\label{appendix:interpretation}

As part of the verification activity, we identified crucial safety constraints that have to be satisfied in order to choose a new curve or accelerate safely.
These constraints are entirely symbolic and summarized in \rref{tab:constraintsummary}.
Next, we analyze the constraints for common values of acceleration force, braking force, control cycle time, and obstacle distance (\ie, door width, corridor width).

\subsection{Safe Distances and Velocities}

\paragraph{Static safety}

Recall safety constraint \eqref{eq:st-safe} from \rref{model:dynamicwindowstatic}, which is justified by \rref{thm:staticsafety} to correctly capture when it is safe to accelerate in the presence of stationary obstacles $\po$.
\begin{equation}
\norm{\pr - \po}{_\infty} > \frac{\vr^2}{2b} + \left(\frac{A}{b}+1\right)\left(\frac{A}{2}\varepsilon^2+\varepsilon \vr\right)
\tag{\ref{eq:st-safe}*}
\end{equation}

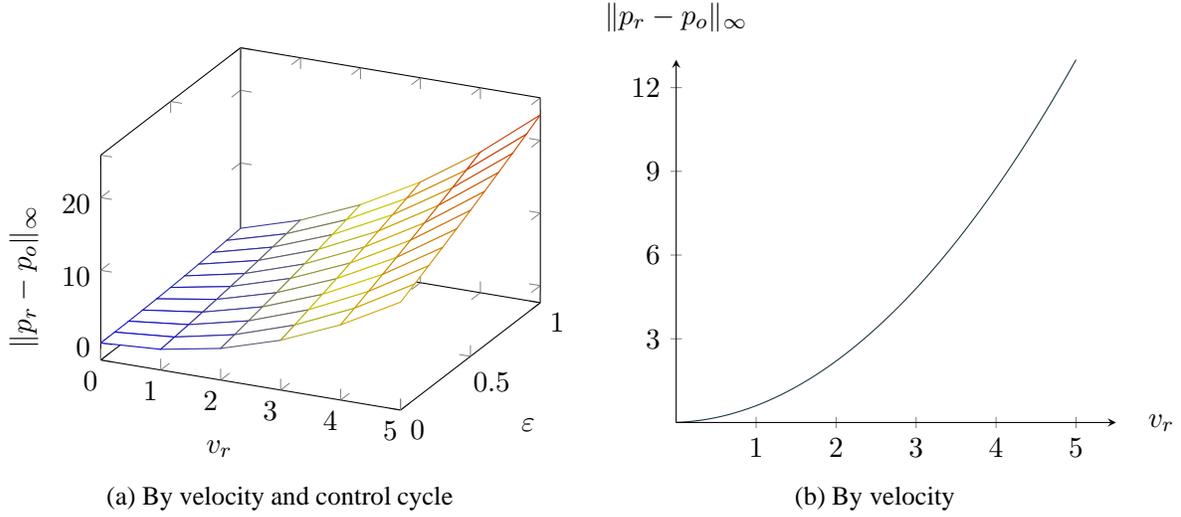
\begin{figure}[htb]
\centering
\begin{subfigure}[b]{0.45\columnwidth}
\begin{tikzpicture}
\begin{axis}[mesh/ordering=y varies,
width=\textwidth,
xlabel=\(\vr\),
ylabel=\(\varepsilon\),
zlabel=\(\norm{\pr-\po}{_\infty}\),
xtick={0,1,...,5},
ytick={0,0.5,...,1},
ztick={0,10,...,20}
]
\addplot3[surf,fill=white,mesh/rows=11] file {./fig/data/1_static/safedistvep.dat};
\end{axis}
\end{tikzpicture}
\caption{By velocity, control cycle}
\label{fig:safetydist_static_v-ep}
\end{subfigure}
\quad
\begin{subfigure}[b]{0.45\columnwidth}
\centering
\begin{tikzpicture}
\begin{axis}[
width=.9\textwidth,
xmin=0, xmax=5.5,
ymin=0, ymax=13,
axis x line=middle,
axis y line=middle,
every axis x label/.style={
    at={(ticklabel* cs:1.05)},
    anchor=west,
},
every axis y label/.style={
    at={(ticklabel* cs:1.05)},
    anchor=south,
},
xlabel=\(\vr\),
ylabel=\(\norm{\pr-\po}{_\infty}\),
xtick={0,1,...,5},
ytick={0,3,...,12},
axis on top
]
\addplot[color=lsblue,solid] plot table {./fig/data/1_static/safedist.dat};
\end{axis}
\end{tikzpicture}
\caption{By velocity}
\label{fig:safetydist_static_v}
\end{subfigure}
\caption{Safety distance for static safety}
\label{fig:safetydist_static}
\end{figure}

The constraint links the current velocity $\vr$ and the distance to the nearest obstacle through the design parameters $A$ (maximum acceleration), $b$ (maximum braking force), and $\varepsilon$ (maximal controller cycle time).
\rref{tab:interpretation_static} lists concrete choices for these parameters and the minimum safety distance identified by \rref{eq:st-safe} in \rref{model:dynamicwindowstatic}.
\begin{table}[htb]
\small\sf\centering
\caption{Static safety: minimum safe distance and maximum velocity for select configurations}
\label{tab:interpretation_static}
\begin{subfigure}[t]{.48\columnwidth}
\caption{Minimum safe distance}
\begin{tabularx}{\columnwidth}{
	X
	X
	X
	X
	c
	}
\toprule
	   $\vr\left[\tfrac{m}{s}\right]$
	& $A\left[\tfrac{m}{s^2}\right]$
	& $b\left[\tfrac{m}{s^2}\right]$
	& $\varepsilon\left[s\right]$
	& $\mathbf{\norm{\pr - \po}}\left[m\right]$\\
\midrule
1    & 1    & 1    & 0.05   & \textbf{0.61}\\
0.5 & 0.5 & 0.5 & 0.025 & \textbf{0.28}\\
2    & 2    & 2    & 0.1     & \textbf{1.42}\\
1    & 1    & 2    & 0.05   & \textbf{0.33}\\
1    & 2    & 1    & 0.05   & \textbf{0.66}\\
\bottomrule
\end{tabularx}
\end{subfigure}
\hfill
\begin{subfigure}[t]{.48\columnwidth}
\caption{Maximum velocity through corridors and doors}
\label{tab:interpretation_static_maxvel}
\begin{tabularx}{\columnwidth}{
    p{1.3cm}
	X
	X
	X
	X
	}
\toprule
	& $A\left[\tfrac{m}{s^2}\right]$
	& $b\left[\tfrac{m}{s^2}\right]$
	& $\varepsilon\left[s\right]$
	& $\mathbf{\vr}\left[\tfrac{m}{s}\right]$\\
\midrule
\parbox[t]{2mm}{\multirow{5}{*}{\rotatebox[origin=c]{90}{\parbox{1.5cm}{\centering Corridor\\$\norm{\pr - \po}$\\ $= 1.25m$}}}}
& 1    & 1    & 0.05   & \textbf{1.48}\\
& 0.5 & 0.5 & 0.025 & \textbf{1.09}\\
& 2    & 2    & 0.1     & \textbf{1.85}\\
& 1    & 2    & 0.05   & \textbf{2.08}\\
& 2    & 1    & 0.05   & \textbf{1.43}\\
\midrule
\parbox[t]{2mm}{\multirow{5}{*}{\rotatebox[origin=c]{90}{\parbox[c]{1.5cm}{\centering Door\\$\norm{\pr - \po}$\\$ = 0.25m$}}}}
& 1    & 1    & 0.05   & \textbf{0.61}\\
& 0.5 & 0.5 & 0.025 & \textbf{0.47}\\
& 2    & 2    & 0.1     & \textbf{0.63}\\
& 1    & 2    & 0.05   & \textbf{0.85}\\
& 2    & 1    & 0.05   & \textbf{0.56}\\
\bottomrule
\end{tabularx}
\end{subfigure}
\end{table}
All except the third robot configuration (whose movement and acceleration capabilities outperform its reaction time) lead to a reasonable performance in in-door navigation environments.
\rref{fig:safetydist_static} plots the minimum safety distance that a specific robot configuration requires in order to avoid stationary obstacles, obtained from \eqref{eq:st-safe} by instantiating the parameters $A$, $b$, $\varepsilon$ and the current velocity $\vr$.

\rref{tab:interpretation_static_maxvel} turns the question around and lists concrete choices for these parameters and the resulting maximum safe velocity of the robot that \rref{eq:st-safe} identifies.

\paragraph{Moving obstacles}

\begin{table}[htb]
\small\sf\centering
\caption{Passive safety: minimum safe distance and maximum velocity for select configurations}
\begin{subfigure}[t]{.48\columnwidth}
\caption{Minimum safe distance}
\label{tab:interpretation_passive}
\begin{tabularx}{\columnwidth}{
	X
	X
	X
	X
	X
	c
	}
\toprule
	   $\vr\left[\tfrac{m}{s}\right]$
	& $A\left[\tfrac{m}{s^2}\right]$
	& $b\left[\tfrac{m}{s^2}\right]$
	& $V\left[\tfrac{m}{s}\right]$
	& $\varepsilon\left[s\right]$
	& $\mathbf{\norm{\pr - \po}}\left[m\right]$\\
\midrule
1    & 1    & 1    & 1    & 0.05   & \textbf{0.61}\\
0.5 & 0.5 & 0.5 & 0.5 & 0.025 & \textbf{0.28}\\
2    & 2    & 2    & 2    & 0.1     & \textbf{1.42}\\
1    & 1    & 2    & 1    & 0.05   & \textbf{0.33}\\
1    & 2    & 1    & 2    & 0.05   & \textbf{0.66}\\
\bottomrule
\end{tabularx}
\end{subfigure}
\hfill
\begin{subfigure}[t]{.48\columnwidth}
\caption{Maximum velocity through corridors and doors}
\label{tab:interpretation_passive_maxvel}
\begin{tabularx}{\columnwidth}{
    p{1.5cm}
	X
	X
	X
	X
	X
	}
\toprule
	& $A\left[\tfrac{m}{s^2}\right]$
	& $b\left[\tfrac{m}{s^2}\right]$
	& $V\left[\tfrac{m}{s}\right]$
	& $\varepsilon\left[s\right]$
	& $\mathbf{\vr}\left[\tfrac{m}{s}\right]$\\
\midrule
\parbox[t]{2mm}{\multirow{5}{*}{\rotatebox[origin=c]{90}{\parbox[c]{1.5cm}{\centering Corridor\\$\lVert \pr - \po \rVert$\\$ = 1.25m$}}}}
& 1    & 1    & 1    & 0.05   & \textbf{0.77}\\
& 0.5 & 0.5 & 0.5 & 0.025 & \textbf{0.69}\\
& 2    & 2    & 2    & 0.1     & \textbf{0.61}\\
& 1    & 2    & 1    & 0.05   & \textbf{0.4}\\
& 2    & 1    & 2    & 0.05   & \textbf{1.3}\\
\midrule
\parbox[t]{2mm}{\multirow{5}{*}{\rotatebox[origin=c]{90}{\parbox[c]{1.5cm}{\centering Door\\$\norm{\pr - \po}$\\$ = 0.25m$}}}}
& 1    & 1    & 1    & 0.05   & \textbf{0.12}\\
& 0.5 & 0.5 & 0.5 & 0.025 & \textbf{0.18}\\
& 2    & 2    & 2    & 0.1     & \textbf{0}\\
& 1    & 2    & 1    & 0.05   & \textbf{0.26}\\
& 2    & 1    & 2    & 0.05   & \textbf{1}\\
\bottomrule
\end{tabularx}	
\end{subfigure}
\end{table}

Below, we repeat the control constraint \eqref{eq:pf:3-2} from \rref{model:dynamicwindowpassive} for accelerating or choosing a new curve in the presence of movable obstacles.
The constraint introduces a new parameter $V$ for the maximum velocity of obstacles.

\begin{equation}
\norm{\pr - \po}{_\infty} > \frac{\vr^2}{2b} + V\frac{\vr}{b} + \left(\frac{A}{b}+1\right)\left(\frac{A}{2}\varepsilon^2+\varepsilon (\vr + V)\right)
\tag{\ref{eq:pf:3-2}*}
\end{equation}

\rref{fig:safetydist_passive} plots the minimum safety distance that the robot needs in order to maintain passive safety in the presence of moving obstacles.
The maximum velocity in presence of movable obstacles can drop to zero when the obstacles move too fast, the controller cycle time or the maximum acceleration force are too large, or when the maximum available braking force is too small.

\begin{figure}[htb]
\centering
\begin{subfigure}[b]{0.45\columnwidth}
\begin{tikzpicture}
\begin{axis}[mesh/ordering=y varies,
width=\textwidth,
xlabel=\(\vr\),
ylabel=\(\varepsilon\),
zlabel=\(\norm{\pr-\po}{_\infty}\),
xtick={0,1,...,5},
ytick={0,0.5,...,1},
ztick={0,10,...,30}
]
\addplot3[surf,fill=white,mesh/rows=11] file {./fig/data/2_moving/safedistvep.dat};
\end{axis}
\end{tikzpicture}
\caption{By velocity and control cycle}
\label{fig:safetydist_passive_v-ep}
\end{subfigure}
\quad
\begin{subfigure}[b]{0.45\columnwidth}
\centering
\begin{tikzpicture}
\begin{axis}[
width=.9\textwidth,
xmin=0, xmax=5.5,
ymin=0, ymax=21,
axis x line=middle,
axis y line=middle,
every axis x label/.style={
    at={(ticklabel* cs:1.05)},
    anchor=west,
},
every axis y label/.style={
    at={(ticklabel* cs:1.05)},
    anchor=south,
},
xlabel=\(\vr\),
ylabel=\(\norm{\pr-\po}{_\infty}\),
xtick={0,1,...,5},
ytick={0,5,...,20},
axis on top
]
\addplot[color=lsblue,solid] plot table {./fig/data/2_moving/safedist.dat};
\end{axis}
\end{tikzpicture}
\caption{By velocity}
\label{fig:safetydist_passive_v}
\end{subfigure}
\caption{Safety distance for passive safety}
\label{fig:safetydist_passive}
\end{figure}

\rref{fig:maxvel_static} compares the maximum velocity that the robot can travel in order to avoid stationary vs. moving obstacles.
The maximum velocity is obtained from \eqref{eq:st-safe} and from \eqref{eq:ps-safe} by instantiating the parameters $A$, $b$, $\varepsilon$ and the distance to the nearest obstacle $\norm{\pr - \po}$.
This way of reading the constraints \eqref{eq:st-safe}--\eqref{eq:ps-safe} makes it possible to adapt the maximal desired velocity of the robot safely based on the current spatial relationships.

\begin{figure}[htb]
\centering
\begin{subfigure}[b]{0.24\columnwidth}
\centering
\begin{tikzpicture}
\begin{axis}[
width=\textwidth,
xmin=1, xmax=5.5,
ymin=1.2, ymax=3.2,
axis x line=middle,
axis y line=middle,
every axis x label/.style={
    at={(ticklabel* cs:1.05)},
    anchor=west,
},
every axis y label/.style={
    at={(ticklabel* cs:1.05)},
    anchor=south,
},
xlabel=\(\norm{\pr-\po}{_\infty}\),
ylabel=\(\vr\),
xtick={1,...,5},
ytick={1.5,2,...,3},
axis on top
]
\addplot[color=lsblue,solid] plot table {./fig/data/1_static/maxvdist.dat};
\path (axis cs:1,0) -- (axis cs:5,0);
\end{axis}
\end{tikzpicture}
\caption{Static safety: by distance for $\varepsilon=0.05$}
\label{fig:maxvel_static_dist}
\end{subfigure}
\begin{subfigure}[b]{0.24\columnwidth}
\centering
\begin{tikzpicture}
\begin{axis}[
width=\textwidth,
xmin=0, xmax=1.1,
ymin=0, ymax=1.7,
axis x line=middle,
axis y line=middle,
every axis x label/.style={
    at={(ticklabel* cs:1.05)},
    anchor=west,
},
every axis y label/.style={
    at={(ticklabel* cs:1.05)},
    anchor=south,
},
xlabel=\(\varepsilon\),
ylabel=\(\vr\),
xtick={0,0.3,...,1},
ytick={0,0.5,...,1.5},
axis on top
]
\addplot[color=lsblue,solid] plot table {./fig/data/1_static/maxvep.dat};
\path (axis cs:0,0) -- (axis cs:1,0);
\end{axis}
\end{tikzpicture}
\caption{Static safety: by control cycle time for $\norm{\pr-\po}{_\infty}=1$}
\label{fig:maxvel_static_ep}
\end{subfigure}
\begin{subfigure}[b]{0.24\columnwidth}
\begin{tikzpicture}
\begin{axis}[
width=\textwidth,
xmin=1, xmax=5.5,
ymin=0, ymax=2.5,
axis x line=middle,
axis y line=middle,
every axis x label/.style={
    at={(ticklabel* cs:1.05)},
    anchor=west,
},
every axis y label/.style={
    at={(ticklabel* cs:1.05)},
    anchor=south,
},
xlabel=\(\norm{\pr-\po}{_\infty}\),
ylabel=\(\vr\),
xtick={1,...,5},
ytick={0.5,1,...,2},
axis on top
]
\addplot[color=lsblue,solid] plot table {./fig/data/2_moving/maxvdist.dat};
\path (axis cs:1,0) -- (axis cs:5,0);
\end{axis}
\end{tikzpicture}
\centering \caption{Passive safety: by distance for $\varepsilon=0.05$}
\label{fig:maxvel_passive_dist}
\end{subfigure}
\begin{subfigure}[b]{0.24\columnwidth}
\centering
\begin{tikzpicture}
\begin{axis}[
width=\textwidth,
xmin=0, xmax=1.1,
ymin=0, ymax=1.1,
axis x line=middle,
axis y line=middle,
every axis x label/.style={
    at={(ticklabel* cs:1.05)},
    anchor=west,
},
every axis y label/.style={
    at={(ticklabel* cs:1.05)},
    anchor=south,
},
xlabel=\(\varepsilon\),
ylabel=\(\vr\),
xtick={0,0.3,...,1},
ytick={0,0.5,...,1},
axis on top
]
\addplot[color=lsblue,solid] plot table {./fig/data/2_moving/maxvep.dat};
\path (axis cs:0,0) -- (axis cs:1,0);
\end{axis}
\end{tikzpicture}
\caption{Passive safety: by control cycle time for $\norm{\pr-\po}{_\infty}=1$}
\label{fig:maxvel_passive_ep}
\end{subfigure}
\caption{Comparison of safe velocities for static/passive safety with acceleration $A=1$ and braking $b=1$}
\label{fig:maxvel_static}
\end{figure}

\section{Monitoring for Compliance At Runtime}
\label{sec:monitoring}

The previous sections discussed models of obstacle avoidance control and of the physical behavior of ground robots in their environment, and we proved that these models are guaranteed to possess crucial safety and liveness properties.
The proofs present absolute mathematical evidence of the correctness of the models. 
If the models used for verification are an adequate representation of the real robot and its environment, these proofs transfer to the real system.
But any model necessarily deviates from the real system to some extent.

In this section, we discuss how to use ModelPlex \cite{DBLP:journals/fmsd/MitschP16} to bridge the gap between models and reality by verification.
The idea is to provably detect and safely respond to deviations between the model and the real robot in its environment by monitoring appropriate conditions at runtime.
ModelPlex complements offline proofs with runtime monitoring.
It periodically executes a \emph{monitor}, which is systematically synthesized from the verified models by an automatic proof of correctness, and checks input from sensors and output to actuators for compliance with the verified model.
If a deviation is detected, ModelPlex initiates a fail-safe action, e.g. stopping the robot or cutting its power to avoid actively running into obstacles, and, by that, ensure that \emph{safety proofs} from the model carry over to the real robot.
Of course, such fail-safe actions need to be triggered early enough to make sure the robot stops on time, which is what the monitors synthesized by ModelPlex ensure.

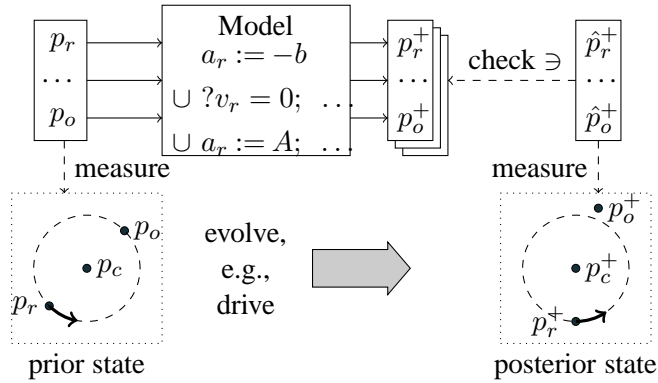
\begin{wrapfigure}{r}{.5\textwidth}
\centering
\begin{tikzpicture}

\begin{scope}[shift={(-1.2,-3.5)}]
\draw[color=black,dotted] (-0.5,-0.5) rectangle (1.5,1.5);
\draw[color=black,anchor=north] node at (0.5,-0.5) {prior state};
\draw[fill=lsblue] (1,1) circle (1.5pt);
\draw[color=black,anchor=west] node at (1,1) {$\po$};
\draw[fill=lsblue] (0,0) circle (1.5pt);
\node[color=black,anchor=east] at (0,0) {$\pr$};
\draw[color=black,dashed] (0.5,0.5) circle (0.707);
\draw[->,color=black,very thick] (0,0) arc (225:260:0.707);
\draw[fill=lsblue] (0.5,0.5) circle (1.5pt);
\draw[color=black,anchor=west] node at (0.5,0.5) {$\pc$};
\end{scope}

\begin{scope}[shift={(4.7,-3.5)}]
\draw[color=black,dotted] (-0.5,-0.5) rectangle (1.5,1.5);
\draw[color=black,anchor=north] node at (0.5,-0.5) {posterior state};
\draw[fill=lsblue] (0.8,1.3) circle (1.5pt);
\draw[color=black,anchor=west] node at (0.8,1.3) {$\po^+$};
\draw[fill=lsblue] (0.5,-0.207) circle (1.5pt);
\node[color=black,anchor=east] at (0.5,-0.207) {$\pr^+$};
\draw[color=black,dashed] (0.5,0.5) circle (0.707);
\draw[->,color=black,very thick] (0.5,-0.207) arc (270:305:0.707);
\draw[fill=lsblue] (0.5,0.5) circle (1.5pt);
\draw[color=black,anchor=west] node at (0.5,0.5) {$\pc^+$};
\end{scope}

\begin{scope}[shift={(2.3,-3)}]
\draw[color=black,anchor=east,text width=1.5cm,align=center] node at (0,0) {evolve, e.g., drive};
\draw[color=black,fill=black!20] (0,-0.25)--(0,0.25)--(1,0.25)--(1,0.35)--(1.3,0)--(1,-0.35)--(1,-0.25) -- cycle;
\end{scope}

\draw[color=black] (0,-1) rectangle (2.5,1);
\draw[color=black,anchor=north] node at (1.25,1) {Model};
\draw[color=black,text width=2cm,anchor=west] node at (0,0) {
\begin{minipage}{\textwidth}
\begin{align*}
& \humod{\ar}{-b}\\
\cup~ & \ptest{\vr=0};~\ldots\\
\cup~ & \humod{\ar}{A};~\ldots
\end{align*}
\end{minipage}
};

\draw[color=black] (-1.7,-0.8) rectangle (-1,0.8);
\draw[color=black,->] (-1,-0.5) -- (0,-0.5);
\draw[color=black,->] (-1,0) -- (0,0);
\draw[color=black,->] (-1,0.5) -- (0,0.5);
\draw[color=black,anchor=east] node at (-1,-0.5) {$\po$};
\draw[color=black,anchor=east] node at (-1,0) {\ldots};
\draw[color=black,anchor=east] node at (-1,0.5) {$\pr$};

\draw[color=black,fill=white] (3.2,-1) rectangle (3.8,0.6);
\draw[color=black,fill=white] (3.1,-0.9) rectangle (3.7,0.7);
\draw[color=black,fill=white] (3,-0.8) rectangle (3.6,0.8);
\draw[color=black,->] (2.5,-0.5) -- (3,-0.5);
\draw[color=black,->] (2.5,0) -- (3,0);
\draw[color=black,->] (2.5,0.5) -- (3,0.5);
\draw[color=black,anchor=west] node at (3,-0.5) {$\po^+$};
\draw[color=black,anchor=west] node at (3,0) {\ldots};
\draw[color=black,anchor=west] node at (3,0.5) {$\pr^+$};

\draw[color=black,fill=white] (5.5,-0.8) rectangle (6.1,0.8);
\draw[color=black,anchor=west] node at (5.5,-0.5) {$\hat{\po}^+$};
\draw[color=black,anchor=west] node at (5.5,0) {\ldots};
\draw[color=black,anchor=west] node at (5.5,0.5) {$\hat{\pr}^+$};

\draw[color=black,->,dashed] (5.5,0) -- (3.8,0);
\draw[color=black,anchor=south] node at (4.7,0) {check $\ni$};

\draw[color=black,dashed,->] (-1.3,-0.8) -- (-1.3,-2);
\draw[color=black,dashed,->] (5.8,-0.8) -- (5.8,-2);
\draw[color=black,anchor=west] node at (-1.3,-1.5) {measure};
\draw[color=black,anchor=east] node at (5.8,-1.5) {measure};
\end{tikzpicture}
\caption{The principle behind a ModelPlex monitor: can the model reproduce or explain the observed real-world behavior?}
\label{fig:modelplexillustration}
\end{wrapfigure}

A monitor checks the actual evolution of the real robot implementation to discover failures and mismatches with the verified model. 
The acceleration chosen by the robot's control software implementation must fit to the current situation.
For example, accelerate only when the verified model considers it safe.
And the chosen curve must fit to the current orientation.
No unintended change to the robot's speed, position, orientation has happened, and no violations of the assumptions about the obstacles have occurred.
This means, any variable that is allowed to change in the model must be monitored.
In the examples here, these variables include the robot's position $\pr$, longitudinal speed $\vr$, rotational speed $\omegar$, acceleration $\ar$, orientation $\dr$, curve $\rcurve$, obstacle position $\po$ and velocity $\vo$.

A ModelPlex monitor is designed for periodic sampling.
For each variable there will be two observed values, one from the previous sample time (for example, previous robot position $\pr$) and one from the current sample time (for example, next robot position $\pr^+$).
It is not important for ModelPlex that the values are measured exactly at the sampling period, but merely that there is an upper bound $\varepsilon$ on the amount of time that passed between two samples.
A ModelPlex monitor checks in a provably correct way whether the evolution observed in the difference of the sampled values can be explained by the model.
If it does, the current behavior fits to a verified behavior and is, thus, safe.
If it does not, the situation may have become unsafe and a fail-safe action is initiated to mitigate safety hazards.

\rref{fig:modelplexillustration} illustrates the principle behind a ModelPlex monitor.
The values from the previous sample time serve as starting state for executing the model. The values produced by executing the model are then compared to the values observed in the current sample time by the monitor.

\begin{figure*}
\begin{align}
\textit{monitor} & \equiv \textit{mon}_o \wedge \textit{mon}_\textit{dyn} \wedge \left(\textit{mon}_b \vee \textit{mon}_s \vee \textit{mon}_a\right) \label{eq:monitor-all}\\
\textit{mon}_o & \equiv \lVert \vo^+ \rVert \leq V \label{eq:monitor-obstacle}\\
\textit{mon}_\textit{dyn} & \equiv 0 \leq \varepsilon \wedge \vr \geq 0 \wedge t^+ = 0 \label{eq:monitor-dyn}\\
\textit{mon}_b & \equiv \po^+=\po \wedge \pr^+ = \pr \wedge \dr^+ = \dr \wedge \vr^+ = \vr \wedge \omegar^+ = \omegar \wedge \ar^+ = -b \wedge \rcurve^+ = \rcurve \label{eq:monitor-brake}\\
\textit{mon}_s & \equiv \vr = 0 \wedge \po^+ = \po \wedge \pr^+ = \pr \wedge \dr^+ = \dr \wedge \vr^+ = \vr \wedge \omegar^+ = 0 \wedge \ar^+ = 0 \wedge \rcurve^+ = \rcurve \label{eq:monitor-stay}\\
\textit{mon}_a & \equiv \ar^+ = A \wedge \rcurve^+ \neq 0 \wedge \omegar^+ \rcurve^+ = \vr \wedge \pr^+ = \pr \wedge \dr^+ = \dr \wedge \vr^+ = \vr \label{eq:monitor-acc}\\
& \phantom{\equiv} \wedge \lVert \pr - \po^+ \rVert_\infty > \frac{\vr^2}{2 b} + V \frac{\vr}{b} + \left(\frac{A}{b}+1\right) \left(\frac{A}{2} \varepsilon^2 + \varepsilon (\vr+V) \right) \label{eq:monitor-safedist}
\end{align}
\caption{Synthesized safety conditions.
The generated monitor captures conditions on obstacles $\textit{mon}_o$, on dynamics $\textit{mon}_\textit{dyn}$, and on the robot's decisions on braking $\textit{mon}_b$, staying stopped $\textit{mon}_s$, and accelerating $\textit{mon}_a$.
The monitor distinguishes two observed values per variable, separated by a controller run (for example, $\pr$ denotes the position before running the controller, whereas $\pr^+$ denotes the position after running the controller).
\label{fig:safety}}
\end{figure*}

The verified models themselves are too slow to execute, because they involve nondeterminism and differential equations.
Hence, provably correct monitor expressions in real arithmetic are synthesized from a model using an offline proof in \KeYmaeraX.
These expressions capture the behavior of the models, projected onto the pairwise comparisons of sampled values that are needed at runtime.

Monitor techniques for model compliance monitors and state prediction monitors are reported in \cite{DBLP:journals/fmsd/MitschP16}.
Here, we focus on a controller monitor expression synthesized from \rref{model:dynamicwindowpassive}, which captures all possible control decisions of the robot that are verified to be safe.
The controller monitor checks the decisions of an (unverified) controller implementation for consistency with the verified discrete model without differential equations.
ModelPlex automatically extracts the discrete model by a proof with the ordinary differential equation (ODE) being conservatively over-approximated by its evolution domain.
The resulting condition $\textit{monitor}$ \eqref{eq:monitor-all} in \rref{fig:safety}, which is synthesized by an automatic proof in \KeYmaeraX, mimics the structure of the model: it captures the assumptions on the obstacle $\textit{mon}_o$, the evolution domain from dynamics $\textit{mon}_\textit{dyn}$, as well as the specification for each of the three controller branches (braking $\textit{mon}_b$, staying stopped $\textit{mon}_s$, or accelerating $\textit{mon}_a$).

The obstacle monitor part $\textit{mon}_o$ in \eqref{eq:monitor-obstacle}, checks that the measured obstacle velocity $\vo^+$ must not exceed the assumptions made in the model about their maximum velocity.
The dynamics monitor part $\textit{mon}_\textit{dyn}$ in  \eqref{eq:monitor-dyn} checks the evolution domain of the ODE and that the controller did reset its clock ($t^+=0$).
The braking monitor $\textit{mon}_b$ in \eqref{eq:monitor-brake} defines that emergency brakes can only hit the brakes and do not change anything else (acceleration $\ar^+ = -b$, while everything else is of the form $x^+=x$ meaning that no change is allowed).
Monitor $\textit{mon}_s$ in \eqref{eq:monitor-stay} expresses that staying stopped is possible if the speed is zero ($\vr=0$) and the controller must have chosen no acceleration and no rotation ($\ar=0$ and $\omegar=0$), while everything else is unchanged.
Finally, the acceleration monitor $\textit{mon}_a$ in \eqref{eq:monitor-acc}--\eqref{eq:monitor-safedist} says that, if the distance is safe, the robot can choose maximum acceleration $\ar^+ = A$, a new non-spinning steering $\rcurve^+ \neq 0$ that fits to the current speed $\omegar^+ \rcurve^+ = \vr$; position, orientation, and speed must not be set by the controller (those follow from the acceleration and steering choice).

\section{Conclusion and Future Work}
\label{sec:conclusion}

Robots are modeled by hybrid systems, because they share continuous physical motion with advanced computer algorithms controlling their behavior.
We demonstrate that this understanding also helps proving robots safe.
We develop hybrid system models for collision avoidance algorithms for autonomous ground vehicles and prove that the algorithms guarantee static safety for static obstacles and both passive safety and passive friendly safety in the presence of moving obstacles.

We augment the models and safety proofs with robustness for localization uncertainty and imperfect actuation.
Incremental revision of models and proofs helps reducing the verification complexity, since in lower-fidelity models the safety-critical effects of computations on the physical behavior are easier to predict and characterize in control conditions. 
Additional details can then be understood incrementally as extensions to these previously found control conditions (\eg, it helps to first understand the safety margins for perfect sensing, and later add the impact of uncertainty on the behavior of the robot).
All parameters in our models---such as those for maximum obstacle velocity and sensor/actuator uncertainty---are fully symbolic and can be instantiated arbitrarily, including bounds from probabilistic models (\eg, assume the $2\sigma$ confidence interval of the distribution of obstacle velocities as maximum obstacle velocity).
In this case, our verified safety guarantees translate into a probability of safety.

Theorems~\ref{thm:staticsafety}--\ref{thm:passivesafety-nonsync}, \ref{thm:waypoint}, and \ref{thm:cx-liveness} were proved with significant automation and succinct reusable proof tactics in the \dL theorem prover \KeYmaeraX.
All theorems were additionally proved in its predecessor \KeYmaera.
The most important insight in all proofs were the loop and differential invariants.
Some proofs needed simple insights on how to eliminate variables to reduce the complexity of the resulting arithmetic.

\begin{wraptable}{r}{.55\textwidth}
\caption{Proof statistics in \KeYmaeraX}\label{tab:proofstatistics}
\begin{tabularx}{.55\columnwidth}{Xr@{\hspace{.2em}}r@{\hspace{.4em}}r@{\hspace{.4em}}r@{\hspace{.2em}}r}
\toprule
\textbf{Theorem} & \multicolumn{2}{c}{\textbf{Tactic size}} & \multirowcell{2}{\textbf{Proof}\\ \textbf{steps}} & \multicolumn{2}{c}{\textbf{Time} $[s]$} \\
\cmidrule(r){2-3}\cmidrule(lr){5-6}
& LOC & Steps & & QE & Total\\
\midrule
\multicolumn{6}{l}{\textbf{Safety proofs}}\\
\ref{thm:staticsafety}: Static
& 12 & 71 & 30355 & 74 & 89
\\
\ref{thm:passivesafety}: Passive
& 12 & 73 & 51956 & 229 & 268
\\
\ref{thm:passivefriendlysafety}: Passive-friendly
& 45 & 140 & 68620 & 342 & 407
\\
\ref{thm:passiveorientationsafety}: Orientation
& 15 & 108 & 173989 & 934 & 1006
\\
\midrule
\multicolumn{6}{l}{\textbf{Passive safety extensions}}\\
\ref{thm:passivesafetyactuala}: Acceleration 
& 16 & 84 & 67604 & 405 & 463
\\
\ref{thm:passivesafety-locationuncertainty}: Location
& 12 & 73 & 57775 & 445 & 485
\\
\ref{thm:passivesafety-actuatorperturbation}: Perturbation
& 21 & 120 & 56297 & 254 & 299
\\
\ref{thm:passivesafety-velocityuncertainty}: Velocity
& 12 & 94 & 54601 & 359 & 404
\\
\ref{thm:passivesafety-nonsync}: Async Control
& 42 & 122 & 61772 & 284 & 335
\\
\midrule
\multicolumn{6}{l}{\textbf{Liveness proofs}}
\\
\ref{thm:waypoint}: Reach waypoint
& 32 & 93 & 46530 & 69 & 125
\\
\ref{thm:cx-liveness}: Pass intersection
& 234 & 440 & 61878 & 83 & 182
\\
\bottomrule
\end{tabularx}
\end{wraptable}

Overall, the tactics follow a similar structure across all theorems, with only minor variation.
The tactics use proof automation for symbolic execution of programs, for proving differential invariants, and for simplifying arithmetic, which all perform a large number of internal steps automatically to turn the proof hints provided by users into actual proofs.
\rref{tab:proofstatistics} summarizes the proof statistics: the tactic size characterizes the manual effort of the user (mostly proof hints on differential invariants and minor arithmetic simplifications), while the proof steps are the corresponding internal steps taken by \KeYmaeraX to fill in gaps in the proof hints with automated proof search and justify the proof from the axioms of \dL.
As a performance indicator, we list the total time needed to run the proofs on a $2.4\text{GHz}$ Intel Core i7 with $16\text{GB}$ memory, most of which is spent in external tools for handling real arithmetic with quantifier elimination (QE time column).

As part of the verification activity, we identified crucial safety constraints that have to be satisfied in order to choose a new curve or accelerate safely.
These constraints are entirely symbolic and summarized in \rref{tab:constraintsummary}.
The static safety invariant is equivalent to the admissible velocities identified in \cite{DBLP:journals/ram/FoxBT97}, which assumes instantaneous control. 
Our proofs identified the invariants required for safety in the presence of moving obstacles, sensor uncertainty and coverage, and actuator perturbation, as well as the additional margins in column ``safe control'' that account for the reaction time of the robot.
When instantiated with concrete numerical values of a robot design, these safety constraints can be used for design decision tradeoffs and to get an intuition about how conservative or aggressive our robot can drive, such as: 
\begin{itemize}[noitemsep]
\item how fast can the robot pass through a narrow door?
\item how fast can the robot drive on a given corridor?
\end{itemize}

The analyzed constraints for common values of acceleration force, braking force, control cycle time, and obstacle distance (\ie, door width, corridor width) illustrate that the verified collision avoidance protocol is suitable for indoor navigation at reasonable speeds.

\begin{sidewaystable}
\small\sf\centering
\caption{Invariant and safety constraint summary. Safe control margins account for the reaction time of the robot.}
\label{tab:constraintsummary}
\begin{tabularx}{\textheight}{Xr@{\hspace{.2em}}l}
\toprule
\textbf{Safety} & \textbf{Invariant} & $+$\textbf{Safe Control} \\
\midrule
Static \hfill(\rref{model:dynamicwindowstatic}, \rref{thm:staticsafety}) & $\norm{\pr - \po}{_\infty} > \frac{\vr^2}{2b}$ & $+\left(\frac{A}{b}+1\right)\left(\frac{A}{2}\varepsilon^2 + \varepsilon \vr\right)$ \\
Passive \hfill(\rref{model:dynamicwindowpassive}, \rref{thm:passivesafety}) & $\vr \neq 0 \limply \norm{\pr - \po}{_\infty} > \frac{\vr^2}{2b} + V\frac{\vr}{b}$ & $+ \left(\frac{A}{b}+1\right)\left(\frac{A}{2}\varepsilon^2 + \varepsilon (\vr + V)\right)$
\\
Passive friendly \hfill(\rref{model:dynamicwindowpassivefriendly2}+\ref{model:obstacle}, \rref{thm:passivefriendlysafety}) & $\norm{\pr - \po}{_\infty} > \frac{\vr^2}{2b} + V\left(\frac{\vr}{b} + \tau\right) + \frac{V^2}{2b_o}$ & $+ \left(\frac{A}{b}+1\right)\left(\frac{A}{2}\varepsilon^2+\varepsilon(\vr+V)\right)$\\
Passive orientation \hfill(\rref{model:fieldOfView}, \rref{thm:passiveorientationsafety}) & $\inView > 0 \limply \norm{\pr - \po}{_\infty} > \frac{\vr^2}{2b} + V\frac{\vr}{b}$ & $+\left(\frac{A}{b}+1\right)\left(\frac{A}{2}\varepsilon^2 + \varepsilon(\vr+V)\right)$\\ 
& and $\gamma \abs{r} > \frac{\vr^2}{2b}$ & $+ \left(\frac{A}{b}+1\right)\left(\frac{A}{2}\varepsilon^2 + \varepsilon \vr\right)$\\
\midrule
Extensions (passive safety examples)\\
\hspace{1em} with actual acceleration \hfill(\rref{model:dynamicwindowpassiveactuala}, \rref{thm:passivesafetyactuala}) & $\vr \neq 0 \limply \norm{\pr - \po}{_\infty} > 
\begin{cases}
-\frac{\vr^2}{2\ar} - V\frac{\vr}{\ar} \\
\frac{\vr^2}{2b} + V\frac{\vr}{b}
\end{cases}
$ & $
\begin{aligned}
& \quad \text{if } \vr+\ar\varepsilon < 0\\
+ \left(\frac{\ar}{b}+1\right)\left(\frac{\ar}{2}\varepsilon^2 + \varepsilon(\vr + V)\right) & \quad \text{otherwise}
\end{aligned}
$
\\
\hspace{1em} $+$ location uncertainty \hfill(\rref{model:positionuncertainty}, \rref{thm:passivesafety-locationuncertainty}) & $\vr \neq 0 \limply \norm{\hat{\pr} - \po}{_\infty} > \frac{\vr^2}{2b} + V\frac{\vr}{b}$ & $+ \left(\frac{A}{b}+1\right)\left(\frac{A}{2}\varepsilon^2 + \varepsilon (\vr + V)\right) + \Delta_p$ \\
\hspace{1em} $+$ actuator perturbation \hfill(\rref{model:motionuncertainty}, \rref{thm:passivesafety-actuatorperturbation}) & $\vr \neq 0 \limply \norm{\pr - \po}{_\infty} > \frac{\vr^2}{2b\Delta_a} + V\frac{\vr}{b\Delta_a}$ & $+ \left(\frac{A}{b\Delta_a}+1\right)\left(\frac{A}{2}\varepsilon^2 + \varepsilon (\vr + V)\right)$ \\
\hspace{1em} $+$ velocity uncertainty \hfill(\rref{model:velocityuncertainty}, \rref{thm:passivesafety-velocityuncertainty}) & $\vr \neq 0 \limply \norm{\pr - \po}{_\infty} > \frac{(\hat{\vr} + \Delta_v)^2}{2b} + V\frac{\hat{\vr}+\Delta_v}{b}$ & $+\left(\frac{A}{b}+1\right)\left(\frac{A}{2}\varepsilon^2+\varepsilon(\hat{\vr}+\Delta_v+V)\right)$\\
\hspace{1em} asynchronous robot and obstacle control \hfill(\rref{model:mchange1}, \rref{thm:passivesafety-nonsync}) & see Passive safety\\
\hspace{1em} arbitrary many obstacles \hfill(\rref{model:multipleObstacles}, \rref{thm:passivesafety-qdl}) & $\vr \neq 0 \limply \forall i{\in}O\, \norm{\pr - \po(i)}{_\infty} > \frac{\vr^2}{2b} + V\frac{\vr}{b}$ & $+ \left(\frac{A}{b}+1\right)\left(\frac{A}{2}\varepsilon^2 + \varepsilon (\vr + V(i))\right)$
\\
\bottomrule
\end{tabularx}
\end{sidewaystable}

Future work includes exploiting more kinematic capabilities (\eg, going sideways with omni-drive) and explicit accounts for distance measurement uncertainty, which is, however, easier than location uncertainty.

\paragraph{Acknowledgments}
This material is based upon work supported by NSF CAREER Award CNS-1054246, NSF EXPEDITION CNS-0926181, NSF CNS-1446712, by DARPA FA8750-12-2-0291, AFOSR FA9550-16-1-0288, and by Bosch.
This project is funded in part by Carnegie Mellon University's Technologies for Safe and Efficient Transportation, the National USDOT University Transportation Center for Safety (T-SET UTC) which is sponsored by the US Department of Transportation.
This work was also supported by the Austrian BMVIT under grant FIT-IT 829598, FFG BRIDGE 838526, and FFG Basisprogramm 838181.

\bibliographystyle{plainnat}
\bibliography{references}

\end{document}